 %
\input harvmac
\input epsf.tex
%
%
 %
\catcode`@=11
\def\rlx{\relax\leavevmode}                   
 %
 %
 %
\font\eightrm=cmr8 \font\eighti=cmmi8 \font\eightsy=cmsy8
\font\eightbf=cmbx8 \font\eightit=cmti8 \font\eightsl=cmsl8
\skewchar\eighti='177 \skewchar\eightsy='60
\def\eightpoint{\def\rm{\fam0\eightrm}
\textfont0=\eightrm \scriptfont0=\sixrm \scriptscriptfont0=\fiverm
\textfont1=\eighti \scriptfont1=\sixi \scriptscriptfont1=\fivei
\textfont2=\eightsy \scriptfont2=\sixsy \scriptscriptfont2=\fivesy
\textfont\itfam=\eighti
\def\it{\fam\itfam\eightit}\def\sl{\fam\slfam\eightsl}%
\textfont\bffam=\eightbf \def\bf{\fam\bffam\eightbf}\rm}
 %
\font\tenmib=cmmib10
\font\sevenmib=cmmib10 at 7pt 
\font\fivemib=cmmib10 at 5pt  
\font\tenbsy=cmbsy10
\font\sevenbsy=cmbsy10 at 7pt 
\font\fivebsy=cmbsy10 at 5pt  
\def\BMfont{\textfont0\tenbf \scriptfont0\sevenbf
                              \scriptscriptfont0\fivebf
            \textfont1\tenmib \scriptfont1\sevenmib
                               \scriptscriptfont1\fivemib
            \textfont2\tenbsy \scriptfont2\sevenbsy
                               \scriptscriptfont2\fivebsy}
\def\BM#1{\rlx\ifmmode\mathchoice
                      {\hbox{$\BMfont#1$}}
                      {\hbox{$\BMfont#1$}}
                      {\hbox{$\scriptstyle\BMfont#1$}}
                      {\hbox{$\scriptscriptstyle\BMfont#1$}}
                 \else{$\BMfont#1$}\fi}
 %
 %
 %
 %
\def\inbar{\vrule height1.5ex width.4pt depth0pt}
\def\sinbar{\vrule height1ex width.35pt depth0pt}
\def\ssinbar{\vrule height.7ex width.3pt depth0pt}
\font\cmss=cmss10
\font\cmsss=cmss10 at 7pt
\def\ZZ{\rlx\leavevmode
             \ifmmode\mathchoice
                    {\hbox{\cmss Z\kern-.4em Z}}
                    {\hbox{\cmss Z\kern-.4em Z}}
                    {\lower.9pt\hbox{\cmsss Z\kern-.36em Z}}
                    {\lower1.2pt\hbox{\cmsss Z\kern-.36em Z}}
               \else{\cmss Z\kern-.4em Z}\fi}
\def\Ik{\rlx{\rm I\kern-.18em k}}  
\def\IC{\rlx\leavevmode
             \ifmmode\mathchoice
                    {\hbox{\kern.33em\inbar\kern-.3em{\rm C}}}
                    {\hbox{\kern.33em\inbar\kern-.3em{\rm C}}}
                    {\hbox{\kern.28em\sinbar\kern-.25em{\sevenrm C}}}
                    {\hbox{\kern.25em\ssinbar\kern-.22em{\fiverm C}}}
             \else{\hbox{\kern.3em\inbar\kern-.3em{\rm C}}}\fi}
\def\IP{\rlx{\rm I\kern-.18em P}}
\def\IR{\rlx{\rm I\kern-.18em R}}
\def\Ione{\rlx{\rm 1\kern-2.7pt l}}
 %
 %

 %

\def\intem#1{\par\leavevmode%
              \llap{\hbox to\parindent{\hss{#1}\hfill~}}\ignorespaces}
 %


 %
\newskip\humongous \humongous=0pt plus 1000pt minus 1000pt   
\def\caja{\mathsurround=0pt}
\newif\ifdtup
 %
\def\cmath#1{\,\vcenter{\openup2\jot \caja
     \ialign{\strut \hfil$\displaystyle{##}$\hfil\crcr#1\crcr}}\,}
 %
\def\eqalign#1{\,\vcenter{\openup2\jot \caja
     \ialign{\strut \hfil$\displaystyle{##}$&$
      \displaystyle{{}##}$\hfil\crcr#1\crcr}}\,}
 %
\def\twoeqsalign#1{\,\vcenter{\openup2\jot \caja
     \ialign{\strut \hfil$\displaystyle{##}$&$
      \displaystyle{{}##}$\hfil&\hfill$\displaystyle{##}$&$
       \displaystyle{{}##}$\hfil\crcr#1\crcr}}\,}
 %
\def\panorama{\global\dtuptrue \openup2\jot \caja
     \everycr{\noalign{\ifdtup \global\dtupfalse
      \vskip-\lineskiplimit \vskip\normallineskiplimit
      \else \penalty\interdisplaylinepenalty \fi}}}
 %

 %
\def\eqalignno#1{\panorama \tabskip=\humongous
     \halign to\displaywidth{\hfil$\displaystyle{##}$
      \tabskip=0pt&$\displaystyle{{}##}$\hfil
       \tabskip=\humongous&\llap{$##$}\tabskip=0pt\crcr#1\crcr}}
 %
\def\eqalignnotwo#1{\panorama \tabskip=\humongous
     \halign to\displaywidth{\hfil$\displaystyle{##}$
      \tabskip=0pt&$\displaystyle{{}##}$
       \tabskip=0pt&$\displaystyle{{}##}$\hfil
        \tabskip=\humongous&\llap{$##$}\tabskip=0pt\crcr#1\crcr}}
 %
\def\twoeqsalignno#1{\panorama \tabskip=\humongous
     \halign to\displaywidth{\hfil$\displaystyle{##}$
      \tabskip=0pt&$\displaystyle{{}##}$\hfil
       \tabskip=0pt&\hfil$\displaystyle{##}$
        \tabskip=0pt&$\displaystyle{{}##}$\hfil
         \tabskip=\humongous&\llap{$##$}\tabskip=0pt\crcr#1\crcr}}
 %

 %
 %
 %
 %
   \let\SS=\S       
\let\ii=\i          
\def\,{\hskip1.5pt}           
 %
\let\a=\alpha
\let\b=\beta
\let\c=\chi
\let\d=\delta       \let\vd=\partial             \let\D=\Delta
\let\e=\epsilon     
\let\f=\phi         \let\vf=\varphi              \let\F=\Phi

\let\i=\iota
\let\j=\psi                                      \let\J=\Psi
\let\k=\kappa
\let\l=\lambda                                   \let\L=\Lambda
\let\m=\mu
\let\n=\nu
\let\p=\pi          \let\vp=\varpi               \let\P=\Pi
\let\q=\theta       \let\vq=\vartheta            \let\Q=\Theta
\let\r=\rho         \let\vr=\varrho
\let\s=\sigma       \let\vs=\varsigma            \let\S=\Sigma
\let\t=\tau
\let\w=\omega                                    \let\W=\Omega
\let\x=\xi                                       \let\X=\Xi
\let\y=\upsilon                                  \let\Y=\Upsilon
\let\z=\zeta
 %
 %
\def\Box{{\sqcap\mkern-12mu\sqcup}}
\def\lapp{\lower.4ex\hbox{\rlap{$\sim$}} \raise.4ex\hbox{$<$}}
\def\gapp{\lower.4ex\hbox{\rlap{$\sim$}} \raise.4ex\hbox{$>$}}
\def\con{\ifmmode\raise.1ex\hbox{\bf*}
          \else\raise.1ex\hbox{\bf*}\fi}
\let\iff=\leftrightarrow
\let\Iff=\Leftrightarrow
\let\from=\leftarrow
\let\To=\Rightarrow

\def\dual{\relax\leavevmode\lower.9ex\hbox{\titlerms*}}
\def\define{\buildrel\rm def\over =}
\let\id=\equiv
\let\8=\otimes
 %
 %
 %
 %
\let\ba=\overline
\let\2=\underline

\let\Tw=\widetilde
 %
\def\dt#1{{\buildrel{\smash{\lower1pt\hbox{.}}}\over{#1}}}
\def\pd#1#2{{\partial#1\over\partial#2}}

\def\6(#1){\relax\leavevmode\hbox{\eightrm(}#1\hbox{\eightrm)}}
\def\0#1{\relax\ifmmode\mathaccent"7017{#1}     
                \else\accent23#1\relax\fi}      
\def\7#1#2{{\mathop{\null#2}\limits^{#1}}}      
\def\5#1#2{{\mathop{\null#2}\limits_{#1}}}      
 %
\def\bra#1{\left\langle #1\right|}
\def\ket#1{\left| #1\right\rangle}
\def\V#1{\langle#1\rangle}

 %

 %

 %

 %
\newbox\t@b@x
\def\rightarrowfill{$\m@th \mathord- \mkern-6mu
     \cleaders\hbox{$\mkern-2mu \mathord- \mkern-2mu$}\hfill
      \mkern-6mu \mathord\rightarrow$}
\def\tooo#1{\setbox\t@b@x=\hbox{$\scriptstyle#1$}%
             \mathrel{\mathop{\hbox to\wd\t@b@x{\rightarrowfill}}%
              \limits^{#1}}\,}
\def\leftarrowfill{$\m@th \mathord\leftarrow \mkern-6mu
     \cleaders\hbox{$\mkern-2mu \mathord- \mkern-2mu$}\hfill
      \mkern-6mu \mathord-$}
\def\froo#1{\setbox\t@b@x=\hbox{$\scriptstyle#1$}%
             \mathrel{\mathop{\hbox to\wd\t@b@x{\leftarrowfill}}%
              \limits^{#1}}\,}
 %
\def\frac#1#2{{#1\over#2}}
\def\frc#1#2{\relax\ifmmode{\textstyle{#1\over#2}} 
                    \else$#1\over#2$\fi}           
\def\inv#1{\frc{1}{#1}}                            
 %
\def\Claim#1#2#3{\bigskip\begingroup%
                  \xdef #1{\secsym\the\meqno}%
                   \writedef{#1\leftbracket#1}%
                    \global\advance\meqno by1\wrlabeL#1%
                     \noindent{\bf#2}\,#1{}\,:~\sl#3\vskip1mm\endgroup}

\def\QED{\rlx\hfill$\Box$\kern-7pt\raise3pt\hbox{$\surd$}\bigskip}
 %
 %
\def\1{\raise1pt\hbox{,}}     

\def\:{\buildrel!\over=}

\def\CP#1{\rlx\ifmmode\IP^{#1}\else\IP$^{#1}$\fi}
\def\cP#1{\rlx\ifmmode\IC{\rm P}^{#1}\else$\IC{\rm P}^{#1}$\fi}

\def\sll#1{\rlx\rlap{\,\raise1pt\hbox{/}}{#1}}
\def\Sll#1{\rlx\rlap{\,\kern.6pt\raise1pt\hbox{/}}{#1}\kern-.6pt}

\let\SSS=\scriptstyle
\let\ttt=\textstyle

 %
\def\eg{\hbox{\it e.g.}}        
\def\ie{\hbox{\it i.e.}}        
\def\etc{\hbox{\it etc.}}       
\def\topic#1{\bigskip\noindent$\2{\hbox{#1}}$\nobreak\vglue0pt%
              \noindent\ignorespaces}

\def\CY{Calabi-\kern-.2em Yau}

\def\3{\ifmmode\ldots\else$\ldots$\fi}
\def\\{\hfill\break}
\def\Z{\hfil\break\rlx\hbox{}\quad}
\def\3{\ifmmode\ldots\else$\ldots$\fi}

 %
 %
\def\I#1{{\it ibid.\,}{\bf#1\,}}

\def\PR#1{{\it Phys.\,Rev.\,}{\bf#1\,}}
\def\NP#1{{\it Nucl.\,Phys.\,}{\bf#1\,}}
\def\PL#1{{\it Phys.\,Lett.\,}{\bf#1\,}}
\def\PRp#1{{\it Phys.\,Rep.\,}{\bf#1\,}}

\def\MPL#1{{\it Mod.\,Phys.\,Lett.\,}{\bf#1\,}}
\def\PRL#1{{\it Phys.\,Rev.\,Lett.\,}{\bf#1\,}}

\def\IJMP#1{{\it Int.\,J.\,Mod.\,Phys.\,}{\bf#1\,}}
\baselineskip=13.0861pt plus2pt minus1pt
\parskip=\medskipamount
\let\ft=\foot
\noblackbox
 %
\def\Afour{\ifx\answ\bigans
            \hsize=16.5truecm\vsize=24.7truecm
             \else
              \hsize=24.7truecm\vsize=16.5truecm
               \fi}
 %
 %
\def\SaveTimber{\abovedisplayskip=1.5ex plus.3ex minus.5ex
                \belowdisplayskip=1.5ex plus.3ex minus.5ex
                \abovedisplayshortskip=.2ex plus.2ex minus.4ex
                \belowdisplayshortskip=1.5ex plus.2ex minus.4ex
                \baselineskip=12pt plus1pt minus.5pt
 \parskip=\smallskipamount
 \def\ft##1{\unskip\,\begingroup\footskip9pt plus1pt minus1pt\setbox%
             \strutbox=\hbox{\vrule height6pt depth4.5pt width0pt}%
              \global\advance\ftno by1
               \footnote{$^{\the\ftno)}$}{\ninepoint##1}%
                \endgroup}
 \def\listrefs{\footatend\vfill\supereject\immediate\closeout\rfile%
                \writestoppt\baselineskip=10pt%
                 \centerline{{\bf References}}%
                  \bigskip{\frenchspacing\parindent=20pt\escapechar=` %
                   \rightskip=0pt plus4em\spaceskip=.3333em%
                    \input refs.tmp\vfill\eject}\nonfrenchspacing}}
\catcode`@=12
%
%
%
 %
\SaveTimber  
 %
 %
\def\rd{{\rm d}}
\def\hc{\hbox{\it h.c.}}
\def\Zp{\big|}
\def\Rf{\big\rfloor}
\def\Rc{\big\rceil}
\def\@{{\ttt\cdot}}
\def\pp{{\mathchar'75\mkern-9mu|\mkern3mu}} 
\def\mm{{=}}                                
\def\dvd{\!{\mathop{\smash\partial\vrule height1.3ex width0pt}%
          \limits^{\,_{\SSS\iff}}}\mkern-4mu}
\def\]#1{\mkern#10mu}
\def\[#1{\mkern-#10mu}
\def\ab{{\bar\alpha}}
\def\bb{{\bar\beta}}
\def\Bb{{\bar B}}
\def\cb{{\bar\chi}}
\def\db{{\bar\partial}}

\def\fb{{\bar\phi}}

\def\jb{{\bar\j}}
\def\lb{\bar\lambda}
\def\bl{{\mathchar"16\mkern-10mu\lambda}} 
\def\mb{{\bar\mu}}
\def\nb{{\bar\nu}}

\def\bp{\bar p}
\def\pb{\bar\p}
\def\pB{\relax\leavevmode\hbox{$\BMfont p$\kern-.45em
               \vrule height1.35ex depth-1.25ex width4pt}\kern1pt}
\def\qB{\relax\leavevmode\hbox{$\BMfont q$\kern-.45em
               \vrule height1.35ex depth-1.25ex width4pt}\kern1pt}
\def\rb{{\bar\rho}}

\def\vsb{\bar\varsigma}
\def\bt{\bar t}
\def\tb{\bar\t}

\def\qb{\bar\theta}
\def\xb{\bar\xi}
\def\bx{\skew3\bar x}
\def\bz{\mathchar"420\mkern-3mu\zeta}
\def\Cb{\relax\leavevmode\hbox{$C$\kern-.53em
               \vrule height1.9ex depth-1.8ex width5pt}\kern.5pt}
\def\Db{\relax\leavevmode\hbox{$D$\kern-.6em
               \vrule height1.9ex depth-1.8ex width5pt}\kern.5pt}
\def\rDb{\relax\leavevmode\hbox{D\kern-.7em
               \vrule height1.9ex depth-1.8ex width5pt}\kern1.5pt}
\def\Fb{\relax\leavevmode\hbox{$F$\kern-.55em
               \vrule height1.9ex depth-1.8ex width4.5pt}\kern1pt}
\def\FB{\relax\leavevmode\hbox{$\Phi$\kern-.6em
               \vrule height1.9ex depth-1.8ex width4.5pt}\kern1.5pt}
\def\Gb{\relax\leavevmode\hbox{$G$\kern-.55em
               \vrule height1.9ex depth-1.8ex width4.5pt}\kern1pt}

\def\JB{\relax\leavevmode\hbox{$\Psi$\kern-.7em
               \vrule height1.9ex depth-1.8ex width6pt}\kern1.5pt}
\def\LB{\relax\leavevmode\hbox{$\Lambda$\kern-.6em
               \vrule height1.9ex depth-1.8ex width5pt}\kern1.5pt}
\def\Mb{\relax\leavevmode\hbox{$M$\kern-.8em
               \vrule height1.9ex depth-1.8ex width7pt}\kern.8pt}
\def\Pb{\relax\leavevmode\hbox{$P$\kern-.55em
               \vrule height1.9ex depth-1.8ex width4.5pt}\kern.5pt}
\def\PB{\relax\leavevmode\hbox{$\P$\kern-.6em
               \vrule height1.9ex depth-1.8ex width4.5pt}\kern.5pt}
\def\Qb{\relax\leavevmode\hbox{$Q$\kern-.6em
               \vrule height1.9ex depth-1.8ex width5.5pt}\kern.5pt}
\def\QB{\relax\leavevmode\hbox{$\Q$\kern-.65em
               \vrule height1.9ex depth-1.8ex width5.0pt}\kern.5pt}
\def\Tb{\relax\leavevmode\hbox{$T$\kern-.55em
               \vrule height1.9ex depth-1.8ex width4.5pt}\kern.5pt}
\def\Xb{\relax\leavevmode\hbox{$X$\kern-.67em
               \vrule height1.9ex depth-1.8ex width6pt}\kern.5pt}
\def\XB{\relax\leavevmode\hbox{$\X$\kern-.65em
               \vrule height1.95ex depth-1.85ex width6pt}\kern.5pt}
\def\YB{\relax\leavevmode\hbox{$\Upsilon$\kern-.6em
               \vrule height1.9ex depth-1.8ex width4.5pt}\kern1.5pt}
\def\Wb{\relax\leavevmode\hbox{$W$\kern-.9em
               \vrule height1.9ex depth-1.8ex width7pt}\kern-.1pt}

\def\cDb{\relax\leavevmode\hbox{$\cal D$\kern-.6em
               \vrule height1.9ex depth-1.8ex width5pt}\kern.5pt}
\def\ac{{\check a}}
\def\bc{{\check b}}
\def\ic{{\check\imath}}
\def\jc{{\check\jmath}}
\def\ad{{\dot\a}}
\def\bd{{\dot\b}}

\def\ah{{\hat a}}
\def\bh{{\hat b}}
\def\ch{{\hat c}}
\def\dh{{\hat d}}
\def\ih{{\hat\imath}}

\def\ra{{\rm a}}

\def\rQb{\relax\leavevmode\hbox{Q\kern-.65em
               \vrule height1.9ex depth-1.8ex width5pt}\kern1.5pt}
\def\bQb{\relax\leavevmode\hbox{{\bf Q}\kern-.65em
               \vrule height1.9ex depth-1.8ex width5pt}\kern1.5pt}
\def\Mt{\widetilde{M}}
\def\mt{{\tilde\mu}}
\def\Nt{\widetilde{N}}

\def\CBp{\,\BM{C}\!_+}
\def\CBm{\,\BM{C}\!_-}
\def\ssl#1{\rlx\rlap{\,\raise1pt\hbox{$\backslash$}}{#1}}
\def\Ssl{\rlap{\kern1.2pt\raise1pt\hbox{\rm/}}{\hbox{$S$}}}
\def\sT{{/\mkern-9mu\vs}}
\def\sW{{\backslash\mkern-9mu\vs}}
\def\sL{{\vs^{{}_\mm}}}
\def\sR{{\vs^{{}_\pp}}}
\def\LL{{\!_L}}
\def\RR{{\!_R}}

\font\ff=cmff10 at 11pt
\def\FF#1{\relax\hbox{\ff#1}}
 %
\def\PixCap#1#2#3{\midinsert\vbox{\centerline{\epsfbox{#1}}%
                   \noindent\narrower{\bf Figure~#2}.~#3}\endinsert}
 %
 %
\Title{\rightline{hep-th/yymmddd}}
      {\vbox{\centerline{Haploid (2,2)-Superfields}
              \vskip1mm
             \centerline{In 2-Dimensional Spacetime}}}
\centerline{\titlerms Tristan H\"ubsch\footnote{$^{\spadesuit}$}{On leave
            from the ``Rudjer Bo\v skovi\'c'' Institute, Zagreb, Croatia.}}
                                                             \vskip0mm
 \centerline{\it Department of Physics and Astronomy}        \vskip-.5mm
 \centerline{\it Howard University, Washington, DC~20059}    \vskip-.5mm
 \centerline{\tt thubsch\,@\,howard.edu}
\vfill

\centerline{ABSTRACT}\vskip2mm
\vbox{\narrower\narrower\baselineskip=12pt\noindent
Superfields in 2-dimensional (2,2)-superspacetime which are independent of
(some) half of the fermionic coordinates are discussed in a
hopefully both comprehensive and comprehensible manner. An embarrassing
abundance of these simplest `building blocks' makes it utterly impossible
to write down the `most general Lagrangian'.
 With some {\it ad hoc\/} but perhaps plausible restrictions, a rather
general Lagrangian is found, which exhibits many of the phenomena that have
been studied recently, and harbors many more.
 In particular, it becomes patently obvious that the (2,2)-supersymmetric
2-dimensional field theory target space geometries (many of which are
suitable for (super)string propagation) are far more general than K\"ahler
manifolds with holomorphic bundles.}

\Date{December '98. \hfill}  
\footline{\hss\tenrm--\,\folio\,--\hss}
 %
 %
 %
\lref\rCBAbWo{E.M.C.~Abreu and C.~Wotzasek: 
       \PR{D58}(1998)101701, hep-th/9805043.}

\lref\rAKMRV{D.~Amati, K.~Konishi, Y.~Meurice, G.C.~Rossi and G.~Veneziano:
       \PRp{162C}(1988)170, and references therein.}

\lref\rBatFra{I.A.~Batalin and E.S.~Fradkin: \NP{B279}(1987)514--528.}

\lref\rCec{S.~Cecotti: \IJMP{A6}(1991)1749, \NP{B355}(1991)755.}

\lref\rCeGiPa{S.~Cecotti, L.~Girardello and A.~Pasquinucci:\Z
       \NP{B328}(1989)701, \IJMP{A6}(1991)2427.}

\lref\rClaHal{M.~Claudson and M.B.~Halpern: \NP{B250}(1985)689.}

\lref\rdAdda{A.~d'Adda, A.C.~Davis, P.~Di~Vecchia and P.~Salomonson:
       \NP{B222}(1983)45.}

\lref\rDixon{L.~Dixon: in {\it Superstrings, Unified Theories and Cosmology
       1987},\,p.67--127, eds.~G.~Furlan et al.\ (World Scientific,
       Singapore, 1988).}

\lref\rEliSch{S.~Elizur and A.~Schwimmer: \NP{B226}(1983)109.}

\lref\rJimNew{S.J.~Gates, Jr.: \NP{B238}(1984)349.}

\lref\rTwJim{S.J.~Gates, Jr.: \PL{B352}(1995)43--49.}

\lref\rNMJiDeo{S.J.~Gates, Jr.\ and B.B.Deo: \NP{B254}(1985)187-200.}

\lref\rGGRS{S.J.~Gates, Jr., M.T.~Grisaru, M.~Ro\v cek and
       W.~Siegel: {\it Superspace}\Z (Benjamin/Cummings Pub.\ Co.,
       Reading, Massachusetts, 1983).}

\lref\rGGW{S.J.~Gates, Jr., M.T.~Grisaru and M.E.~Wehlau:
       \NP{B460}(1996)579--614.}

\lref\rUseNM{S.J.~Gates, Jr., M.T.~Grisaru, M.B.~Knutt-Wehlau, M.~Ro\v{c}ek
       and O.~Solovev:\Z \PL{B396}(1997)167--176\semi
       S.J.~Gates, Jr.\ and S.~M.~Kuzenko: The CNM-Hypermultiplet
       Nexus.\Z hep-th/9810137.}

\lref\rCYHS{S.~J.~Gates, Jr.\ and T.~H\"ubsch: \PL{B226}(1989)100,
       \NP{B343}(1990)741.}

\lref\rGHR{S.J.~Gates, Jr., C.M.~Hull and M.~Ro\v cek: \NP{B248}(1984)157.}

\lref\rAleph{S.J.~Gates, Jr.\ and L.~Rana:
       \PL{B352}(1995)50--58, hep-th/9504025;
       \PL{B369}(1996)262--268, hep-th/9510151.}

\lref\rNMJiWa{S.J.~Gates, Jr.\ and W.~Siegel: \NP{B187}(1981)389.}

\lref\rCBJiWa{S.J.~Gates, Jr.\ and W.~Siegel: \PL{B206}(1988)631.}

\lref\rSevTro{A.~Sevrin and J.~Troost: \NP{B492}(1997)623--646\semi
       M.T.~Grisaru, A.~Massar, A.~Sevrin and J.~Troost:
       {\it Fortsch.\,Phys.\,}{\bf47}(1999)301--307.}

\lref\rMarjD{T.~H\"ubsch: \MPL{A6}(1991)1553--1559.}

\lref\rSingS{T.~H\"ubsch: \MPL{A6}(1991)207--216.}

\lref\rBeast{T.~H\"ubsch: {\it \CY\ Manifolds---A Bestiary for
      Physicists}\Z (World Scientific, Singapore, 1992).}

\lref\rMorPle{D.R.~Morrison and M.R~.~Plesser: 
       \NP{A46}(Proc.\,Supl.,~1996)177--186, hep-th/9508107.}

\lref\rNSVZ{V.~Novikov, M.~Shifman, A.~Vainshtein and V.A.~Zakharov:
       \NP{B229}(1993)381, \NP{B229}(1993)407.}

\lref\rJoeD{J.~Polchinski: TASSI Lectures on D-Branes.
       hep-th/9611050}

\lref\rSriCB{P.P.~Srivastava: \PRL{63}(1989)2791.}

\lref\rNSEW{N.~Seiberg and E.~Witten:
       \NP{B426}(1994)19--52, Erratum: \I{B430}(1994)485--486;
       \NP{B431}(1994)484--550.}

\lref\rCBSieg{W.~Siegel: \NP{B238}(1984)307.}

\lref\rAuxWS{W.~Siegel: {\it Physica\/}{\bf D15}(1985)208--212.}

\lref\rVafa{C.~Vafa: \MPL{A4}~(1989)~1615.}

\lref\rChiRi{C.~Vafa and N.~Warner: \PL{B218}(1989)51\semi
      W.~Lerche, C.~Vafa and N.~Warner: \NP{B324}(1989)427.}

\lref\rWB{J.~Wess and J.~Bagger: {\it Supersymmetry and Supergravity}\Z
      (Princeton University Pub., Princeton NJ, 1983).}

\lref\rSuSyM{E.~Witten: 
      {\it J.\,Diff.\,Geom.\,\bf17}(1982)661--692.}

\lref\rWAB{E.~Witten: 
      in {\it Essays on Mirror Manifolds}, p.120, Ed.~S.-T.~Yau
      (International Press, Hong Kong, 1992).}

\lref\rSuSyM{E.~Witten: 
      {\it J.\,Diff.\,Geom.\,\bf17}(1982)661--692.}

\lref\rWitInd{E.~Witten: 
       \NP{B202}(1982)253--316.}

\lref\rPhases{E.~Witten: \NP{B403}(1993)159--222.}

\lref\rZumino{B.~Zumino: \PL{B87}(1979)203.}

 %
 %
\newsec{Introduction}\noindent
Superfields are used extensively in 2-dimensional supersymmetric field
theories. Alas! the diversity of conventions obscures the straightforward.
In particular, these building blocks are frequently obtained by dimensional
reduction from the 4-dimensional case, which (by far) does not provide for
full generality~\rGHR: contrary to some claims in the literature (even in
book form!) in 2-dimensional spacetimes, {\it(2,2)- (a.k.a.\ $N{=}2$)
supersymmetry does not imply K\"ahlerness of the field space!}

An intrinsically 2-dimensional analysis, `from scratch', is presented here,
and provides generalizations to the results in the recent
literature; see, \eg, Refs.~\refs{\rPhases,\rTwJim,\rGGW,\rSevTro}.

This article is organized as follows: The remaining part of this
introductory section presents some basic facts about 2-dimensional
(2,2)-superspace and sets up the notation; further details are found in
Appendix~A. Section~2 presents a rather rich set of constrained superfields
in a systematic way, and provides several directions for generalizations.
The `most general' Lagrangian for the superfields of \SS\,2 (subject to
a few hopefully plausible restrictions) is given in Section~3, using
`implicitly constrained' superfields, with the details of component field
expansions deferred to the Appendix~B. An alternate, `explicitly
constrained' formalism is discussed in Section~4. Section~5 then turns to
some relations to geometry, in particular that of the target (field) space.
Section~6 summarizes the presented results and discusses some further
topics.

\subsec{2-dimensional spacetime}\noindent
The crucial peculiarity of supersymmetric models in 2-dimensional spacetime
stems from the fact that the Lorentz group, $SO(1,1)$, is Abelian and so
has only 1-dimensional irreducible representations: for example, the
coordinate 2-vector $(\s^0,\s^1)$ decomposes into the (light-cone)
characteristic coordinates $\s^{\pm\pm}{\define}\inv2(\s^0{\pm}\s^1)$.
These, in fact, are eignefunctions of the (Lorentz boost) generator\ft{The
eigenvalue of the Lorentz boost operator,
$\BM{B}\define(\s^1\vd_0{+}\s^0\vd_1)$, extended to total angular momentum
in the usual way, equals the spin projection, and will therefore be denoted
by $j_3$ although it does not stem from its 4-dimensional namesake.
Moreover, since all representations of the 2-dimensional Lorentz group are
1-dimensional, there is no need to distinguish between `spin' and `spin
projection', whence we will call $j_3$ simply `spin'.},
$\BM{B}\define(\s^1\vd_0{+}\s^0\vd_1)$, of $SO(1,1)$:
\eqn\eXXX{ \BM{B}\,\>\s^{\pm\pm}
 =j_3(\s^{\pm\pm})\,\s^{\pm\pm}=\pm\,\s^{\pm\pm}
}
so that the group element, $\BM{U}\!_\a\,{\define}e^{i\a\BM{B}}$, acts
diagonally: $\BM{U}\!_\a\,\s^{\pm\pm}=e^{\pm i\a}\s^{\pm\pm}$.
  For brevity, we also write $\s^\pp{\define}\s^{++}$ and
$\s^\mm{\define}\s^{--}$. Note that upon the analytic continuation
$\s^0\to i\s^0$, $(\s^\pp,\s^\mm)\to(z,-\bar{z})$: the light-cone structure
becomes a complex structure.
 Similar arguments show that all other tensors (spinors) decompose into
1-component objects. Upon the analytic continuation to imaginary time,
$\s^0\to i\s^0$, $\BM{U}\!_\a$ becomes the winding number (holomorphic
homogeneity) operator. The sub- and superscripts ``$\pm$'' then simply
denote the winding number (spin in real time) in units of
$\inv2$($\hbar$). Functions depending on only $\s^\pp$ or only $\s^\mm$ are
called left- and right-movers, respectively, and become holomorphic
(complex-analytic) and anti-holomorphic (complex-antianalytic) functions
upon analytic continuation to imaginary time.

\subsec{(2,2)-superbasics}\noindent
The (2,2)-supersymmetry algebra involves the supersymmetry charges
$Q_\pm$ and $\Qb_\pm$, which satisfy (adapting from
Refs.~\refs{\rWB,\rPhases}; comparison with Refs.~\refs{\rGGRS,\rGGW} is
provided in appendix~A):
\eqn\eSusyQ{ \big\{\, Q_\pm \,,\, \Qb_\pm \,\big\}~~
             = ~~-2i\vd_{\pm\pm}~ = ~2(H \pm p)~.}
Here $\vd_{++}{\id}\vd_\pp{\define}\pd{}{\s^\pp}{=}(\vd_0{+}\vd_1)$ and
$\vd_{--}{\id}\vd_\mm{\define}\pd{}{\s^\mm}{=}(\vd_0{-}\vd_1)$.
$p{=}-i\vd_1{=}-\frc{i}2(\vd_\pp{-}\vd_\mm)$ is the linear momentum
operator and $H{=}-i\vd_0{=}-\frc{i}2(\vd_\pp{+}\vd_\mm)$ is the Hamiltonian
(energy) operator. Note that $p$ has the usual sign as in Quantum Mechanics,
while $H$ has the opposite sign; with these conventions, $(-E,E)$ is a
light-like 2-vector.

Equipping the world-sheet with anticommuting fermionic coordinates,
$\vs^\pm,\vsb^\pm$, the supercharges are realized as differential
operators on the (super) world-sheet\ft{Since
$(\vd_{\pm\pm})^\dagger=-\vd_{\pm\pm}$, for $(Q\pm)^\dagger=\Qb_\pm$, it
must be that $(\vs^\pm)^\dagger=-\vsb^\pm$ and
$(\vd_\pm)^\dagger=-\db_\pm$.}:
\eqn\eQs{
 Q_\pm~ \define ~\vd_\pm+i\vsb^\pm\vd_{\pm\pm}~,
 \qquad
 \Qb_\pm~ \define ~-\db_\pm-i\vs^\pm\vd_{\pm\pm}~. }
Supersymmetry transformations are generated by the $Q,\Qb$'s, and
implemented (effected) by unitary operator
\eqn\eSuSy{ \BM{U}\!_{\e,\bar\e}\> \define
 \exp\big\{i(\e^\pm Q_\pm+\bar\e^\pm\Qb_\pm)\big\}~. }
which, in fact, is a quartic multinomial in the $Q,\Qb$'s, by virtue of
their nilpotency. The spinorial derivatives
\eqn\eDs{
 D_\pm~ \define ~\vd_\pm-i\vsb^\pm\vd_{\pm\pm}~,
 \qquad
 \Db_\pm~ \define ~-\db_\pm+i\vs^\pm\vd_{\pm\pm}~, }
are covariant with respect to $\BM{U}\!_{\e,\bar\e}$. Easily,
$\{Q_\pm,D_\pm\}=0=\{\Qb_\pm,D_\pm\}$, whereupon
$[\BM{U}\!_{\e,\bar\e},D_\pm]=0=[\BM{U}\!_{\e,\bar\e},\Db_\pm]$.

 Finally the $D,\Db$'s close virtually the same algebra~\eSusyQ,
as do the $Q,\Qb$'s:
\eqn\eSusyD{
 \big\{\, D_\pm \,,\, \Db_\pm \,\big\}~
 = ~2i\vd_{\pm\pm}~= ~-2(H \pm p)~.}
All anticommutators among the $Q_\pm,\Qb_\pm,D_\pm,\Db_\pm$, other
than~\eSusyQ\ and~\eSusyD, vanish.

The Berezin superintegrals (over supercoordinates) are by definition
equivalent to partial superderivatives, and up to total (world-sheet)
spacetime derivatives\ft{Hereafter, `total derivative' will stand for
`total (world-sheet) spacetime derivative'.} (which we ignore, assuming
world-sheets without boundaries) equivalent to covariant
superderivatives~\refs{\rWB,\rGGRS}. We use:
\eqn\eDI{ \int\rd^4\vs~[\3] ~\define~
 \inv4[D_-,\Db_-][D_+,\Db_+][\3]\Zp \id~D^4[\3]~, }
where ``$|$'' means setting $\vs^\pm=0=\vsb^\pm$.
 Integration over a fermionic subspace is formally achieved by inserting a
fermionic Dirac delta-function, so that:
\eqna\eHFI
 $$ \eqalignno{
 \int\rd^2\vs~[\3]  &\id \inv2\big[D_-,D_+\big][\3]\Zp~,\quad
 \int\rd^2\vsb~[\3]  \id \inv2\big[\Db_+,\Db_-\big][\3]\Zp~;&\eHFI{a,b}\cr
 \int\rd^2\sT~[\3]  &\id \inv2\big[D_-,\Db_+\big][\3]\Zp~,\quad
 \int\rd^2\sW~[\3]   \id \inv2\big[D_+,\Db_-\big][\3]\Zp~;  &\eHFI{c,d}\cr
 \int\rd^2\sR\,[\3] &\id \inv2\big[D_-,\Db_-\big][\3]\Zp~,\quad
 \int\rd^2\sL\,[\3]  \id \inv2\big[D_+,\Db_+\big][\3]\Zp~.  &\eHFI{e,f}\cr}
 $$
Clearly, the definitions~\eHFI{c,d} may also be obtained from~\eHFI{a,b} by
swapping $\vs^+\iff\vsb^+$; see below. Similarly, the
definitions~\eHFI{e,f} follow from~\eHFI{c,d} upon $\vsb^+\iff\vsb^-$.
 As expected then, for example,
\eqn\eXXX{\int\!\rd^4\vs~[\3]~=~\int\!\rd^2\vs\int\!\rd^2\vsb~[\3]~,
                           ~=~-\int\!\rd^2\sT\int\!\rd^2\sW~[\3]~.}
Note the negative sign in the second identity, and that the integral
measures in~\eHFI{e,f} are not (pseudo)scalars; they have spin $\pm1$.
Also, a few {\it operatorial\/} identities collected in Appendix~A will be
useful.

\subsec{Superspace symmetries}\noindent
\subseclab\ssSuSpSym
In more than 2-dimensional spacetimes, Lorentz transformations leave
the integrals~\eHFI{} unchanged. However, they rotate
Eqs.~\eHFI{c}$\iff$\eHFI{d} and~\eHFI{e}$\iff$\eHFI{f}.

It is of interest then to examine the discrete symmetries acting on the
four-fermionic coordinate system, or alternatively, the four covariant
superdifferentials, $D_-,\Db_-,D_+,\Db_+$. Since the Lorentz
transformations act diagonally on the $D,\Db$'s, they can be treated as
independent objects in a Lorentz-invariant fashion. Thus, the maximal
discrete group acting on the $D_\pm,\Db_\pm$ system by swapping them is
$S_4$, the group of permutations of four distinguishable objects. The
simplest are, of course, the swaps:
\eqna\eSwp
 $$\twoeqsalignno{
 \CBp:   &D_+{\iff}\Db_+~, \quad&\quad
 \CBm:   &D_-{\iff}\Db_-~,   &\eSwp{a,b}\cr
 \BM{p}: &D_+{\iff}D_-~, \quad&\quad
 \pB:    &\Db_+{\iff}\Db_-~, &\eSwp{c,d}\cr
 \BM{q}: &\Db_-{\iff}D_+~, \quad&\quad
 \qB:    &D_-{\iff}\Db_+~. &\eSwp{e,f}\cr
}$$
where each of these operations only swaps the indicated superderivatives
and leaves the other two intact. Also, it should be clear that
$[\CBp,\CBm]{=}0$, $[\BM{p},\pB]{=}0$ and $[\BM{q},\qB]{=}0$. Furthermore,
\eg, $\BM{p}\CBm\BM{p}=\CBm\BM{p}\CBm=\qB$, and
$\pB\CBp\pB=\CBp\pB\CBp=\qB$.

There are additional discrete transformations involving time-reversal,
\BM{T}, which acts by $\BM{T}(\vd_\pp)=-\vd_\mm$ and
$\BM{T}(\vd_\mm)=-\vd_\pp$. Since time-reversal squares to the identify,
$\BM{T}^2=\Ione$, it will `double' the group of discrete transformations of
the $D_\pm,\Db_\pm$ system, and we call this group $2S_4$.

Notice that the familiar (ambidexterous) complex conjugation,
\BM{C}, may be expressed as $\BM{C}{=}\CBm\CBp$. This operation exchanges
the integrals~\eHFI{a}$\iff$\eHFI{b} and~\eHFI{c}$\iff$\eHFI{d}, but
leaves~\eHFI{e} and~\eHFI{f} intact.

Similarly, the full parity (spatial reflection), $\BM{P}$, may be expressed
as $\BM{P}{=}\BM{p}\pB$, and it is easy to show that $[\BM{C},\BM{P}]{=}0$,
as should be the case. \BM{P} exchanges the
integrals~\eHFI{c}$\iff$\eHFI{d} and~\eHFI{e}$\iff$\eHFI{f}, but
leaves~\eHFI{a} and~\eHFI{b} intact.

Also, the combined parity and complex conjugation, $\BM{CP}=\BM{q}\qB$,
exchanges the integrals~\eHFI{a}$\iff$\eHFI{b} and~\eHFI{e}$\iff$\eHFI{f},
but leaves those in~\eHFI{c} and~\eHFI{d} intact.

It is very important to realize that the operators~\eSwp{} are defined there
to act on the `soul' of the $(2,2)$-super-Riemann superface, \ie, the
fermionic coordinates and so the superderivatives acting on them. However,
the relation between the superderivatives and the world-sheet
derivatives~\eSusyD\ induces an action on the `body' too, \ie, on the
world-sheet coordinates. In the often used Euclidean incarnation,
$(\s^\pp,\s^\mm)\to(z,-\bar{z})$, what was called `parity' (the
unidexterous \BM{p} and \pB, and the ambidexterous \BM{P}) becomes
(another!) complex conjugation. While the induced action of $\BM{C}_\pm$
and \BM{C} on the world-sheet coordinates $(\s^\pp,\s^\mm)$ is trivial
(since the supersymmetry relations~\eSusyD\ are Hermitian), the induced
action of \BM{p}, \pB\ and \BM{P} is not! These do become the (unidexterous
and ambidexterous, respectively) complex conjugations when acting on the
Euclideanized `body', \ie, bosonic world-sheet coordinates, but remain
parity (spin-flip) in the `soul', \ie, the fermionic coordinates. Foremost,
they are independent from the $\BM{C}_\pm$ and \BM{C}. The unaware Reader
should be cautioned that the careful distinction between these operations is
sometimes muddled in the literature.

Finally, the Lorentz boost group element (generalized, as usual, to
include spin) $\BM{U}\!_\a=e^{i\a\BM{B}}$ satisfies
$\BM{U}\!_\a\BM{U}\!_\b=\BM{U}\!_{(\a+\b)}$ so
$[\BM{U}\!_\a,\BM{U}\!_\b]=0$. Also, $[\BM{U}\!_\a,\BM{C}\!_\pm]=0$, but
$\BM{U}\!_\a\BM{P}=\BM{PU}\!_\a^{-1}$. Thus, the usual super-Poincar\'e
group is extended by the various discrete symmetries~\eSwp{} and
time-reversal, \BM{T}, so that the discrete subroup is $2S_4$.

The discrete symmetry, $S_4$, generated by the swaps~\eSwp{}, is part of the
continuous $GL(4,\IC)$, the $GL_-(2,\IC){\times}GL_+(2,\IC)$ supgroup of
which then commutes with the Lorentz symmetry. I defer the study of such
continuous transformations to a later opportunity.

\newsec{Haploid Superfields}\noindent
\seclab\sHSF
General superfields are (super)functions of $(\s^0,\s^1;\vs^\pm,\vsb^\pm)$.
Akin to complex-analytic (holomorphic) functions depending only on the
complex-analytic half of complex variables, we define {\it haploid}
superfields to depend only on (some) half of the four
fermionic coordinates $\vs^\pm,\vsb^\pm$. Models that involve such
`super-holomorphic' superfields exhibit a much better quantum behavior, not
unrelated to the special features of complex-analytic
functions~\refs{\rWB,\rGGRS}.

\subsec{First-order constrained haploid superfields}\noindent
\subseclab\ssFrst
As the Lorentz transformations in 1+1-dimensional spacetime leave the four
$\vs^\pm,\vsb^\pm$ independent of each other, there exist {\it six}
distinct types of hapliod superfields (as ${4\choose2}=6$), all of which
are annihilated by (some) half of the superderivatives:
\eqna\eHSF
 $$ \twoeqsalignno{
 \[61. & \hbox{\bf Chiral}:
 &\qquad  & \Db_+ \F~ = ~0~ = ~\Db_- \F~, &\eHSF{a} \cr
 \[62. & \hbox{\bf Antichiral}:
 &\qquad  & D_+\FB~ = ~0~ = ~D_-\FB~, &\eHSF{b} \cr
 \[63. & \hbox{\bf Twisted-chiral}:
 &\qquad  & D_+ \X~ = ~0~ = ~\Db_- \X~,   &\eHSF{c} \cr
 \[64. & \hbox{\bf Twisted-antichiral}:
 &\qquad  & \Db_+\XB~= ~0~ =~D_-\XB~, &\eHSF{d} \cr
 \[65. & \hbox{\bf Lefton}:
 &\qquad  & D_-\L~ = ~0~ = ~\Db_-\L~,     &\eHSF{e} \cr
 \[66. & \hbox{\bf Righton}:
 &\qquad  & D_+\Y~ = ~0~ = ~\Db_+\Y~.     &\eHSF{f} \cr}
 $$
The first four pairs of superconstraints simply restrict $\F,\FB,\X,\XB$ to
be independent of two of the fermionic coordinates $\vs,\vsb$, up to total
derivatives.

By contrast, of the last two superconstraint pairs, each one
{\it includes\/} also a spacetime derivative constraint (see below):
\eqn\eSeCon{ \hbox{Eqs.}~\eHSF{e}~\ni~\big\{\vd_\mm\L=0\big\}~, \qquad
             \hbox{Eqs.}~\eHSF{f}~\ni~\big\{\vd_\pp\Y=0\big\}~. }
 As a result, the component fields in the lefton superfield
$\L$ are functions of only $\s^\pp$ (`holomorphic', \ie, `left-movers'),
whereas those in the righton superfield $\Y$ are functions of only $\s^\mm$ 
(`anti-holomorphic', \ie, `right-movers'). We return to this below, after
defining the component fields.

We will also encounter an even more restricted type of superfields, which
depend only on one out of the four $\vs,\vsb$'s, and so are annihilated by
{\it three\/} superderivatives:
\eqna\eQSF
$$ \twoeqsalignno{
 1.&\hbox{\bf Chiral Lefton}:
   &\qquad& \Db_+\F_\LL=\Db_-\F_\LL=D_-\F_\LL=~0~,&\eQSF{a}\cr
 2.&\hbox{\bf Antichiral Lefton}:
   &\qquad& D_+\FB_\LL=\Db_-\FB_\LL=D_-\FB_\LL=~0~, &\eQSF{b} \cr
 3.&\hbox{\bf Chiral Righton}:
   &\qquad& \Db_-\F_\RR=\Db_+\F_\RR=D_+\F_\RR=~0~,&\eQSF{c}\cr
 4.&\hbox{\bf Antichiral Righton}:
   &\qquad& D_-\FB_\RR=\Db_+\FB_\RR=D_+\FB_\RR=~0~. &\eQSF{d} \cr}
$$
Depending only on some quarter of the supercoordinates $\vs,\vsb$, we call
them {\it quartoid} superfields. Clearly, they may be considered as special
cases of the haploid ones~\eHSF{}: \eg, $\F_\LL$ may be regarded as a
$D_-$-annihilated chiral superfield, a $\Db_-$-annihilated twisted
antichiral superfield, or a $\Db_+$-annihilated (\ie, chiral) lefton. We
adopted this last nomenclature in virtue of its simplicity and the fact
that it also alludes to the implicit spacetime constraints
$\vd_\mm\F_\LL=0=\vd_\mm\FB_\LL$ and $\vd_\pp\F_\RR=0=\vd_\pp\FB_\RR$. The
Reader may prefer the obvious twisted analogues: $\X_\LL=\FB_\LL$,
$\X_\RR=\F_\RR$, $\XB_\LL=\F_\LL$ and $\XB_\RR=\FB_\RR$.

A disclaimer is in order: our purpose thus far was merely to list the
logical possibilities. The quantum theory of the unidexterous fields
defined in Eqs.~\eHSF{e,f} and~\eQSF{a{-}d} may indeed seem curiously
cumbersome, owing to the ocurrence of the spacetime constraints as
exemplified in Eqs.~\eSeCon. For unidexterous spin-0 fields (`chiral
bosons') and their (1,0)- and (1,1)-supersymmetrizations see
Refs.~\refs{\rCBSieg,\rCBJiWa}; Ref.~\rCBAbWo\ offers some welcome
clarifications about the dynamics and symmetries involved and an updated
reference list on the topic. We merely remark that the constraints may be
regarded as generators of a gauge symmetry, with the transformations:
\eqn\eChiGau{\eqalign{
 \vd_\mm\L=0\quad&\to\quad \d\L=\k^\mm(\vd_\mm\L)~, \cr
 \vd_\pp\Y=0\quad&\to\quad \d\Y=\k^\pp(\vd_\pp\Y)~. \cr }}
This gauge invariance then ensures the proper decoupling of the unwanted
degrees of freedom. Thus, a well-defined quantum field theory model
involving unidexterous (super)fields must be invariant with respect
to the chiral gauge symmetry~\eChiGau.

Seeking generality, our approach will then be as follows: at first, we
ignore the unidexterous quantization issues. We then return to these
issues in a simple setting and discuss some of the implications.
Relying on the general and powerful machineries such as the operatorial
quantization (see Ref.~\rBatFra\ and the references therein), we defer the
(careful tedium of {\it Ausarbeitung\/} of the) covariantization of (a
suitable subset of) the present results with respect to~\eChiGau\ and a full
treatment of the (super)constraint effects to a later effort.

\subsec{Superfield symmetries}\noindent
In more than 2-dimensional spacetimes, Lorentz transformations leave
the pair of superconstraint superderivatives~\eHSF{a,b} (and so also
$\F,\FB$) unchanged. However, they mix those in Eqs.~\eHSF{c}$\iff$\eHSF{d}
and~\eHSF{e}$\iff$\eHSF{f}. So, for example, requiring a superfield to
satisfy Eq.~\eHSF{c} implies, upon a suitable Lorentz transformation, that
{\it the same\/} superfield must also satisfy \eHSF{d}, whereupon it
follows that it must be a constant. The same applies to the
Eqs.~\eHSF{e,f}. Only chiral superfields and their conjugates remain viable
haploid superfields when spacetime has more than two dimensions.
Conversely, $\X,\XB,\L,\Y$ cannot be obtained by dimensional reduction;
their existence is exceptional to (2,2)-supersymmetric ($\leq$)2-dimensional
spacetime.

Now, complex conjugation\ft{In 2-dimensional spacetime, hermitian
conjugation is complex conjugation and transposition (including derivative
action to the left), {\it without\/} parity or time-reversal.}, \BM{C},
leaves the superconstraints~\eHSF{e,f} invariant, while exchanging
\eHSF{a}$\iff$\eHSF{b} and \eHSF{c}$\iff$\eHSF{d}, and so $\F{\iff}\FB$ and
$\X{\iff}\XB$. Thus, $\F,\FB$ and $\X,\XB$ are complex conjugate pairs,
while $\L$ and $\Y$ remain unrelated. So, while in general complex, the
unidexterous superfields $\L$ and $\Y$ {\it may be chosen\/} to be real or
imaginary. Note that the `left-handed conjugation', $\CBp$, exchanges
$\F{\iff}\X$ and $\FB{\iff}\XB$, while the `right-handed conjugation',
$\CBm$ exchanges $\F{\iff}\XB$ and $\FB{\iff}\X$, and both leave $\L,\Y$
intact.

On the other hand, parity, $\BM{P}$, reflects spin ($+{\iff}-$). Therefore,
it exchanges Eqs.~\eHSF{c}$\iff$\eHSF{d} and~\eHSF{e}$\iff$\eHSF{f}, and so
$\X{\iff}\XB$ and $\L{\iff}\Y$, but leaves Eqs.~\eHSF{a,b} unchanged. That
is, $\X,\XB$ and $\L,\Y$ are paired by the left-right parity, \BM{P}, while
$\F$ and $\FB$ remain unrelated. Note that the `unconjugate parity',
$\BM{p}$, exchanges $\X{\iff}\L$ and $\XB{\iff}\Y$, while the
`conjugate parity', $\pB$ exchenges $\X{\iff}\Y$ and $\XB{\iff}\L$, and
both leave $\F,\FB$ intact.

However, since \BM{C} commutes with \BM{P}, in \BM{C}-invariant models
$\F$ and $\FB$ must transform identically under the action of \BM{P} and
any other symmetry that commutes with \BM{C}. Similarly, in
`left-right-symmetric' (parity-invariant) models, $\L$ and
$\Y$ must transform identically under the action of \BM{C} and any other
symmetry that commutes with \BM{P}.

Besides the $S_4$ group of discrete symmetries~\eSwp{} and time-reversal
\BM{T}, there may also exist discrete or continuous transformations
operating entirely in the (super)field space. The fact that the unidexterous
superfields~\eHSF{e,f} and~\eQSF{} obey a world-sheet spacetime constraint
sets them apart form the others~\eHSF{a{-}d}, and there can be no
transformation in the superfield space which would mix the ambidexterous
superfields~\eHSF{a{-}d} and the unidexterous ones~\eHSF{e,f}
and~\eQSF{}. This also means that we cannot require of models to be
\BM{p},- $\pB$,- \BM{q},- or $\qB$-invariant.
 Further symmetry considerations (and especially gauging) are beyond the
scope of this paper, but we note that \BM{C}-invariance ensures that
$(\F,\FB)$, $(\X,\XB)$, $(\F_\LL,\FB_\LL)$ and $(\F_\RR,\FB_\RR)$ all come
in (conjugate) pairs. It is $\CBp$- (or $\CBm$-) invariance that would also
require that the (anti)chiral and twisted (anti)chiral superfields are
paired. No such restriction ensues for the unidexterous superfields $\L,\Y$.

\subsec{Component fields}\noindent
\subseclab\ssCFlds
The fermionic coordinates $\vs^\pm,\vsb^\pm$ being anticommutative, they
are also nilpotent, and so only non-negative powers make sense. Moreover,
all Taylor series over $\vs^\pm,\vsb^\pm$ must terminate with the
$\vsb^-\vsb^+\vs^+\vs^-$ term. Thus, superfields are always analytic, in
fact, multinomials of finite (up to quartic) order over $\vs^\pm,\vsb^\pm$.

\topic{Superfield components}
The component fields are defined as the super-Taylor expansion coefficients
\eqna\eCps
 $$
   \f\define\F|~,\qquad
   \j_\pm\define \inv{\sqrt{2}}D_\pm\F|~,\qquad
   F\define \inv2D_-D_+\F|~; \eqno\eCps{a}
 $$
 $$
   \fb\define\FB|~,\qquad
   \bar\j_\pm\define \inv{\sqrt{2}}\Db_\pm\FB|~,\qquad
   \Fb\define \inv2\Db_+\Db_-\FB|~; \eqno\eCps{b}
 $$
are the component fields of the chiral and antichiral spin-0 superfield.
Note that both $F,\Fb$ change sign with respect to the parity, \BM{P}; that
is, $F,\Fb$ are pseudo-scalars if the lowest components, $\f,\fb$, are
scalars. Complex conjugation on all components of a chiral superfield may
(and herein will) be carried only by the index. That is, we write
$\fb\to\f^\mb$, $\jb_\pm\to\j_\pm^\mb$, $\Fb\to F^\mb$. Component field
content is also specified by $\F=(\f;\j_\pm;F)$ and
$\FB=(\fb;\jb_\pm;\Fb)$, leaving the numerical tedium of $\sqrt2$ factors
and such up to the Reader's preference.

Next,
 $$
   x\define\X|~,\quad
   \x_-\define \inv{\sqrt{2}}D_-\X|~,\quad
   \c_+\define \inv{\sqrt{2}}\Db_+\X|~,\quad
   X\define \inv2D_-\Db_+\X|~; \eqno\eCps{c}
 $$
 $$
  \bx\define\XB|~,\quad
   \xb_-\define \inv{\sqrt{2}}\Db_-\XB|~,\quad
   \cb_+\define \inv{\sqrt{2}}D_+\XB|~,\quad
   \Xb\define \inv2D_+\Db_-\XB|~; \eqno\eCps{d}
 $$
are the component fields of a twisted chiral and a twisted antichiral spin-0
superfield. The fields $X,\Xb$ change sign (are pseudoscalar) with respect
to the combined parity and complex conjugation, \BM{CP}, if $x,\bx$ are
scalars with respect to it. Complex conjugation on the components of a
twisted chiral superfield may again be carried only by the index,
at the expense of labeling the $\vs^-$- and the $\vsb^+$-components
by different characters\ft{The alternative is to meticulously retain the
`dot-over' on spin indices stemming from the $\Db$'s, as was done in
Ref.~\rGGW; for some further details, see the appendix~A.}, as was done
in~\eCps{c,d}. Appending indices is then straightforward: $\bx\to x^\ab$,
$\xb_-\to\x_-^\ab$, $\cb_+\to\c_+^\ab$ and $\Xb\to X^\ab$. As with the
chiral superfields, we may write $\X=(x;\x_-,\c_+;X)$ and
$\XB=(\bx;\xb_-,\cb_+;\Xb)$.

Finally, we define
 $$
   \ell\define\L|~,\quad
   \l_+\define \inv{\sqrt{2}}D_+\L|~,\quad
   \bl_+\define \inv{\sqrt{2}}\Db_+\L|~,\quad
   L_\pp\define \inv4\big[D_+,\Db_+\big]\L|~; \eqno\eCps{e}
 $$
and
 $$
   r\define\Y|~,\quad
   \r_-\define \inv{\sqrt{2}}D_-\Y|~,\quad
   \vr_+\define \inv{\sqrt{2}}\Db_-\Y|~,\quad
   R_\mm\define \inv4\big[D_-,\Db_-\big]\Y|~; \eqno\eCps{f}
 $$
are the component fields in the lefton and righton superfields. Note that
both ${\rm L}_\pp,{\rm R}_\mm$ change sign under complex conjugation if
$\ell,r$ do not: ${\rm L}_\pp,{\rm R}_\mm$ are imaginary if $\ell,r$ are
real. In general, however, these superfields and their components are
complex. Components of the conjugate superfields are defined as hermitian
conjugates of~\eCps{e,f}, and so
 $$
   \bar\ell\define\LB|~,\quad
   \lb_+\define \inv{\sqrt{2}}\Db_+\LB|~,\quad
   \bar\bl_+\define \inv{\sqrt{2}}D_+\LB|~,\quad
   \bar{L}_\pp\define \inv4\big[D_+,\Db_+\big]\LB|~; \eqno\eCps{\bar e}
 $$
and
 $$
   \bar{r}\define\YB|~,\quad
   \rb_-\define \inv{\sqrt{2}}\Db_-\YB|~,\quad
   \bar\vr_+\define \inv{\sqrt{2}}D_-\YB|~,\quad
   \bar{R}_\mm\define \inv4\big[D_-,\Db_-\big]\YB|~; \eqno\eCps{\bar f}
 $$
 Again, we may write $\L=(\ell;\l_+,\bl_+;L_\pp)$ and
$\Y=(r;\r_-,\vr_-;R_\mm)$, and $\LB=(\bar\ell;\lb_+,\bar\bl_+;\bar{L}_\pp)$
and $\YB=(\bar{r};\rb_-,\bar\vr_-;\bar{R}_\mm)$.

A remark is perhaps in order. The astute Reader must have wondered about the
seemingly unequal definitions of $L_\pp,R_\mm$ in~\eCps{e,f}, as
compared to those of $F,\Fb,X,\Xb$ in Eqs.~\eCps{a,b,c,d}. Since, e.g,
$[D_-,D_+]=2D_-D_+$, the original definition of $F$ in~\eCps{a} is merely
a minor simplification of $F=\inv4[D_-,D_+]\F|$, and the analogous applies
to the definitions~\eCps{b{-}d}. However, since
$[D_+,\Db_+]=2D_+\Db_+{-}2i\vd_\pp$, the definitions of $L_\pp,R_\mm$ in
Eqs.~\eCps{e,f} imply
\eqna\eAux
 $$ \eqalignno{
 D_+\Db_+\L|&=i\vd_\pp\L|+\inv2\big[D_+,\Db_+\big]\L|=i\vd_\pp\ell+2L_\pp~,
  & \eAux{a}\cr
 D_-\Db_-\Y|&=i\vd_\mm\Y|+\inv2\big[D_-,\Db_-\big]\Y|=i\vd_\mm r+2R_\mm~.
  & \eAux{b}\cr}
 $$
These definitions will produce more symmetry in derivations of Lagrangian
densities.

\topic{Component diagrammatics}
In analyzing the component field content of constrained superfields, the
Reader may find it convenient to use the diagram in Fig.~1. A component is
obtained by acting on the superfield with one of the operators from the
diagram, and then setting $\vs^\pm{=}0{=}\vsb^\pm$.
\PixCap{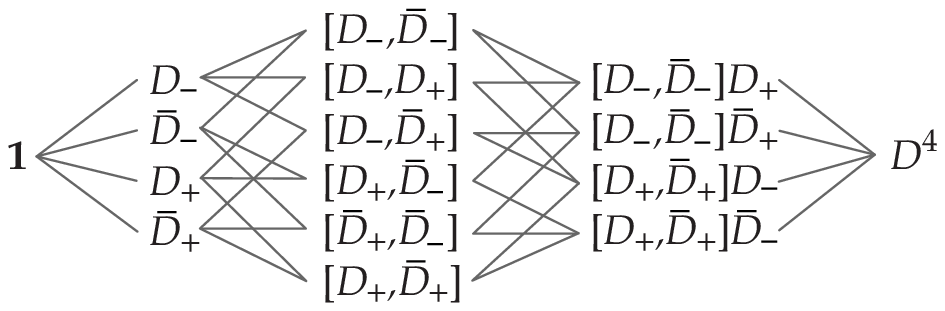}{1}{The sequence of multiple superderivatives used in
defining component fields. Component fields of conjugate superfields are
found using the hermitian conjugates of the operators shown here.}
Applying further (super)derivatives merely produces total derivatives
of the already obtained component fields. In this sense, the  component
fields, onto which the operators in the diagram in Fig.~1 project (upon
setting $\vs^\pm{=}0{=}\vsb^\pm)$, are {\it independent\/}.
 The Reader of a more mathematical persuasion may appreciate the fact that
the diagram in Fig.~1 shows the projection operators (with the setting of
$\vs{=}0{=}\vsb$ understood) on the components of $\wedge^*V$, where $V$ is
a 4-dimensional vector space spanned by $D_\pm,\Db_\pm$. The diagram itself
then represents the exact sequence
\eqn\eXXX{ 1 \to V \to \wedge^2V \to \wedge^3V \to \wedge^4V~. }

 Using the diagram in Fig.~1 as a guide, the components of a general,
unconstrained complex superfield {\bf A} may be listed as
\eqn\eXXX{ {\bf A}
 = \left(\quad\matrix{a\cr}\quad
         \matrix{\a_-\cr \ra_-\cr\a_+\cr \ra_+\cr}\quad
         \matrix{A_\mm\cr A~~\cr A_\mp\cr A_\pm\cr\FF{A}~\cr A_\pp\cr}\quad
         \matrix{@_-\cr \Tw{@}_-\cr @_+\cr \Tw{@}_+\cr}\quad
         \matrix{\D\cr}\quad
         \right) }
Note that the subscripts on $A_\mp,A_\pm$ indicate that they were obtained
from $[D_-,\Db_+]{\bf A}|$ and $[D_+,\Db_-]{\bf A}|$, respectively. Finally,
the numerical coefficients, as in~\eCps{a}, are herein set: $\inv{\sqrt2}$
for every superderivative, and $\inv2$ for every commutator. Thus, for
example,
\eqn\eXXX{{\cmath{
 \ra\define\inv{\sqrt2}\Db_-{\bf A}\Zp~,\qquad
 A\define\inv2 D_-D_+{\bf A}\Zp~,\qquad
 A_\pm\define\inv2D_+\Db_-{\bf A}\Zp~,\cr
 \FF{A}\define\inv2\Db_+\Db_-{\bf A}\Zp~,\qquad
 @_+\define\inv{4\sqrt2}[D_+,\Db_+]D_-{\bf A}\Zp~,\quad
 \hbox{\it etc.}\cr
 }}}
The minor simplifying departure in Eqs.~\eCps{c,d} from this convention,
which dictates $X_\mp,\Xb_\pm$ instead of the simpler $X,\Xb$,
hopefully causes no undue confusion.

\topic{Elimination of component fields}
Consider, for example, a superfield annihilated by $\Db_+$. The component
fields obtained by projection after acting with some superderivatives
including $\Db_+$ either vanish or become total derivatives of
`lower' components. All these components are connected by the criss-cross
lines in Fig.~1, following to the right of $\Db_+$; they are shown in the
shaded area of the diagram in Fig~2.
\PixCap{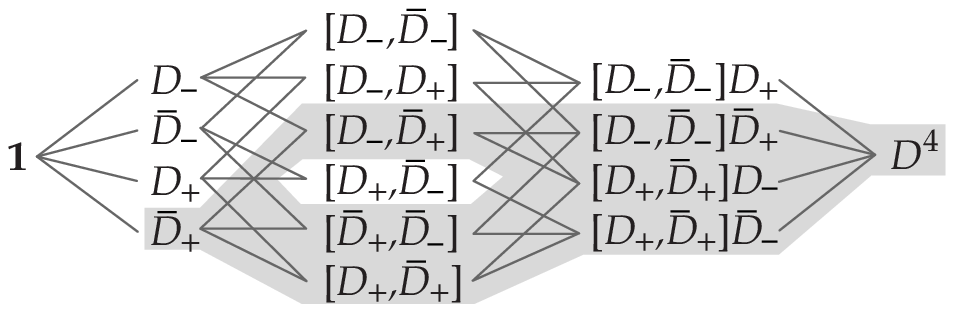}{2}{The (independent) component fields eliminated by
the superconstraint $\Db_+\J{=}0$ are obtained using the derivatives in the
shaded area. Thus, the superfield $\J$, satisfying $\Db_+\J{=}0$ retains
the independent components obtained by projection with the unshaded
superderivatives.}
The superconstraint of being annihilated by $\Db_-$ on the other hand
eliminates the (independent) component fields obtained by projection with
the derivatives in the shaded area of the diagram in Fig.~3.
\PixCap{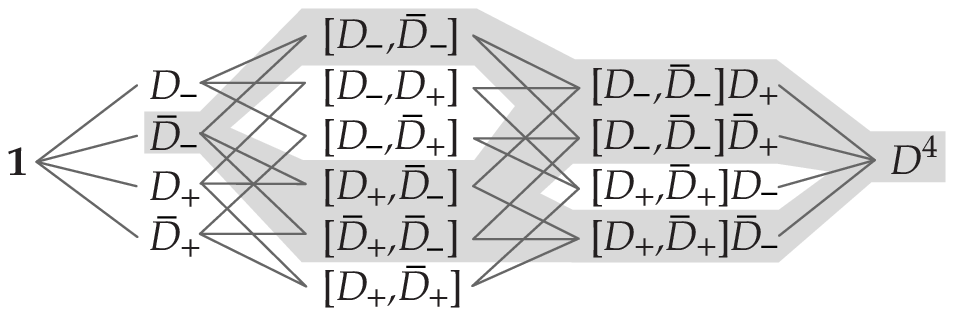}{3}{The (independent) component fields eliminated by
the superconstraint $\Db_-\J{=}0$ are obtained using the derivatives in the
shaded area.}

The chiral superfield, $\F$, satisfies both $\Db_+\F{=}0$ and
$\Db_-\F{=}0$, so the (independent) component fields eliminated by both
superconstraints are in the {\it union\/} of the shaded areas of Figs.~2
and~3; this is shown in Fig.~4.
\PixCap{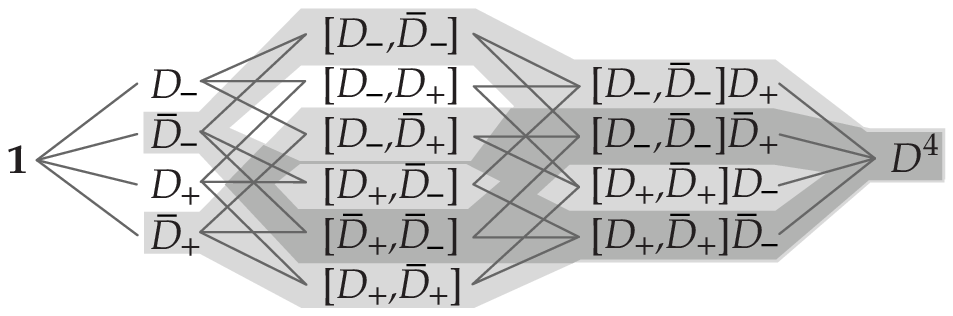}{4}{The (independent) component fields eliminated by
the superconstraint pair $\Db_-\F{=}0{=}\Db_+\F$ are obtained using the
derivatives in the shaded area; their overlap is shaded darker. Unshaded
remain the operators which project on the independent components of the
chiral superfield, $\F$: see Eqs.~\eCps{a}.}
Note that there is an overlap of the two shaded areas: the derivatives
projecting onto the component fields that are elliminated both by 
$\Db_+\F{=}0$ {\it and\/} by $\Db_-\F{=}0$. Presently, this is of no
concern, but will turn up again later.

Clearly, the component field content of all constrained superfields~\eHSF{}
and~\eQSF{} can be treated in much the same way. Unfortunately, while this
does give a quick indication of the {\it independent\/} component field
content, it does not specify the dependences. For example, such diagrams
{\it do\/} indicate that the right-hand side of Eq.~\eAux{a} is a linear
combination of $L_\pp$ and a derivative (clearly, $\vd_\pp$) of a component
field two columns to the left in the diagram (clearly, $\ell$). Such
diagrams, {\it by themselves\/}, do not however specify the numerical
contstants of the linear combination, here `2' and `$i$', respectively.
These details, while not difficult to guess, should be verified by direct
calculation.

\topic{Unidexterity superconstraints}
We have noted before that the pair of superconstraints~\eHSF{e} {\it
includes\/} the spacetime constraint, $\vd_\mm\L=0$, and that~\eHSF{f}
similarly {\it includes\/} $\vd_\pp\Y=0$.
 The quick argument based on
\eqn\eSeCon{ \left.\matrix{D_-\L=0\cr\Db_-\L=0\cr}\right\}
        \To\left.\matrix{\Db_-D_-\L=0\cr D_-\Db_-\L=0\cr}\right\}
        \To\big\{D_-,\Db_-\big\}\L=2i\vd_\mm\L=0~, }
is somewhat misleading in suggesting that $(\vd_\mm\L){=}0$ is somehow a
{\it consequence\/} of~\eHSF{e}, and so a secondary constraint; a
clarification may thus be welcome. To that end, let us study the content of
Eqs.~\eHSF{e} on the component fields. Start with an unconstrained
(general) superfield, $\0\L$, and define `$\rfloor$' to mean setting
$\vs^-{=}0{=}\vsb^-$, but leaving $\vs^+,\vsb^+$ intact. Then, we define
the ({\it still $\vs^+,\vsb^+$-dependent!\/})
$\vs^-,\vsb^-$-component (0,2)-superfields:
\eqn\eLex{ \L\define\0\L\Rf~,\quad
           \0\l_-\define\inv{\sqrt2}D_-\0\L\Rf~,\quad
           \0\bl_-\define\inv{\sqrt2}\Db_-\0\L\Rf~,\quad
           \0L_\mm\define\inv4\big[D_-,\Db_-\big]\0\L\Rf~, }
or $\0\L=(\L;\0\l_-,\0\bl_-;\0L_\mm)$.

 Now project on the $\vs^-,\vsb^-$-component constraints (components of
$D_-\0\L\:0$):
\eqna\eLLng
 $$\eqalignnotwo{
 0&\:(D_-\0\L)\Rf &= \sqrt2\0\l_-~,                           &\eLLng{a}\cr
 0&\:\inv{\sqrt2}\Db_-(D_-\0\L)\Rf
  &= \frc{i}{\sqrt2}(\vd_\mm\L)+\sqrt2\0L_\mm~,               &\eLLng{b}\cr
 0&\:\inv4[D_-,\Db_-](D_-\0\L)\Rf&= \frc{i}{\sqrt2}(\vd_\mm\0\l_-)~,
                                                               &\eLLng{c}\cr
 }$$
and of $\Db_-\0\L\:0$:
 $$\eqalignnotwo{
 0&\:(\Db_-\0\L)\Rf&= \sqrt2\0\bl_-~,                          &\eLLng{d}\cr
 0&\:\inv{\sqrt2}D_-(\Db_-\0\L)\Rf
  &= \frc{i}{\sqrt2}(\vd_\mm\L)-\sqrt2\0L_\mm~,                &\eLLng{e}\cr
 0&\:\inv4[D_-,\Db_-](\Db_-\0\L)\Rf&= \frc{i}{\sqrt2}(\vd_\mm\0\bl_-)~,
                                                               &\eLLng{f}\cr
}$$
where $\:$ is the equality of the enforced (super)constraint. We have used
only the {\it operatorial\/} identity
\eqn\eXXX{ \{D_-,D_-\}=0~,\quad\To\quad
           [D_-,\Db_-]D_-\id\{D_-,\Db_-\}D_-~, }
and its conjugate.

Clearly, the sum of Eqs.\eLLng{b} and~\eLLng{e} equals the spacetime
constraint, $\vd_\mm\L\:0$, for {\it all\/} of the haploid
(0,2)-superfield $\L$. Their difference produces $\0L_\mm\:0$---which is in
point of fact also {\it contained\/} both in~\eLLng{a} and in~\eLLng{d}.
 To summarize then, the superconstraints~\eHSF{e} {\it contain\/} the
(simple) constraints:
\eqn\eSeC{(\vd_\mm\L)\:0~,\quad\hbox{and}\quad\0\l_-,\0\bl_-,\0L_\mm\:0~.}
That is, the vanishing of three quarters ($\0\l_-,\0\bl_-,\0L_\mm$) of the
unconstrained superfield $\0\L$ and the vanishing of the right-moving modes
of all the component fields in the haploid superfield $\L$ are all {\it
contained} in Eqs.~\eHSF{e}; their content is {\it identical\/} to that of
Eqs.~\eSeC!

Thus, the various stages (columns) of the
`derivation'~\eSeCon\ are {\it consistent\/}, but also unnecessary.
Eqs.~\eSeC\ merely expand the original superdifferential system~\eHSF{e},
which contains all the (super)constraints; there are no further, induced,
(super)constraints to be derived in the Lagrangian formalism used here. The
strategy in the Hamiltonian (Dirac, BRST, \etc) formalism will depend
crucially on the conjugate momenta, and so on the choice of the Lagrangian
density. This again is beyond our present scope, as we seek to present the
(reasonably) general case rather than belabor the details---and consequent
restrictions---of specific models.

\topic{Quartoid superfield components}
The component fields of the quartoid superfields~\eQSF{} are easily
obtained, by truncation, from those of the haploid ones~\eCps{}:
\eqna\eQps
 $$\twoeqsalignno{
 \f_\LL&\define\F_\LL|~, \quad&\quad
 \j_{\LL+}&\define\inv{\sqrt{2}}D_+\F_\LL|~; &\eQps{a}\cr
 \fb_\LL&\define\FB_\LL|~, \quad&\quad
 \jb_{\LL+}&\define\inv{\sqrt{2}}\Db_+\FB_\LL|~; &\eQps{b}\cr
 \f_\RR&\define\F_\RR|~, \quad&\quad
 \j_{\RR-}&\define\inv{\sqrt{2}}D_-\F_\RR|~; &\eQps{c}\cr
 \fb_\RR&\define\FB_\RR|~, \quad&\quad
 \jb_{\RR-}&\define\inv{\sqrt{2}}\Db_-\FB_\RR|~. &\eQps{d}\cr
 }$$

All the superfields defined in Eqs.~\eHSF{} and~\eQSF{} are spin-0
superfields, in that their lowest components~\eCps{} are all spin-0 fields.
For these to be standard physical (pseudo)scalar fields, we fix their
canonical dimension to be the standard 0. Thence, the canonical dimensions
of all spinors in~\eCps{} is the standard $+\inv2$.

\subsec{Spinning superfields and superdifferential relations}\noindent
\subseclab\sDSFlds
The Eqs.~\eHSF{} define {\it scalar\/} superfields, \ie, the lowest
component fields of $\F,\FB,\X,\XB,\L,\Y$ are scalar fields. It is easy to
endow the whole superfield with (overall) additional spin, say $\F\to\F^+$,
so that the lowest component $\F^+|=\f^+$ has spin $+\inv2$ (see
appendix~A). Next, it is also possible to relate the (anti)chiral and the
twisted (anti)chiral superfields via a superdifferential relation. For
example, let $\F^+$ be a spin-up (overall spin $+\inv2$) chiral superfield;
then $\X'\define(D_+\F^+)$ is a scalar (overall spin 0) twisted
chiral superfield, since
\eqn\eXXX{ D_+(D_+\F^+)~=~0~=~\Db_-(D_+\F^+)~. }
In particular, note that although $\Db_+\F^+=0$,
\eqn\eXXX{ \Db_+(D_+\F^+) = (2i\vd_\pp-D_+\Db_+)\F^+
 = 2i\vd_\pp\F^+\neq0~. }
This then implies that there will exist (differential) relations among the
component fields of $\F^+$ and $\X'\define(D_+\F^+)$, as follows:
 $$ \eqalign{
   x'&\define\X'|=\big(D_+\F^+\big)|=\sqrt2\j_+^+~,\cr
 \x'_-&\define\inv{\sqrt{2}}D_-\X'|
            =\inv{\sqrt{2}}\big(D_-D_+\F^+\big)|=\sqrt2F^+~,\cr
 \x'_+&\define\inv{\sqrt{2}}\Db_+\X'|
            =\inv{\sqrt{2}}\big(\Db_+D_+\F^+\big)|=\sqrt2i(\vd_\pp\f^+)~,\cr
    X'&\define\inv2D_-\Db_+\X'|
            =\inv2\big(D_-\Db_+D_+\F^+\big)|=\sqrt2i(\vd_\pp\j_-^+)~.\cr
 }$$

Also, since the canonical (scaling) dimension of $D_\pm$ and
$\Db_\pm$ is $[D_\pm]=[\Db_\pm]=+\inv2$, the component fields in
$\X'\define D_+\F^+$ will have their canonical dimensions shifted by
$+\inv2$ compared to those in $\F^+$. For example, the lowest component of
$\F^+$ is the spinor $\f^+$ and so should (in 2-dimensional spacetime), have
$[\f^+]=\inv2$; then $[\j_\pm^+]=1$ and $[F^+]=\frc32$. Thus, $[x']=1$,
$[\x'_\pm]=\frc32$, and $[X']=2$---none of which are the `correct'
canonical dimensions of physical (propagating) fields in 2-dimensional
spacetime. If a mass/energy parameter, $m$, is available in the model, the
rescaled superfield $\Tw\X\define\inv{m}(D_+\F^+)$ has component fields of
`correct' (physical) canonical dimensions.

Alternatively, with $\X$ a scalar (overall spin-0) twisted chiral
superfield, $\F_+\define(\Db_+\X)$ is a spin $-\inv2$ chiral
superfield since
\eqn\eXXX{ \Db_+(\Db_+\X)~=~0~=~\Db_-(\Db_+\X)~. }
Again, the component fields of $\X$ and $\F_+\define(D_+\X)$ are
related as follows:
 $$ \eqalign{
   \f_+&\define\F_+|=(\Db_+\X)|=\sqrt2\c_+~,\cr
     \j&\define\inv{\sqrt{2}}D_-\F_+|
              =\inv{\sqrt{2}}\big(D_-\Db_+\X\big)|=\sqrt2X~,\cr
 \j_\pp&\define\inv{\sqrt{2}}\Db_+\F_+|
              =\inv{\sqrt{2}}\big(D_+\Db_+\X\big)|=\sqrt2i(\vd_\pp x)~,\cr
    F_+&\define\inv2D_-D_+\F_+|
              =\inv2\big(D_-D_+\Db_+\X\big)|=\sqrt2i(\vd_\pp\x_-)~.\cr
 }$$
Again, the canonical (scaling) dimensions of the component fields of
$\F_+\define(\Db_+\X)$ will each have their canonical dimensions shifted by
$+\inv2$ compared to those in $\X$. So, if all the component fields of $\X$
have `correct' canonical dimensions, not all of those of $\F_+$ do. Whilst
$[\f_+]=[\c_+]=\inv2$ is the correct canonical dimension for a dynamical
spinor, and $[\j_\pp]=[\vd_\pp x]=1$ and $[\j]=[F]=1$ are correct for a
component of a vector and an auxiliary field, respectively,
$[F_+]=[\vd_\pp\x_-]=\frc32$ is too high for a dynamical spinor.

This suggests that superderivatives of superfields generally have fewer
dynamical component fields than the original superfields. Furthermore, the
unphysical canonical dimensions of the component fields of
superderivative superfields allow for higher derivative terms in the
action. Throughout this article and with one noted exception below, we will
include such superderivative superfields only if this does not lead to
higher derivative terms.

Note that no superderivative of either of $\F,\FB,\X,\XB$ satisfies the
unidexterous superconstraints~\eHSF{e,f}. Thus, it is not possible to
construct lefton or righton superfields from the (twisted) (anti)chiral
superfields $\F,\FB,\X,\XB$.

Conversely however, superderivatives of leftons and rightons do satisfy some
of the (twisted) (anti)chiral conditions. In fact, these are annihilated by
{\it three\/} superderivatives and are {\it quartoid\/}, no longer merely
haploid, superfields. Thus:
\item{$\bullet$} $(\Db_+\L)$ is a chiral lefton~\eQSF{a}, and
                 $(D_+\L)$ an antichiral lefton~\eQSF{b},
\item{$\bullet$} $(\Db_-\Y)$ is a chiral righton~\eQSF{c}, and
                 $(D_-\Y)$ an antichiral righton~\eQSF{d}.

\subsec{Second-order constrained superfields}\noindent
\subseclab\ssScnd
The hitherto discussed and thoroughly familiar (anti)chiral
superfields~\eHSF{a,b}, their somewhat less well known twisted
counterparts~\eHSF{c,d}, and the unidexterous fields~\eHSF{e,f}
and~\eQSF{}, are all defined by satisfying a system of {\it first order\/}
superconstraints. Clearly, there is more.

The obvious generalization of the (quite simple) first-order
superconstraints~\eHSF{} and~\eQSF{} are the (equally simple) {\it
second-order\/} superconstraints:
\eqna\eNMF
 $$ \twoeqsalignno{
 \[61. & \hbox{\bf NM-Chiral}:
 &\qquad  & \Db_+\Db_-\Q~ = ~0~, &\eNMF{a} \cr
 \[62. & \hbox{\bf NM-Antichiral}:
 &\qquad  & D_-D_+\QB~ = ~0~~, &\eNMF{b} \cr
 \[63. & \hbox{\bf NM-Twisted-chiral}:
 &\qquad  & D_+\Db_-\P~ = ~0~,   &\eNMF{c} \cr
 \[64. & \hbox{\bf NM-Twisted-antichiral}:
 &\qquad  & D_-\Db_+\PB~= ~0~, &\eNMF{d} \cr
 \[65. & \hbox{\bf NM-(Almost)-Lefton}:
 &\qquad  & [D_-,\Db_-]{\bf A}~=~0~,     &\eNMF{e} \cr
 \[66. & \hbox{\bf NM-(Almost)-Righton}:
 &\qquad  & [D_+,\Db_+]{\bf U}~=~0~.     &\eNMF{f} \cr}
 $$
These are called `linear', or as better befits the extendedness of their
defining superconstraints in comparison with~\eHSF{},
`non-minimal'~\refs{\rNMJiWa,\rGGRS,\rUseNM}. Note that the `minimal'
haploid superfields~\eHSF{} obey the second-order superconstraints~\eNMF{},
but that the non-minimal ones~\eNMF{} do not obey the first-order
superconstraints~\eHSF{}.

Again, the latter two of the superconstraints~\eNMF{} are set apart from the
former four, although the divide is not as sharp as between the `minimal'
unidexterous superfields~\eHSF{e,f} from the ambidexterous
ones~\eHSF{a{-}d}. To see this, let us examine their component field
content.

\topic{Ambidexterous non-minimal superfields}
The component field content is straightorward to obtain, only this time
there are more component fields than in Eqs.~\eCps{}. For example, 
\eqna\eNMCp
 $$\cmath{
 t\define\Q|~,\quad \q_\pm\define\inv{\sqrt2}D_\pm\Q|~,\quad
 \vq_\pm\define\inv{\sqrt2}\Db_\pm\Q|~,\cr
 T\define\inv2D_-D_+\Q|~,\quad
 T_\mp\define\inv2D_-\Db_+\Q|~,\quad
 T_\pm\define\inv2D_+\Db_-\Q|~,\cr
 T_\mm\define\inv4[D_-,\Db_-]\Q|~,\quad
 T_\pp\define\inv4[D_+,\Db_+]\Q|~,\cr
 \t_-\define\inv{4\sqrt2}[D_-,\Db_-]D_+\Q|~,\quad
 \t_+\define\inv{4\sqrt2}[D_+,\Db_+]D_-\Q|~.\cr
 }\eqno\eNMCp{a}$$
Note that the $\pm$ and $\mp$ substripts on $T_\pm,T_\mp$ indicate
{\it single\/} components, whereas the $\pm$ on the fermions
$\q_\pm,\vq_\pm,\t_\pm$ indicates a {\it choice\/} of spin $+\inv2$ and
$-\inv2$. Similarly,
 $$\cmath{
 p\define\P|~,\quad \p_\pm\define\inv{\sqrt2}D_\pm\P|~,\quad
 \vp_\pm\define\inv{\sqrt2}\Db_\pm\P|~,\cr
 P\define\inv2D_-D_+\P|~,\quad
 P_\mp\define\inv2D_-\Db_+\P|~,\quad
 \FF{P}\define\inv2\Db_+\Db_-\P|~,\cr
 P_\mm\define\inv4[D_-,\Db_-]\P|~,\quad
 P_\pp\define\inv4[D_+,\Db_+]\P|~,\cr
 \Tw\vf_-\define\inv{4\sqrt2}[D_-,\Db_-]\Db_+\P|~,\quad
 \vf_+\define\inv{4\sqrt2}[D_+,\Db_+]D_-\P|~.\cr
 }\eqno\eNMCp{b}$$
The component fields of $\QB$ and $\PB$ are then obtained by hermitian
conjugation.
\PixCap{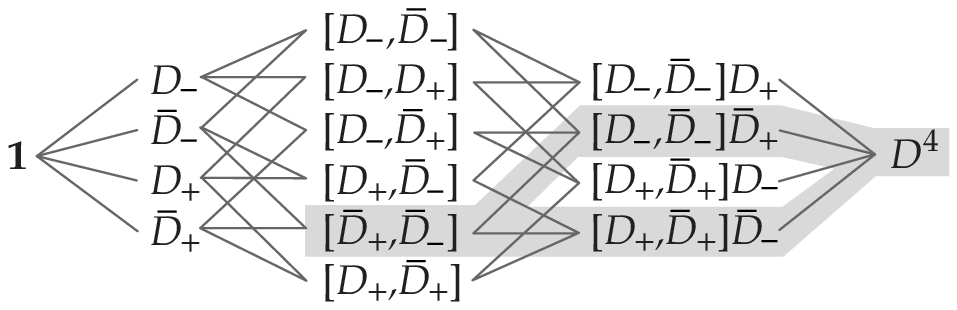}{5}{The (independent) component fields eliminated by
the superconstraint $\Db_+\Db_-\Q{=}\inv2[\Db_+,\Db_-]\Q{=}0$ are obtained
using the derivatives in the shaded area. The unshaded derivatives project
on the independent component fields of $\Q$, as defined by Eqs.~\eNMCp.}
Note that the shaded area of Fig.~5 coincides with the darker shaded area
of Fig.~4. Indeed, trivially so: the former covers the derivatives defining
the component fields eliminated by projection with $\Db_+\Db_-$, the latter
--- with either of $\Db_+,\Db_-$.

Comparing the component field content of corresponding superfields, the
non-minimal superfields~\eNMF{a{-}d} contain three times as many
component fields as do the `minimal' ones~\eHSF{a{-}d}. As it turns
out, however, with rather typical Lagrangian densities for such superfields
(even when interacting with some of the others), the equations of motion
for the `extra' component fields turn out to be algebraic (see, \eg,
Ref.\rGGRS, p.200). This allows an easy ellimination of all the `extra'
component fields, and the remaining component field content of the
non-minimal superfields~\eNMF{a{-}d} turns out to be identical to
their `minimal' counterparts~\eHSF{a{-}d}. To indicate this, we may
list the component field context as
\eqna\eNMCP
 $$\eqalignno{
 \Q &=(t;\q_\pm;T:\bar\vq_\pm;T_\mp,T_\pm,T_\mm,T_\pp;\t_\pm)~,&\eNMCP{a}\cr
 \QB&=(\bt;\qb_\pm;\Tb:\vq_\pm;\Tb_\mp,\Tb_\pm,\Tb_\mm,\Tb_\pp;\tb_\pm)~,
                                                               &\eNMCP{b}\cr
 \P&=(p;\p_-,\vp_+;P:
     \vp_-,\p_+;P_\mp,\FF{P},P_\mm,P_\pp;\Tw\vf_-,\vf_+)~,&\eNMCP{c}\cr
 \PB&=(\bp;\pb_-,\ba\vp_+;\Pb:
     \ba\vp_-,\pb_+;\skew{-4}\bar{\FF{P}},\Pb_\pm,\Pb_\mm,\Pb_\pp;
                             \skew3\Tw{\bar\vf}_-,\bar\vf_+)~, &\eNMCP{d}\cr
 }$$
where the colon separates off the `extra' component fields by which
non-minimal superfields differ from their `minimal' counterparts.

 Thus, the physical component field content warrants the adjective `haploid'
also for the non-minimal superfields, as they depend on the `other half' of
the four supercoordinates, $\vs^\pm,\vsb^\pm$ only by the inclusion of
auxiliary component fields.

Furthermore, the non-minimal superfields~\eNMF{} are {\it dual\/} to their
first-order constrained counterparts~\eHSF{}~(see, \eg, Ref.~\rGGRS,
p.200). This tempts a straightforward dismissal of the non-minimal
superfields. However, while the use of these non-minimal superfields {\it
instead\/} of the `minimal' ones seems to merely harbor unnecessary toil,
there do exist distinct benefits of using the `minimal' and the non-minimal
superfields {\it together\/}~\refs{\rUseNM}. We will therefore include them
in the subsequent analysis, where appropriate.
\PixCap{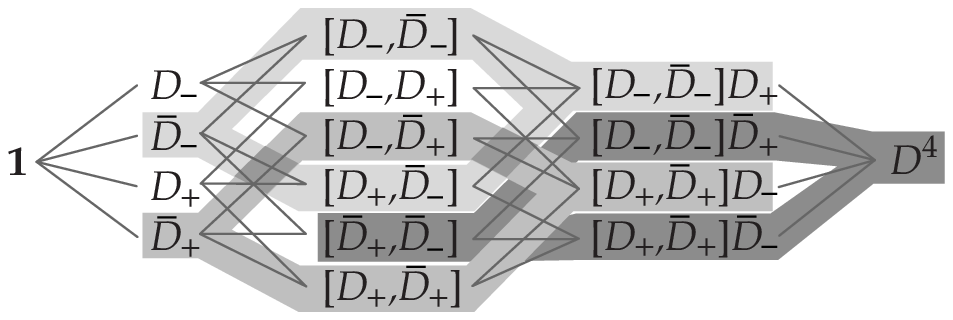}{6}{The darkest shade indicates the derivatives defining
the (independent) component fields eliminated by the superconstraint
$\Db_+\Db_-\Q{=}0$. The medium and lightest shade indicate the two groups
of derivatives which jointly define the `extra' component fields. Compare
this with Fig.~4.}
The above results are displayed diagrammatically in Fig.~6. Note that the
derivatives in the lightest and medium shaded areas separately each have the
structure of the derivatives defining the (independent) component fields of
a `minimal' chiral superfield.
 This motivated the realization that the definition~\eNMF{a} is invariant
with respect to the (rather peculiar) inhomogeneous gauge transformation
\eqna\eSgg
 $$
    \d\Q ~=~ \vsb^-\F_-+\vsb^+\F_+~,\quad\hbox{where}\quad
           \Db_\pm\F_-=0=\Db_\pm\F_+~, \eqno\eSgg{a}
 $$
parametrized by a spin $\pm\inv2$ pair of chiral superfields, $\F_\pm$.
Indeed, the (independent) components of $\vsb^-\F_-$ and $\vsb^+\F_+$ are
projected precisely by those derivatives which are located in the lighly-
and medium-shaded area of Fig.~6---the `extra' component fields of $\Q$. The
component fields in $\F_\pm$ then can be used to gauge away the `extra'
component fields in $\Q$. In other words,
\eqn\eGEq{ \Q~~\hbox{mod}~(\vsb^-\F_-+\vsb^+\F_+) \simeq \F~. }

By the same token, the definitions~\eNMF{b{-}d} are, respectively,
invariant with respect to the following inhomogeneous gauge transformations:
 $$\twoeqsalignno{
 \d\QB &= \vs^-\FB_-+\vs^+\FB_+~,\quad&\hbox{where}\quad
                                  &D_-\FB_\pm=0=D_+\FB_\pm~,   &\eSgg{b}\cr
 \d\P &= \vs^-\X_-+\vsb^+\X_+~,\quad&\hbox{where}\quad
                                  &\Db_-\X_\pm=0=D_+\X_\pm~,   &\eSgg{c}\cr
 \d\PB &= \vsb^-\XB_-+\vs^+\XB_+~,\quad&\hbox{where}\quad
                                  &D_-\XB_\pm=0=\Db_+\XB_\pm~. &\eSgg{d}\cr
}$$
Of course, most Lagrangian densities will explicitly violate this gauge
symmetry~\eSgg{}, and a possible use of this gauge symmetry remains an open
question for a later study.

\topic{(Almost) Unidexterous non-minimal superfields}
Before we specify the component field content of the non-minimal
superfields~\eNMF{e,f}, let us examine the content of the second order
superdifferential contraints~\eNMF{e,f}.
 To this end, consider a previously unconstrained general (2,2)-superfield,
$\0{\bf A}$ with ({\it still $\vs^+,\vsb^+$-dependent!\/}) component
(0,2)-superfield content $\0{\bf A}=({\bf A};\0\a_-,\0a_-;\0A_\mm)$, and
project on the components of $[D_-,\Db_-]\0{\bf A}$:
\eqna\eNMLng
 $$\eqalignnotwo{
 0&\:\big([D_-,\Db_-]\0{\bf A}\big)\Rf
  &=\0A_\mm~,                                                &\eNMLng{a}\cr
 0&\:\big(\inv{\sqrt2}D_-[D_-,\Db_-]\0{\bf A}\big)\Rf
  &=-2i(\vd_\mm\0\a_-)~,                                     &\eNMLng{b}\cr
 0&\:\big(\inv{\sqrt2}\Db_-[D_-,\Db_-]\0{\bf A}\big)\Rf
  &=+2i(\vd_\mm\0a_-)~,                                      &\eNMLng{c}\cr
 0&\:\big(\inv2[D_-,\Db_-][D_-,\Db_-]\0{\bf A}\big)\Rf
  &=-(\vd_\mm^2{\bf A})~.                                    &\eNMLng{d}\cr
 }$$

The component fields are easily defined in analogy with Eqs.~\eNMCp{}, and
we denote them ${\bf A}
 =(\ra;\a_+,a_+;A_\pp:\a_-,a_-;A_\mm,A,A_\mp,A_\pm,\FF{A};@_+,\Tw{@}_+)$.

Just as with the ambidexterous non-minimal superfields, the defining
superconstraint~\eNMF{e} is again invariant with respect to the
inhomogeneous gauge transformation
 $$
    \d{\bf A}~=~ \vs^-\L^1_-+\vsb^-\L^2_-~,\quad\hbox{where}\quad
           D_-\L^i_-=0=\Db_-\L^i_-~,\quad i=1,2~, \eqno\eSgg{e}
 $$
which precisely suffices to elliminate the `extra' component fields,
$\a_-,a_-$, $A_\mm$, $A$, $A_\mp$, $A_\pm$, $\FF{A}$, $@_+,\Tw{@}_+$, all
of which are contained in $\0\a_-$ and $\0a_-$. The remaining component
fields are $(\ra;\a_+,a_+;A_\pp)$, and are contained in ${\bf A}$ but
neither in $\0\a_-$ nor in $\0a_-$. This has the peculiar consequence that
the component fields $(\ra;\a_+,a_+;A_\pp)$ obey the
superconstraint~\eNMLng{d}, but not~\eNMLng{b,c}. That is, the component
fields $(\ra;\a_+,a_+;A_\pp)$ are not purely left-movers, but are linear
functions of $\s^\mm$ and arbitrary functions of $\s^\pp$.

The analogous analysis shows that the non-minimal righton has the following
component fields
${\bf U}=(u;\n_-,\y_-;U_\mm:\n_+,\y_+;U,U_\mp,U_\pm,\FF{U};\m_-,\mt_-)$. Its
definition is invariant with respect to the gauge transformation
 $$
    \d{\bf U}~=~ \vs^+\Y^1_++\vsb^+\Y^2_+~,\quad\hbox{where}\quad
           D_+\Y^i_+=0=\Db_+\Y^i_+~,\quad i=1,2~, \eqno\eSgg{f}
 $$
which elliminates the `extra' component fields,
$\n_+,\y_+,U,U_\mp,U_\pm,\FF{U},\m_-,\mt_-$, contained in $\0\r_+$
and $\0\vr_+$. The remaining component fields are $(u;\n_-,\y_-;U_\mm)$, and
are linear functions of $\s^\pp$ and arbitrary functions of $\s^\mm$.

Finally, we recall that an analytic function of any one type of `minimal'
haploid (quartoid) superfields~\eHSF{} and~\eQSF{} is again a haploid
(quartoid) superfield, and of the same type. For example, any analytic
function $f(\F)$ of chiral superfields, $\F^\m$, is again chiral:
$f(\F)$ satisfies the same superconstraints~\eHSF{a}. In contradistinction,
only linear combinations of non-minimal superfields of the same
type~\eNMF{a{-}f} satisfy the same superconstraint. No {\it
non-linear\/} function of any non-minimal superfields satisfies any one of
the non-minimal superconstraints~\eNMF{a{-}f}---they are all general,
unconstrained superfields.

\topic{Minimization of non-minimal superfields}
As observed in the literature~\rGGW, superderivatives of non-minimal
superfields may become minimal. For example, consider the non-minimal
chiral superfield~\eNMF{a}, $\Q$, satisfying $\Db_+\Db_-\Q=0$. Then both
$(\Db_+\Q)$ and $(\Db_-\Q)$ are minimal chiral superfields; indeed,
\eqn\eXXX{{\twoeqsalign{ 
 \Db_+(\Db_+\Q)&\id~0~, \quad&\quad \Db_-(\Db_+\Q)&=~0~; \cr
 \Db_-(\Db_-\Q)&\id~0~, \quad&\quad \Db_+(\Db_-\Q)&=~0~. \cr }}}
However, $(D_+\Q)$ and $(D_-\Q)$ are not even non-minimal haploid
superfields: they are annihilated by only one of the four superderivatives.

In this way, every non-minimal ambidexterous superfield~\eNMF{a}-\eNMF{d}
provides two minimal haploid superfields~\eHSF{a{-}d} of the
corresponding kind:
\eqna\eNMM
 $$\eqalignnotwo{
 (\Db_+\Q),~(\Db_-\Q)&\qquad&\hbox{are chiral~\eHSF{a}}~,       &\eNMM{a}\cr
 (D_+\QB),~(D_-\QB)&\qquad&\hbox{are antichiral~\eHSF{b}}~,     &\eNMM{b}\cr
 (D_+\P),~(\Db_-\P)&\qquad&\hbox{are twisted-chiral~\eHSF{c}}~, &\eNMM{c}\cr
 (\Db_+\PB),~(D_-\PB)&\qquad&\hbox{are twisted-antichiral~\eHSF{d}}~.
                                                                &\eNMM{d}\cr
}$$

The same is not true for the unidexterous superfields. Consider the
non-minimal lefton, {\bf A}, satisfying $[D_-,\Db_-]{\bf A}=0$. Then,
\eqn\eXXX{{\twoeqsalign{ 
 \Db_-(\Db_-{\bf A})&\id~0~, \quad&\hbox{but}\quad
  D_-(\Db_-{\bf A})&=i(\vd_\mm{\bf A})\neq~0~; \cr
  D_-(D_-{\bf A})&\id~0~, \quad&\hbox{but}\quad
 \Db_-(D_-{\bf A})&=i(\vd_\mm{\bf A})\neq~0~. \cr }}}
However, there are analogous relations between unidexterous haploid and
quartoid superfields. It is easy to check that
\eqna\eHQ
 $$\eqalignnotwo{
 (\Db_+\L)&\qquad &\hbox{is a chiral lefton~\eQSF{a}}~,       &\eHQ{a}\cr
 (D_+\L)&\qquad &\hbox{is an antichiral lefton~\eQSF{b}}~,    &\eHQ{b}\cr
 (\Db_-\Y)&\qquad &\hbox{is a chiral righton~\eQSF{c}}~,      &\eHQ{c}\cr
 (D_-\Y)&\qquad &\hbox{is an antichiral righton~\eQSF{d}}~.   &\eHQ{d}\cr
}$$

\topic{Effectively minimal haploid superfields}
To summarize, the following (superderivative) superfields satisfy the
specified haploid and quartoid superconstraints:
\eqna\eDSF
 $$\eqalignno{
 &\[4\hbox{Chiral, obeying Eq.}~\eHSF{a}:\cr
 &\F; (\Db_+\X),(\Db_-\XB),(\Db_+\L),(\Db_-\Y);(\Db_+\Q),(\Db_-\Q);\cr
 &(\Db_+\Db_-\FB),(\Db_+D_+\L),(\Db_-D_-\Y),(\Db_+D_+\P),(\Db_-D_-\PB)~;
                                                               &\eDSF{a}\cr
 &\[4\hbox{Antichiral, obeying Eq.}~\eHSF{b}:\cr
 &\FB; (D_-\X),(D_+\XB),(D_+\L),(D_-\Y);(D_+\QB),(D_-\QB);\cr
 &(D_-D_+\F),(D_+\Db_+\L),(D_-\Db_-\Y),(D_-\Db_-\P),(D_+\Db_+\PB)~;
                                                               &\eDSF{b}\cr
 &\[4\hbox{Twisted-chiral, obeying Eq.}~\eHSF{c}:\cr
 &\X; (D_+\F),(\Db_-\FB),(D_+\L),(\Db_-\Y);(D_+\P),(\Db_-\P);\cr
 &(D_+\Db_-\FB),(D_+\Db_+\L),(\Db_-D_-\Y),(D_+\Db_+\Q),(\Db_-D_-\QB)~;
                                                               &\eDSF{c}\cr
 &\[4\hbox{Twisted-antichiral, obeying Eq.}~\eHSF{d}:\cr
 &\XB; (\Db_+\FB),(D_-\F),(\Db_+\L),(D_-\Y);(\Db_+\PB),(D_-\PB);\cr
 &(D_-\Db_+\X),(\Db_+D_+\L),(D_-\Db_-\Y),(D_-\Db_-\Q),(\Db_+D_+\QB)~;
                                                               &\eDSF{d}\cr
 &\[4\hbox{Leftons, obeying Eq.}~\eHSF{e}:\cr
 &\L; (D_+\L),(\Db_+\L); (\Db_+D_+\L),(D_+\Db_+\L);            &\eDSF{e}\cr
 &\[4\hbox{Rightons, obeying Eq.}~\eHSF{f}:\cr
 &\Y; (D_-\Y),(\Db_-\Y); (D_-\Db_-\Y),(\Db_-D_-\Y);            &\eDSF{f}\cr
 }$$
and
 $$\eqalignnotwo{
 &\[4\hbox{Chiral leftons, obeying Eq.}~\eQSF{a}:\quad
 &\F_\LL; (\Db_+\L); (\Db_+D_+\L);                             &\eDSF{g}\cr
 &\[4\hbox{Antichiral leftons, obeying Eq.}~\eQSF{b}:\quad
 &\FB_\LL; (D_+\L); (D_+\Db_+\L);                              &\eDSF{h}\cr
 &\[4\hbox{Chiral rightons, obeying Eq.}~\eQSF{c}:\quad
 &\F_\RR; (\Db_-\Y); (\Db_-D_-\Y);                             &\eDSF{i}\cr
 &\[4\hbox{Antichiral rightons, obeying Eq.}~\eQSF{d}:\quad
 &\FB_\RR; (D_-\Y); (D_-\Db_-\Y).                              &\eDSF{j}\cr
 }$$
The order of derivatives has been terminated at two only so as to prohibit
higher derivative and/or negative-dimension coefficient terms in the
Lagrangian density.

Finally, as it will be useful when constructing Lagrangian densities, we
list (superderivative) superfields which are annihilated by at least one of
the superderivatives. Any analytic function of these is then also
annihilated by at least that one superderivative, and forms the
{\it kernel\/} of that superderivative:
\eqna\eKerD
 $$\eqalignno{
 \ker(D_-)&=\Big\{\FB,\XB,\L,\F_\LL,(D_+\QB),(\Db_+\PB),
                  \Big\{{\SSS D_+(\hbox{\eightpoint preceding})\atop
                         \SSS\bar{D}_+(\hbox{\eightpoint preceding})}\Big\},
                  D_-(\hbox{anything})\Big\}~;                &\eKerD{a}\cr
 \ker(D_+)&=\Big\{\FB,\X,\Y,\F_\RR,(D_-\QB),(\Db_-\P),
                  \Big\{{\SSS D_-(\hbox{\eightpoint preceding})\atop
                         \SSS\bar{D}_-(\hbox{\eightpoint preceding})}\Big\},
                  D_+(\hbox{anything})\Big\}~;                &\eKerD{b}\cr
 \ker(\Db_+)&=\Big\{\F,\XB,\Y,\FB_\RR,(\Db_-\Q),(D_-\PB),
                  \Big\{{\SSS D_-(\hbox{\eightpoint preceding})\atop
                         \SSS\bar{D}_-(\hbox{\eightpoint preceding})}\Big\},
                  \Db_+(\hbox{anything})\Big\}~;              &\eKerD{c}\cr
 \ker(\Db_-)&=\Big\{\F,\X,\L,\FB_\LL,(\Db_+\Q),(D_+\P),
                  \Big\{{\SSS D_+(\hbox{\eightpoint preceding})\atop
                         \SSS\bar{D}_+(\hbox{\eightpoint preceding})}\Big\},
                  \Db_-(\hbox{anything})\Big\}~.              &\eKerD{d}\cr
}$$

\subsec{Other (super)constraints}\noindent
\subseclab\ssOthr
Having listed the simple first-order and second-order
superconstraints~\eHSF{} and~\eNMF{}, respectively, it should be clear that
even this is {\it by far\/} not all there is. One may also define
constrained superfields by means of less simple superderivative equations.
Let us denote the sixteen components of a so far unconstrained superfield:
\eqn\eXXX{ \W
 =\left(\quad\matrix{w\cr}\quad
        \matrix{\w_-\cr \vp_-\cr\w_+\cr \vp_+\cr}\quad
        \matrix{W_\mm\cr W~~\cr W_\mp\cr W_\pm\cr\FF{W}~\cr W_\pp\cr}\quad
        \matrix{o_-\cr \Tw{o}_-\cr o_+\cr \Tw{o}_+\cr}\quad
        \matrix{{\cal O}\cr}\quad
         \right)~. }

\topic{Yet higher order superconstraints}
Having examined first and second order (simple) superconstraints in some
detail, the inquisitive Reader will naturally wonder about third order
superconstraints. The second order constraints examined in \SS\,\ssScnd\
seem to produce merely a technically more involved cousins of the `minimal'
haploid superfields~\ssFrst. Whereas the `minimal' superfields are
constrained by {\it two\/} simple, first order superconstraints which set
three quarters of the component fields to zero, the non-minimal ones are
constrained by a single, second-order superconstraint which sets only one
quarter of the component fields to zero; the balance of a half of the
component fields (eight of them) present in a non-minimal superfield but not
in a `minimal' one are the `extra' component fields which in known
applications turn out to be non-physical\ft{Recall that the property of a
component field being physical, \ie, propagating (with a differential
equation of motion) depends most crucially on the choice of the
Lagrangian density!}.

Na{\"\ii}vely then, we expect a cubic (simple) superconstraint to set only
an eighth (two) of the component fields to zero. Also, we expect several of
the non-vanishing components turning out to be auxiliary (with algebraic,
\ie, non-differential equations of motion), the precise identification
possibly strongly depending on the choice of the Lagrangian density.

This turns out to be only partially true; in particular, the first
expectation turns out to be wrong.

Let us examine the case
\eqn\eCbc{ [D_-,\Db_-]D_+\W=0~. }
Easily, by projecting on the lowest component of~\eCbc, this includes the
simple statement that
\eqn\eXXX{ 0\:[D_-,\Db_-]D_+\W\Zp\define4\sqrt2o_-~, }
so $o_-=0$. Projecting with $D_-,\Db_-$ and $\Db_-$ ($D_+$ of course
trivially annihilates the whole superconstraint), we obtain the
next three component constraints:
\eqn\eXXX{ 0\:-4i(\vd_\mm W)~,\qquad 0\:-4(\vd_\mm W_\pm)~,\qquad
           0\:4i(\vd_\pp W_\mm) -8{\cal O}~.}
These imply, respectively, that $W$ and $W_\pm$ are left-movers, and that
the `top' component field, $\cal O$, is not independent, but equals
${\cal O}=\frc i2(\vd_\pp W_\mm)$. Continuing this, with the quadratic and
cubic superdifferential operators from Fig.~1 which do not include $D_+$,
we obtain four more constraints. The net effect of the
superconstraint~\eCbc\ is then summarized by writing (in a hopefully obivous
doublet notation):
\eqna\eCbC
 $$\twoeqsalignno{
 \vd_\mm^2\{\w_+,W_\pp\}&=\{0,\frc i2(\vd_\mm^2\vd_\pp w)\}~, \quad&\quad
 \{w,\vp_+\}&\hbox{is free,}                                   &\eCbC{a}\cr
 \vd_\mm\{W_\pm,\Tw{o}_+\}&=\{0,\frc i2(\vd_\mm\vd_\pp\vp_-)\}~, \quad&\quad
 \{\vp_-,\FF{W}\}&\hbox{is free,}                              &\eCbC{b}\cr
 \vd_\mm\{W,o_+\}&=\{0,\frc i2(\vd_\mm\vd_\pp\w_-)\}~, \quad&\quad
 \{\w_-,W_\mp\}&\hbox{is free,}                                &\eCbC{c}\cr
 \{o_-,{\cal O}\}&=\{0,\frc i2(\vd_\pp W_\mm)\}~, \quad&\quad
 \{W_\mm,\Tw{o}_-\}&\hbox{is free.}                                
&\eCbC{d}\cr
 }$$
Thus, half of the component fields, in the right column of Eqs.~\eCbC{},
are untouched by the superconstraint~\eCbc; a quarter of the component
fields obey a differential relationship to a half of the free component
fields. Finally, an eighth of the component fields ($W_\pm$ and $W$) become
left-movers, one component field ($\w_+$) an `almost left-mover', and only
one component field, $o_-$ outright vanishes. The fact that these
results can be organized into the doublet formation of~\eCbC{} is a simple
consequence that these doublets form $\vsb^+$-dependent $(0,1)$-superfields,
$\Db_+$ being the one superderivative not appearing in Eq.~\eCbc.

The straightforward expectations stated above did not turn out to be true
simply because the superderivative operators do not all (anti)commute,
whereupon the counting of independent degrees of freedom remaining
unconstrained by a superconstraint is far from (that) simple. Obviously,
there are  four (essentially distinct) cubic simple superconstraints, and a
single quartic one.

 Having illustrated this point and with the utility of these superfields
entirely unexplored, we now turn to some other, more radically novel
possibilities.

\topic{Non-simple superconstraints}
The (again) first-order superconstraint
\eqn\eLLL{ (D_--\Db_-)\W = 0 }
defines an independent pair of haploid (in fact, lefton) superfields. To see
this, switch to the (0,2)-superfield notation as in Eq.~\eLex\
and~\eNMLng{}, and define
\eqn\eXXX{ \0w\define\W\Rf~,\quad
           \0\w_-\define \inv{\sqrt2}D_-\W\Rf~,\quad
           \0\vp_-\define\inv{\sqrt2}\Db_-\W\Rf~,\quad
           \0W_\mm\define\inv4[D_-,\Db_-]\W\Rf~, }
where $\rfloor$ denotes setting $\vs^-{=}0{=}\vsb^-$ but leaving
$\vs^+,\vsb^+$ intact. Taking the same projections of the
superconstraint~\eLLL, we obtain the (0,2)-superconstraints, which almost
immediately yield:
\eqn\eXXX{ \0\vp_-=\0\w_-~,\quad \0W_\mm=0~,\quad
           (\vd_\mm\0w)=0~,\quad \hbox{and}\quad (\vd_\mm\0\w_-)=0~. }
Thus, the superfield obeying the superconstraint~\eLLL\ consists of a
spin~0 left-moving (haploid) superfield, $\0w=(w;\w_+,\vp_+;W_\pp)$, and two
identical copies of a spin~$+\inv2$ left-moving (haploid) superfield,
$\0\w_-=(\w_-;W,W_\mp;o_+)$.

Note that the complex conjugate of Eq.~\eLLL\ merely states that $\ba\W$,
the hermitian conjugate of $\W$, satisfies the identical superconstraint,
not that $\W$ is real.

\topic{Multi-field constraints}
Of course, there is no {\it a priori} reason to restrict oneself to
(super)constraints involving only one superfield. For example, given a
chiral supefield, $\F$, and a twisted-chiral one, $\X$, the superconstraint
\eqn\ePar{ D_+\F - \Db_+\X \: 0 }
clearly identifies the fermionic component fields $\j_+=\c_+$ (by
straightforward projection). On projecting with $D_-$:
\eqn\eXXX{ D_-(D_+\F-\Db_+\X)|=2(F-X) \: 0~, }
we find that the `auxiliary' fields $F$ and $X$ are also identified.
Projecting further on the remaining six components (using all the
superderivative operators from Fig.~1 which do not include trivially acting
$\Db_-$), we find that the superconstraint~\ePar\ includes the above
identifications, but also forces all of the components of both $\F$ and
$\X$ to be right-moving! That is, Eq.~\ePar\ {\it includes\/} also the
constraints $(\vd_\pp\F)=0=(\vd_\pp\X)$.

What na{\"\ii}vely started out as identifying only some two subsets of two
superfields, turned out to have a quite more restrictive end. It is thus
imperative to always expand all the superconstraint(s) in their component
constraints, so as to identify their full effect.

\topic{Inhomogeneous superconstraints}
Next, one may wish to attempt mixed-order (inhomogeneous in the
$D$'s) superconstraints. So, consider
\eqn\eDmDp{ \big(\Ione-\inv2D_-D_+\big)\W = 0~, }
which, a bit surprisingly, turns out to annihilate the whole superfield! To
see this, we switch to an `antichiral' (1,1)-superfield notation, expanding
in the $\vs$'s, but keeping the component (1,1)-superfields still dependent
on the $\vsb$'s. Define
\eqn\eBiz{ \0w\define\W\Rc~,\quad
           \0\w_\pm\define \inv{\sqrt2}D_\pm\W\Rc~,\quad
           \0W\define\inv2D_-D_+\W\Rc~, }
where $\rceil$ denotes setting $\vs^\pm{=}0$ but leaving $\vsb^\pm$ intact.
Projecting in the similar fashion the superconstraint~\eBiz, we obtain,
respectively:
\eqn\eXXX{ \0w=\0W~,\quad \0\w_\pm=0~,\quad\hbox{and}\quad\0W=0~. }
Combined, they set the whole superfield obeying~\eBiz\ to zero.

Another easy (and far less trivial) example is
\eqn\eXXX{ \big(\Ione-\inv4[D_-,\Db_-][D_+,\Db_+]\big)\W = 0~, }
which does not annihilate $\W$, but restricts all the component fields to
solutions of simple differential equations:
\eqn\eXXX{{\cmath{ 
 (1-\vd_\mm\vd_\pp)\{\,W_\mp,W_\pm,{\cal O}\,\}~=~0~,\qquad
 (1+\vd_\mm\vd_\pp)\{\,W,\FF{W}\,\}~=~0~,\cr
 (1-\vd^2_\mm\vd^2_\pp)
  \{\,w;\w_\pm,\vp_\pm;W_\mm,W_\pp;o_\pm,\Tw{o}_\pm\,\}~=~0~,\cr
 }}}
where, of course,
$(1-\vd^2_\mm\vd^2_\pp)\id(1-\vd_\mm\vd_\pp)(1+\vd_\mm\vd_\pp)$. In this
case, all component fields are restricted `on the mass shell', their mass
being $\pm1$.

\topic{Families of superconstraints}
We next consider a notion that appears to be novel in supersymmetry, but
has become fairly familiar in superstring theory: we construct families of
constraints.

A small modification of the superconstraint~\eDmDp:
\eqn\eDmDpA{ \big(\Ione-\inv2\a D_-D_+\big)\W = 0~,
              \qquad \a=\hbox{const}.,~\a\neq0,\infty~, }
now involves a parameter $\a$. As it turns out, this superconstraint family
still annihilates the superfield $\W$ and so introduces nothing new. The
limiting value $\a=0$ may thus in fact be included without any change in
the enforced vanishing of $\W$. However, the limiting case $\a=\infty$
turns $\W$ into an non-minimal antichiral superfield---far from vanishing.

On the other hand, consider the pair of superconstraints:
\eqn\eCtC{ \big[\sin(\inv2\b)D_+-\cos(\inv2\b)\Db_+\big]\W~=~0~=\Db_-\W~. }
Clearly, the special cases $\b=0$ and $\b=\p$ constrain $\W$ to be a chiral
or a twisted-chiral superfield, respectively. That is, the family of
superconstraints~\eCtC, parametrized by the angle $\b$, interpolates 
{\it continuously\/} between chiral and twisted-chiral superfields! This
may seem perplexing, since the swap $\CBp:\F{\iff}\X$ was at first
regarded as a {\it discrete\/} transformation, whereas it is now shown to be
a continuous $SO(2)\simeq U(1)$ transformation. In fact, this transformation
may be identified with the $U(1)_\LL$ subgroup of the general $SU(2)_+$
group acting on the doublet of supercoordinates $\vs^+,\vsb^+$, which in
turn is the `left-handed' half of $SU(2)_+{\times}SU(2)_-$, the maximal
compact group of transformations of the supercoordinates $\vs,\vsb$ which
commutes with the Lorentz symmetry; see \SS\,\ssSuSpSym.

 It is then natural to ask what does $\W$ of Eq.~\eDmDpA\ become for angles
$\b\neq0,\p$. Straightforwardly, the superconstraint system~\eCtC\ then
defines a pair of chiral leftons~\eQSF{a}, $(\F_\RR,\F_{\RR+})$. In the
limiting cases, we find
\eqn\eItP{
   \F=\left\{\matrix{(\f,\j_-)&\froo{\phantom{0\from\b}}
                              &\F_\RR=(\f_\RR,\j_{\RR-})
                              &\tooo{\phantom{\b\to\p}}
                              &(x,\x_-)\cr
                              (\j_+,F)&\froo{0\from\b}
                              &\F_{\RR+}=(\f_{\RR+},\j_{\RR+-})
                              &\tooo{\b\to\p}
                              &(\c_+,X)\cr}\right\}=\X~.}
The `upper halves' (as displayed here) of $\F$ and $\X$ are easy to
map to each other, and in fact identify: both pairs of component fields are
projected the same way. In contradistinction, the mapping of the `lower
halves', $(\j_+,F)\from(\f_{\RR+},\j_{\RR+-})\to(\c_+,X)$, is non-trivial
and involves the twist $D_+{\iff}\Db_+$. The relation~\eItP\ then shows how
the pair of chiral rightons interpolates between a chiral and a
twisted-chiral superfield.

\topic{Relation to dualities}
Since the chiral$\iff$twisted-chiral mapping may be identified as the root
of the `mirror map'~\rMorPle, the interpolation~\eItP\ may provide a more
complete description. While a detailed study of this is beyond our present
scope, we cannot help noticing that that is but one element of the discrete
symmetry $2S_4$ identified in \SS\,\ssSuSpSym. Another operation, the
left-handed (or the right-handed, of course) parity, denoted \BM{p}
in~\eSwp{c,d}, but when acting also on the world-sheet coordinates, was
identified in Ref.~\rJoeD\ as the root of the Type~IIA$\iff$Type~IIB
duality. With a total of 47 nontrivial operations, the group of discrete
symmetries $2S_4$ from \SS\,\ssSuSpSym\ seems to be a pretty good candidate
for seeking roots of other dualities in (super)string, M- and F-theories.

\topic{Fibration of superconstraints}
Let us, however, return to the 1-parameter family of
superconstraints~\eDmDpA. In retrospect, it is almost trivial, since all
values of $\a\neq\infty$ lead to $\W=0$. Exceptionally, at $\a=\infty$,
$\W\neq0$ and is in fact a non-minimal antichiral superfield! That is, the
1-dimensional parameter space $\{\a\neq\infty\}\simeq\IR^1$ is fibered with
identical fibres, parametrized by the identical constrained superfield,
$\W=0$. Compactifying $\{\a\neq\infty\}\simeq\IR^1$ into $\{\a\}\simeq
S^1$, by including the $\a=\infty$ point, the fibration becomes quite
non-trivial. One can think of the total space of this fibration as a simple
cover of $S^1$ (parametrized by the constant and trivial value $\W{=}0$),
with one exceptional set, $E$, at $\a=\infty$, where $E$ is parametrized by
the non-minimal antichiral superfield satisfying $D_-D_+\W=0$.

A bit more interestingly perhaps,  the 1-parameter family of
superconstraints~\eCtC\ is clearly non-trivial, since different values of
$\a$ lead to different (constrained) superfields. Again, however, the
parameter space, $\{\a\}\simeq S^1$ is fibred mostly by a pair of chiral
rightons. At the two special points, $\a=0,\p$ however, this
$2+2$-dimensional fibre, $(\F_\RR,\F_{\RR+})$, fuses into a single
4-dimensional one, $\F$ and $\X$, respectively. In fact, it is easy to
turn~\eCtC\ into a fibration over a (complex!) $\IP^1$:
\eqn\eCTC{ (z_1D_+-z_2\Db_+)\W~=~0~=\Db_-\W~, }
where the two complex numbers $(z_1,z_2)\simeq(\l z_1,\l z_2)$ parametrize a
complex projective (parameter) space, $\IP^1$. Note that the invariance
under rescaling by a non-zero finite complex parameter, $\l$, arises as the
obious symmetry of the homogeneous equation $(z_1D_+{-}z_2\Db_+)\W=0$. The
chiral and twisted-chiral superfields, $\F,\X$, are then seen as special
fibres (the North and the South poles, respectively) in a non-trivial
fibration of the righton doublet $(\F_\RR,\F_{\RR+})$ over $\IP^1$.

\topic{Non-linear superconstraints}
Whereas the simple modification~\eDmDpA\ of the superconstraint~\eDmDp\
produced an (almost) trivial outcome, a further but still simple
modification
\eqn\eNLn{ \big(\Ione-\inv2f(\W)D_-D_+\big)\W = 0~, }
produces a rather more interesting situation. To see this, switch back to
the (1,1)-superfield notation, expanding in the $\vs$'s, but leaving the
component fields still superfunctions of the $\vsb$'s. Using the
definitions in~\eBiz\ and writing $\0f^{(n)}\id f^{(n)}(\0w)$, we find that
the superconstraint~\eNLn\ includes
\eqn\eXXX{{\cmath{\0w-\0f\0W\:0~,\qquad
           \0\w_\pm[\0f-\0w\0f']\:0~,\cr
           \0w\big([\0f-\0w\0f']+\0\w_-\0\w_+\0f''\0f\big)\:0~.\cr
 }}}
The first constraint simply sets $\0W=\0w/\0f$. With the latter two, this
then enforces $\W=(\0w;\0\w_\pm;\0w/\0f)$, where:
\eqn\eXXX{
 \0w,\0\w_\pm=\cases{ \hbox{free} & if $f(\W)=\W/W_0$,\cr
                       \noalign{\vglue2mm}
                       0 & if $f(\W)\neq\W/W_0$~,}\qquad
 W_0\define\lim_{\W\to0}\big(\W/f(\W)\big)~.}
Again, we have a (degenerate) fibration: over the {\it parameter space of
the function\/} $f$, we generically have the constant $\W=(0;0,0;W_0)$,
while at the special subset where $f$ becomes a constant multiple of its
argument, we have the unconstrained but only $\frc34$- independent
superfield, $\W=\big(\0w;\0\w_\pm;\0w/\0f\big)$ which, in the limit
$f\to\infty$ becomes a non-minimal antichiral superfield, as found above.
Recall that $\0w,\0\w_\pm$ in turn are otherwise unconstrained
$\vsb$-dependent $(1,1)$-superfields.

Other non-homogeneous superderivative constraints may produce less trivial
results and define other possibly interesting constrained superfields.
Reasons for using such more complicated superconstraints are both far from
obvious and model-dependent\ft{I thank Joe Polchinski for discussions on
this point, and sharing some early results of his work with Simeon
Hellerman on some superconstraint equations generalizing the present
Eq.~\eCTC\ into a fully non-linear (albeit holomorphic) superconstraints in
$\W$. Their work focuses on applications in supersymmetric quantum
mechanics. The present models can be dimensionally reduced to such
theories, albeit with some possible loss of generality, just as dimensional
reduction of 4-dimensional models fails to capture most of the diversity
presented herein.}. Whilst one is tempted to dismiss such more complicated
constrained superfields on grounds of physical equivalence to (some
combination of) `minimal' haploid and quartoid superfields at the cost of
greater technical complexity, it may well be that their {\it joint\/} use
offers some distinct advantage, as does the use of `minimal' {\it
together\/} with non-minimal superfields~\rUseNM. More than a mere mention
of this possibility, however, is beyond our present scope.

\topic{Discrete constraints}
Finally, the imposition of reality as a constraint on a superfield is
probably the best known type of constraint: $V^{\dag}\:V$
defines a real superfield, used for Yang-Mills type gauge
vectors~\refs{\rWB,\rGGRS}. Note that the hermiticity of the entire
superfield {\it does not\/} imply the hermiticity of the component fields.
In fact, $V^{\dag}\id\BM{C}(V)\:V$ implies that, \eg, the lowest component
is real, but the $\inv2D_-D_+V|$ and $\inv2\Db_+\Db_-V|$ components are
imaginary.

In view of our analysis of discrete symmetries of the 2-dimensional
(2,2)-superspacetime, it is clear that there are many, many more such
constraints possible. For example, $\BM{P}(W)\:W$ defines a left-right
(parity) symmetric, \ie, scalar superfield; $\CBm(Y)\:Y$ imposes a reality
condition but only on the $\vs^-,\vsb^-$-components;
\etc\ There being 47 nontrivial elements of $2S_4$, there are 47 such types
of constraints possible in the 2-dimensional (2,2)-superspacetime, and they
can of course be combined with the superdifferential operators to produce
some quite formiddably looking constraints.

Again, an exhaustive analysis of all possibilities is clearly out of hand,
and presently we remain content with the `minimal' haploid superfields, and
a brief involvment of their non-minimal counterparts.

\newsec{Lagrangian Densities}\noindent
\seclab\sLag
Once the superfields have been defined, we can turn to their dynamics, to
be governed by a Lagrangian density. As it turns out, 2-dimensional
(2,2)-supersymmetric models have an unprecedented rich set of choices for
the Lagrangian density, and consequently for the dynamics.

\subsec{Unconstrained formalism}\noindent
Owing to the nilpotency of the superderivatives, the complex ambidexterous
constrained superfields can be expressed as superderivatives of
unconstrained ones~\rGGRS. For example, given an unconstrained complex
superfield, $\W$, the following are `minimal' haploid superfields:
\eqna\eUnc
 $$\twoeqsalignno{
 (\Db_+\Db_-\W)& ~\hbox{is chiral,}\quad&\hbox{\ie, it obeys}
 &\hbox{Eqs.~}\eHSF{a}~,  &\eUnc{a}\cr
 (D_+D_-\W)& ~\hbox{is antichiral,}\quad&\hbox{\ie, it obeys}
 &\hbox{Eqs.~}\eHSF{b}~,  &\eUnc{b}\cr
 (D_+\Db_-\W)& ~\hbox{is twisted-chiral,}\quad&\hbox{\ie, it obeys}
 &\hbox{Eqs.~}\eHSF{c}~,  &\eUnc{c}\cr
 (\Db_+D_-\W)& ~\hbox{is twisted-antichiral,}\quad&\hbox{\ie, it obeys}
 &\hbox{Eqs.~}\eHSF{d}~.  &\eUnc{d}\cr
}$$
Similarly, given {\it two\/} unconstrained complex superfields, $\W^\pm$,
the following define non-minimal haploid superfields:
\eqna\eUnm
 $$\twoeqsalignno{
 (\Db_+\W^++\Db_-\W^-)& ~\hbox{is NM-chiral,}~&\hbox{\ie, it obeys}
 &\hbox{Eqs.~}\eNMF{a}~,  &\eUnm{a}\cr
 (D_+\W^++D_-\W^-)& ~\hbox{is NM-antichiral,}~&\hbox{\ie, it obeys}
 &\hbox{Eqs.~}\eNMF{b}~,  &\eUnm{b}\cr
 (D_+\W^++\Db_-\W^-)& ~\hbox{is NM-twisted-chiral,}~&\hbox{\ie, it obeys}
 &\hbox{Eqs.~}\eNMF{c}~,  &\eUnm{c}\cr
 (\Db_+\W^++D_-\W^-)& ~\hbox{is NM-twisted-antichiral,}~
                                                    &\hbox{\ie, it obeys}
 &\hbox{Eqs.~}\eNMF{d}~.  &\eUnm{d}\cr
}$$
Amusingly, both the `minimal' and the non-minimal {\it unidexterous\/}
superfields elude a similarly easy description in terms of unconstrained
superfields.

This then defines an `unconstrained' formalism where each of the constrained
superfields~\eHSF{a{-}d} and~\eNMF{a{-}d} is meticulously
replaced by its proxy~\eUnc{} and~\eUnm{}, respectively. Quantization is
now straightforward in terms of path-integrals over the components of the
unconstrained superfields $\W$ used in~\eUnc{}, and $\W^\pm$ used
in~\eUnm{}. This approach is however unavoidably beset with a substantially
increased amount of sheer algebra, and will not be pursued herein.

\subsec{Constrained formalism}\noindent
With a collection of several constrained superfields of the types defined in
Eqs.~\eCps{}, we now seek the most general Lagrangian density (subject to
some hopefully plausible restrictions). 
 Consider an analytic function, $f$, of some of the haploid
superfields~\eHSF{}. In general~\refs{\rWB,\rGGRS}, the Berezin integral of
this $f$ over only those $\vs,\vsb$'s on which $f$ does depend will
transform under a general supersymmetry transformation into a total
derivative. A world-sheet integral of such a Berezin integral is then, up
to world-sheet boundary terms, invariant under supersymmetry. When
considering only world-sheets without a boundary, as we do, such Berezin
integrals provide supersymmetric Lagrangian densities.

With the benefit of hindsight, we discuss the details of the various
Lagrangian density terms starting with those which require the most number
of Berezin integrations. Also, for now, the quantization issues pertaining
to the unidexterous (super)fields will be completely ignored, and all
expressions will use fully constrained (super)fields.

Finally, while any superfield can be assigned overall spin as discussed in
\SS\,\sDSFlds, for reasonos that will soon become obvious, we consider
endowing only a subset of the unidexterous superfields with overall spin.
We write $\L=(\L^\ah,\L^{\ac\pp})$, where the superfields counted by the
index $\ah$ remain as defined in~\eHSF{e}, while those indexed by
$\ac$ satisfy $j_3(\L^\pp)=1$, so that
$j_3\{\ell^\pp,\l^\pp_+,\bl^\pp_+,L^\pp_\pp\}=\{1,\inv2,\inv2,0\}$.
A susbset of the rightons is similarly endowed with overall spin $-1$:
$\Y=(\Y^\ih,\Y^{\ic\mm})$.

\topic{The general Lagrangian density}
The most general manifestly supersymmetric Lagrangian density is of the
following form:
\eqna\eMGL
 $$\eqalignno{ L=
 &\inv2\int\rd\vs^-~\J^++\hc~~+~~\inv2\int\rd\vs^+~\J^-+\hc   &\eMGL{a}\cr
 &+\inv2\int\rd^2\vs~W+\hc ~~+~~\inv2\int\rd^2\sT~\S+\hc      &\eMGL{b,c}\cr
 &+\inv2\int\rd^2\sL~N^\pp+\hc
 ~~+~~\inv2\int\rd^2\sR~\Nt^\mm+\hc&\eMGL{d,e}\cr
 &+\inv4\int\rd^2\sL\rd\vs^+~M^++\hc
 ~~+~~\inv4\int\rd^2\sR\rd\vs^-~\Mt^-+\hc                     &\eMGL{f,g}\cr
 &+\inv4\int\rd^4\vs~K+\hc~.                                  &\eMGL{h}\cr
}$$
Recall that fermionic integration is equivalent to (covariant)
superderivative(s), followed by the projection on the `body',
$\vs^\pm{=}0{=}\vsb^\pm$:
\eqna\eSGL
 $$\eqalignno{ L=
 &\inv2\big[D_+\J^+\big]\Zp+\hc
 ~~+~~\inv2\big[D_-\J^-\big]\Zp+\hc                          &\eSGL{a}\cr
 &+\inv2\big[D_-D_+W\big]\Zp+\hc
  ~~+~~\inv2\big[D_-\Db_+\S\big]\Zp+\hc                      &\eSGL{b,c}\cr
 &+\inv4\big[[D_+,\Db_+]N^\pp\big]\Zp+\hc
  ~~+~~\inv4\big[[D_-,\Db_-]\Nt^\mm\big]\Zp+\hc              &\eSGL{d,e}\cr
 &+\inv8\big[[D_+,\Db_+]D_-M^+\big]\Zp+\hc
  ~~+~~\inv8\big[[D_-,\Db_-]D_+\Mt^-\big]\Zp+\hc             &\eSGL{f,g}\cr
 &-\inv{16}\big[[D_-,\Db_-][D_+,\Db_+]K\big]\Zp~.            &\eSGL{h}\cr
 }$$
For the above expressions to be supersymmetric, it is
necessary and sufficient for each integrand to be annihilated by all
covariant superderivatives except those which are equivalent to the
integration operator.
 For example, $\J^+$ should be a chiral lefton, annihilated by all
superderivatives except $D_+$, which appears explicitly in~\eSGL{a},
replacing the integral $\int\rd\vs^-$ in~\eMGL{a}. Similarly, $W$ has to be
chiral, $\S$ twisted-chiral, $M^+$ has to be annihilated by $\Db_-$, {\it
etc}. $K$ is always chosen to be real, so that no `$+\hc$' is necessary.
 
The $\int\rd\vs\,W+\hc$ terms are frequently called the `F-terms' as the
$\int\rd\vs$ integral projects on the $\vs^+\vs^-$-component in $W$,
typically called $F$ in a chiral superfield. We will extend the name to all
other two-fermionic integrals in~\eMGL{b,c}, \ie,~\eSGL{b,c}. The $\vs^4$
component being typically called $D$ in a general superfield, the
$\int\rd^4\vs$-term~\eMGL{e}, \ie,~\eSGL{e} is typically called the
`D-term'. In a similar vein, we call the terms in~\eMGL{a}, \ie,~\eSGL{a}
the `$\j$-terms, and those in~\eMGL{d}, \ie,~\eSGL{d} the `$\l$-terms'.

\subsec{D-term}\noindent
\subseclab\ssD
We begin with the most generally $\vs^\pm,\vsb^\pm$-dependent set of
terms, those that depend on all four of them, and so appear under the
$\int\rd^4\vs$ integral~\eDI. Since non-minimal superfields~\eNMF{} do
depend on all four fermionic coordinates, this is in fact the only type of
terms in which they can appear (without a superderivative acting on them).
The hasty Reader should be reminded that the {\it combined\/} use of
`minimal' and non-minimal haploid superfields does have its distinct
advantages and interest (see, \eg, Ref.~\rUseNM), whence here we include
quartoid and both the `minimal' and the (ambidexterous) non-minimal haploid
superfields. The Reader should then have no problem extending the presented
results by including the unidexterous non-minimal superfields, and also the
`other' superfields discussed in \SS\,\ssOthr, or indeed yet other ones, of
their own design.

The D-term involving the haploid superfields is
\eqn\eLD{ L_K ~=~
 \inv4\int\rd^4\vs~K(\F,\FB,\X,\XB,\L,\Y,\Q,\QB,\P,\PB,\3)~, }
where the ellipses denote non-minimal superfields, and $K(\3)$ is required
to be a real function for the Lagrangian (term) $L_K$ to be hermitian. Note
that this implies, \eg, that
$\ba{K,_{\m\nb}}\define\ba{\vd_\m\vd_\nb K}=K,_{\n\mb}$ and
$\ba{K,_{\a\bb}}=K,_{\b\ab}$ are Hermitian, but $K,_{ab}, K,_{ij}$ are
merely symmetric, and all the mixed matrices are general. We do not,
however, allow $K$ to depend on superderivative superfields, as that would
lead to higher derivative terms in the Lagrangian density. 

Finally, as $\L^\pp$ and $\Y^\mm$ carry overall spin, and since $K$ has
spin-0 and it is not allowed to depend on superderivatives, it must be that
$K$ depends on the $\L^\pp,\Y^\mm$ only through their spinless products,
$(\L^\pp\Y^\mm)$. Such products are characterized by a list of matrices,
$\k^m_{\ac\ic}$, where $m$ indexes all the linearly independent bilinear
products, $B^m\define(\L^{\ac\pp}\k^m_{\ac\ic}\Y^{\ic\mm})$. The matrices
$\k^m_{\ac\ic}$ may be chosen so as to have all entries 0, except for the
$m^{th}$ one (counting say elements in row after row) which then equals 1.
 Then,
\eqn\eXXX{ \pd{}{\L^{\ac\pp}} = \k^m_{\ac\ic}\Y^{\ic\mm}\pd{}{B^m}~,
           \quad\hbox{and}\quad
           \pd{}{\Y^{\ic\pp}} = \L^{\ac\pp}\k^m_{\ac\ic}\pd{}{B^m}~,}
so that
\eqn\eChR{ \pd{K_{\1{\cdots}}}{\L^{\ac\pp}}
          = \k^m_{\ac\ic}\Y^{\ic\mm} K_{\1m{\cdots}}~,
          \quad\hbox{and}\quad
          \pd{K_{\1{\cdots}}}{\Y^{\ic\mm}}
          = \L^{\ac\pp}\k^m_{\ac\ic} K_{\1m{\cdots}}~. }
With these, the Lagrangian density is straightforward to expand into
component fields; see Appendix~B.

\subsec{$\l$-terms}\noindent
\subseclab\ssL
We now seek {\it fermionic} functions which are annihilated by one of
the four $D,\Db$'s. Since $\F$, $\X$, their $D_\pm$-superderivatives, $\L$,
$(D_+\L)$, $(\Db_+\L)$ and $(\Db_-\hbox{anything})$ are all annihilated
by $\Db_-$, the Berezin integral over $\vs^\pm,\vsb^-$ of any analytic
function of these superfields is a suitable Lagrangian density term:
\eqn\eXXX{ L_{M^+}
 =\inv4\int\rd^2\vs\,\rd\vsb^-M^+(\F_\LL,\F_\RR,\FB_\LL;
                                   \F,\X,\L;\3) +\hc~,  }
as supersymmetry transforms it into a total derivative\ft{This general
argument rests on the identities (up to total deivatives)
$Q_\pm\simeq D_\pm$ and $\Qb_\pm\simeq\Db_\pm$. Supersymmetry
transformation is then calculated, up to total derivatives, by
acting with the operator $\d_\e\simeq\e{\cdot}D+\bar\e{\cdot}\Db$.}.
 Recall that the component fields~\eQps{} of the quartoid superfields
$\F_\LL,\F_\RR,\FB_\LL,\FB_\RR$ are obtained from those~\eCps{a,b} of the
chiral and antichiral haploid superfields. We can therefore lump
$\F_\LL,\F_\RR$ into $\F$ and
$\FB_\LL,\FB_\RR$ into
$\FB$ without any loss of generality; the interested Reader should have no
difficulty re-introducing them by splitting the ranges:
$\F\to(\F,\F_\LL,\F_\RR)$ and $\FB\to(\FB,\FB_\LL,\FB_\RR)$, and
remembering that the quartoid superfields have only half the component
fields of a haploid one. Here, the quartoid superfields are explicitly
written only if they do not appear in pairs that comprise a haploid
superfield.

Now, prohibiting as before higher derivative (negative dimensional
coefficient) terms, and ensuring that $M^+$ has spin $+\inv2$, it must be
of the form:
\eqna\eThQ
 $$ \eqalignno{M^+
 &=(D_-\F^\m)M^\LL_\m + (D_-\X^\a)M^\LL_\a
  +(D_+\L^{\ac\pp})M^1_\ac + (\Db_+\L^{\ac\pp})M^2_\ac        &\eThQ{a}\cr
 &+(\Db_-\F^\nb)M_\nb+(\Db_-\X^\bb)M_\bb+(\Db_-\Y^{\ih})M_\ih
  +(\Db_+\F^\mt_\LL)\L^{\ac\pp}M_{\ac\mt}                     &\eThQ{b}\cr
 &+(\Db_-\Q^M)M_M+(\Db_-\Q^{\bar N})M_{\bar N}
  +(\Db_-\P^A)M_A+(\Db_-\P^{\bar B})M_{\bar B}~,              &\eThQ{c}\cr
 &+(D_+\F^\m)\L^{\ac\pp}M_{\ac\m}+(D_+\L^\ah)\L^{\bc\pp}M^1_{\ah\bc}
  +(D_+\P^A)\L^{\bc\pp}M_{A\bc}                               &\eThQ{d}\cr
 &+(\Db_+\X^\a)\L^{\bc\pp}M_{\a\bc}+(\Db_+\L^{\ah})\L^{\bc\pp}M^2_{\ah\bc}
  +(\Db_+\Q^M)\L^{\ac\pp}M_{\ac M}~,                          &\eThQ{e}\cr
 }$$
where the $M_\@$'s are all arbitrary analytic functions of the
$\FB_\LL,\F,\X,\L$, except for $M^\LL_\a,M^\LL_\m$, which are depend only on
$\FB_\LL,\L$. These two terms are supersymmetric since, although they are
not annihilated by $\Db_-$ as the other are, they turn into total
derivatives:
\eqn\eXXX{ \Db_-[(D_-\F^\m)M^\LL_\m]=(2i\vd_\mm\F^\m)M^\LL_\m
 =2i\vd_\mm(\F^\m M^\LL_\m)~, }
which suffices.

The quick Reader may notice that the $(\Db_-\3)M_\@$-type terms
in~\eThQ{b,c} may be rewritten as D-terms (since
$(\Db_-\F^\nb)M_\nb=\Db_-(\F^\nb M_\nb)$, \etc), and could argue for their
discarding here on the grounds that they are effectively included in~\eLD.
This would however induce a serious omission. Rather than advance a general
and possibly obtuse argument, consider the following simple example, with
the first term in~\eThQ{b} where $M_\nb=\F^\m M_{\m\nb}(\F)$:
\eqn\eExpl{{\eqalign{ 
 \inv8&\big\{[D_+,\Db_+]D_-\big(M_\nb(\F)(\Db_-\F^\nb)\big)
 +[\Db_+,D_+]D_-\big((D_-\F^\m)\Mb_{\nb\m}\F^\nb\big)\big\}\Zp\cr
 &=\inv2D^4\big(\F^\m M^H_{\m\nb}\F^\nb\big)\Zp
 -\inv4\vd_\mm\big([D_+,\Db_+](\F^\m M^A_{\m\nb}\F^\nb)\big)\Zp~, \cr
 }}}
where $\Mb_{\nb\m}\define(M_{\n\mb})^{\dag}$ and
\eqn\eParts{ M^H_{\m\nb}\define\inv2(M_{\m\nb}+\Mb_{\nb\m})~,\qquad
           M^A_{\m\nb}\define\inv{2i}(M_{\m\nb}-\Mb_{\nb\m}) }
are hermitian and anti-hermitian, respectively.
 The first term in~\eExpl\ contains, upon projection, the familiar `kinetic'
terms
\eqn\eFam{ \inv2[(\vd_\pp\f^\m)(\vd_\mm\f^\nb)
                +(\vd_\mm\f^\m)(\vd_\pp\f^\nb)]M^H_{\m\nb}~, }
whereas the second term in~\eExpl\ includes the `torsion' term:
\eqn\eXXX{ \inv2[(\vd_\pp\f^\m)(\vd_\mm\f^\nb)
                -(\vd_\mm\f^\m)(\vd_\pp\f^\nb)]M^A_{\m\nb}~. }
The antihermitian matrix $M^A_{\m\nb}$ appearing here need not
necessarily be associated with the `metric torsion', which would be a
derivative of $M^A_{\m\nb}$. In fact, $M^A_{\m\nb}$ may well be
constant. However, it does produce an important term upon world-sheet
integration:
\eqn\eTpT{ \int\rd^2\s~\inv2[(\vd_\pp\f^\m)(\vd_\mm\f^\nb)
                -(\vd_\mm\f^\m)(\vd_\pp\f^\nb)]M^A_{\m\nb}
 =\int\rd\f^\m{\wedge}\rd\f^\nb M^A_{\m\nb}\id\int\f^*M^A_{[2]}~, }
the integral of the (the pull-back by the map $\f$ of the) 2-form
$M^A_{[2]}\id M^A_{\m\nb}\rd\f^\m{\wedge}\rd\f^\nb$ over the (image)
of the world-sheet. This type of terms relates to {\it global\/}
information about the field space and is essential in many superstring
applications of 2-dimensional (supersymmetric) field theories. 
 
Since total derivatives are routinely dropped, this important
term would normally be lost. Thus, instead of rewriting the
$\Db_-(\3)$-terms in~\eThQ{} as D-terms and then meticulously salvaging
such and similarly interesting terms from the total
derivative debris {\it en route\/}, it would seem preferable to include
them manifestly in the Lagrangian density via the terms such as those
in~\eThQ{}.

Notice also that the `metric' $M^H_{\m\nb}$ defined in the first one of
Eqs.~\eParts\ stemms from a simple {\it choice\/} of
$M_\nb(\FB_\LL,\F,\X,\L)$. In the general case, the `metric' in the
`kinetic' term~\eFam\ is then a hermitian and analytic but otherwise
unrestricted matrix function of the bosonic fields $\fb_\LL,\f,x,\ell$ and
their conjugates.

Thus, the $(\Db_-\3)M_\@$-type terms in~\eThQ{b,c} play a dual r\^ole:
(1)~they modify the D-terms, and (2)~produce `topological' terms such
as~\eTpT. The remaining eleven terms in~\eThQ{a{-}e} however cannot
be reduced to D-terms and/or total derivatives; they produce totally new
terms for the Lagrangian density.

By the same token, the other batch of $\l$-terms is obtained from
 $$ \eqalignno{ \Mt^-
 &=(D_+\F^\m)\Mt^\RR_\m+(D_+\X^\bb)\Mt^\RR_\bb
  +(D_-\Y^{\ic\mm})\Mt^1_\ic + (\Db_-\Y^{\ic\mm})\Mt^2_\ic    &\eThQ{f}\cr
 &+(\Db_+\F^\nb)\Mt_\nb+(\Db_+\X^\a)\Mt_\a+(\Db_+\L^{\ah})\Mt_\ah
  +(\Db_-\F_\RR^\mt)\Y^{\ic\mm}\Mt_{\ic\mt}                   &\eThQ{g}\cr
 &+(\Db_+\Q^M)\Mt_M+(\Db_+\Q^{\bar N})\Mt_{\bar N}
  +(\Db_+\P^A)\Mt_A+(\Db_+\P^{\bar B})\Mt_{\bar B}~,          &\eThQ{h}\cr
 &+(D_-\F^\m)\Y^{\ic\mm}\Mt_{\ic\m}+(D_-\Y^\ih)\Y^{\jc\mm}\Mt^1_{\ih\jc}
  +(D_-\P^\Bb)\Y^{\jc\mm}\Mt_{\Bb\jc}                         &\eThQ{i}\cr
 &+(\Db_-\X^\bb)\Y^{\jc\mm}\Mt_{\bb\jc}
  +(\Db_-\Y^{\ih})\Y^{\jc\mm}\Mt^2_{\ih\jc}
  +(\Db_-\Q^M)\Y^{\ic\mm}\Mt_{\ic M}~,                        &\eThQ{j}\cr
 }$$
where now the $\Mt_\@$'s are all arbitrary analytic functions of
the $\FB_\RR,\F,\XB,\Y$, except for $\Mt^\RR_\m$ and $\Mt^\RR_\bb$ which
only depend on $\FB_\RR,\Y$.

With these two functions~\eThQ{},
\eqna\eLQ
 $$\eqalignno{
 L_M &=~\inv8\big[[D_+,\Db_+]D_-M^+\big]\Zp+\hc  &\eLQ{a}\cr
     &~~+\inv8\big[[D_-,\Db_-]D_+\Mt^-\big]\Zp+\hc &\eLQ{b}\cr }
 $$
is the most general set of $\l$-terms for the Lagrangian density; see also
Appendix~B.

\subsec{F-terms}\noindent
\subseclab\ssF
`Superpotential terms'~\eMGL{b} are a familiar type of candidates for the
Lagrangian density. For example, any analytic function of only the chiral
superfields, $\F$'s, is itself a chiral superfield: $\Db_\pm W(\F){=}0$.

\topic{Ambidexterous superpotentials}
Since all the superfields in~\eDSF{a} satisfy the chiral pair of
superconstraints~\eHSF{a}, any analytic function of the arguments listed
in~\eDSF{a},
 $$
   W\big(\F,\F_\LL,\F_\RR;(\Db_+\X),\ldots\big)~,
 $$
is than also chiral. Its hermitian conjugate, $\ba{W}$, is antichiral.
Similarly,
 $$
   \S\big(\X,\F_\RR,\FB_\LL;(D_+\F),\ldots\big)~,
 $$
is than also twisted chiral, and its hermitian conjugate, $\ba\S$, is
twisted antichiral. Again, we lump $\F_\LL,\F_\RR$ into $\F$
and $\FB_\LL,\FB_\RR$ into $\FB$, and write out explicitly the dependence
on quartoid superfields only if this cannot be done, as in specifying the
dependence of the coefficient functions in~\eThQ{}.

However, a few further restrictions are needed. For the putative Lagrangian
density terms
\eqn\eXXX{ \int\rd^2\vs~W+\hc\qquad\hbox{and}\qquad\int\rd^2\sT~\S+\hc~, }
to be scalar densities, $W$ and $\S$ have to be scalar densities themselves.
Therefore, each term in their expansion over the superderivative superfields
must have an even number of superderivatives. Finally, to prevent higher
derivative and/or negative dimensional coefficient terms, we write
$W$ in the following form:
\eqna\eThW
 $$
 W~=~W_0(\F) + \inv2 W_>\big(\F;\3\big)~,\eqno\eThW{a}
 $$
and truncate $W_>$ to terms quadratic in superderivatives.

Now, among all the possible terms with a quadratic superderivative, \eg, the
term $\int\rd^2\vs\,(\Db_+\Db_-\F^\nb)W_\nb+\hc$ can be rewritten, without
any loss of generality, as a corresponding $\l$-term~\eLQ{}:
\eqn\eXXX{{\eqalign{\int\rd^2\vs~(\Db_+\Db_-\F^\nb)W_\nb(\F)
 &\id D_-D_+(\Db_+\Db_-\F^\nb)W_\nb(\F)\Zp
  =D_-D_+\Db_+\,(\Db_-\F^\nb)W_{\nb}(\F)\Zp~,\cr
 &\simeq\inv2[D_+,\Db_+]D_-\,\big[(\Db_-\F^\nb)W_\nb(\F)\big]\Zp~,\cr
 }}}
where we have dropped the total derivative
$i\vd_\pp D_-[(\Db_-\F^\nb)W_{\nb}]|$.
 In fact, the corresponding $\l$-term in~\eThQ{b},
$\int\rd^2\vs\rd\vsb^-(\Db_-\F^\nb)M_\nb$, is {\it more\/} general, as the
coefficient function $M_\nb$ may depend on $\FB_\LL,\F,\X,\L$, whereas
$W_\nb$ depends only on $\F$. A similar fate befalls many of the tentative
F-terms, and only the following remain, not being reducible without loss of
generality to any of the terms in~\eLD\ or~\eLQ{}:
 $${{\eqalign{ W_>=
 &(\Db_+\L^{\ac\pp})(\Db_-\Y^{\ic\mm})W_{\ac\ic}
  +(\Db_+\L^{\ac\pp})(\Db_+\Q^M)W_{\ac M}\cr
 &+(\Db_-\Y^{\ic\mm})(\Db_-\Q^M)W_{\ic M}
  +(\Db_+\Q^M)(\Db_-\Q^N)W_{MN}~,\cr}}} \eqno\eThW{b}
 $$
where the $W_{\@\@}$'s are arbitrary analytic functions of the $\F$ only.
 The canonical dimensions of $W_0(\F)$ and the coefficient functions in
$W_>(\F)$ are all non-negative: $[W_0]{=}1$ (mass parameters) and
$[W_{\@\@}]{=}0$ (dimensionless couplings). The omited higher order
superderivative terms would have had coefficient functions with negative
dimensions. With the superpotential~\eThW{},
\eqn\eLW{L_W=\inv2\int\rd^2\vs~W(\F;\3)~+\hc=\inv2[D_-D_+W(\F;\3)]\Zp+\hc}
is the most general superpotential Lagrangian density. In the absence of
unidexterous superfields, $L_W$ reduces to the well-known superpotential
terms involving only the holomorphic function $W_0(\F)$ in~\eThW{}.
 
Similarly,
\eqn\eLS{L_\S=\inv2\int\rd^2\sT~\S(\X;\3)+\hc
              =\inv2[D_-\Db_+\S(\X;\3)]\Zp+\hc}
is the most general twisted superpotential Lagrangian density, with the
twisted superpotential being
\eqna\eThS
 $$
 \S~=~\S_0(\F) + \inv2 \S_>\big(\X;\3\big)~,\eqno\eThS{a}
 $$
where
 $${{\eqalign{ \S_>=
 &(D_+\L^{\ac\pp})(\Db_-\Y^{\ic\mm})\S_{\ac\ic}
  +(\Db_+\P^A)(D_+\L^{\bc\pp})\S_{A\bc}\cr
 &+(D_-\P^A)(\Db_-\Y^{\ic\mm})\S_{A\ic}
  +(D_+\P^A)(\Db_-\P^B)\S_{AB}~,\cr}}} \eqno\eThS{b}
 $$
where the $\S_{\@\@}$'s are arbitrary analytic functions of the $\X$ only.
 The canonical dimensions of $\S_0(\F)$ and the coefficient functions in
$\S_>(\F)$ are all non-negative: $[\S_0]{=}1$ (mass parameters) and
$[\S_{\@\@}]{=}0$ (dimensionless couplings).

\topic{Unidexterous superpotentials}
 By contrast, no (higher) superderivative of either of $\F,\FB,\X,\XB$
satisfies the superconstraints~\eHSF{e,f}. Therefore, $L_L$ ($L_R$), the
lefton (righton) analogue of $L_W$ and $L_\S$, is constructed entirely from
lefton (righton) superfields. Using the list~\eDSF{e}, we find that
\eqn\eLL{\eqalign{L_L&=\inv2\int\rd^2\sL~N^\pp\big(\L;\3\big)+\hc\cr
                     &=\inv4\big[[D_+,\Db_+]N^\pp(\L;\3)\big]\Zp+\hc\cr}}
is the lefton `superpotential' term for the Lagrangian density. For this to
be a scalar density, the function $N^\pp$ must transform as a vector of spin
$+1$, and so must depend on the $\L^\pp$'s (this is the reason for their
introduction). Restricting again to terms in which the coefficient
functions have non-negative dimensions, we find (omitting indices for
brevity):
\eqna\eThN
 $${\eqalign{ N^\pp=~
 &\L^\pp N_\@
  +\inv2\L^\pp(D_+\Db_+\L^\pp)N^1_{\@\@}
  +\inv2\L^\pp(\Db_+D_+\L^\pp)N^2_{\@\@}\cr
 &+\inv2(D_+\L^\pp)(D_+\L^\pp)N^3_{\@\@}
 +\inv2(D_+\L^\pp)(\Db_+\L^\pp)N^4_{\@\@}\cr
 &+\inv2(\Db_+\L^\pp)(\Db_+\L^\pp)N^5_{\@\@}
  +\inv2\L^\pp\L^\pp(D_+\Db_+\L)N^1_{\@\@\@}\cr
 &+\inv2\L^\pp\L^\pp(\Db_+D_+\L)N^2_{\@\@\@}
  +\inv2\L^\pp(D_+\L^\pp)(D_+\L)N^3_{\@\@\@}\cr
 &+\inv2\L^\pp(D_+\L^\pp)(\Db_+\L)N^4_{\@\@\@}
  +\inv2\L^\pp(\Db_+\L^\pp)(D_+\L)N^5_{\@\@\@}\cr
 &+\inv2\L^\pp(\Db_+\L^\pp)(\Db_+\L)N^6_{\@\@\@}
  +\inv2\L^\pp\L^\pp(D_+\L)(D_+\L)N^1_{\@\@\@\@}\cr
 &+\inv2\L^\pp\L^\pp(D_+\L)(\Db_+\L)N^2_{\@\@\@\@}
  +\inv2\L^\pp\L^\pp(\Db_+\L)(\Db_+\L)N^3_{\@\@\@\@}~,\cr}} \eqno\eThN{a}
 $$
where the $N_\@,\3,N_{\@\@\@\@}$'s are all arbitrary functions of the
$\L$. Clearly, some of these have certain symmetries, implicit in the above
expressions; \eg,
$N^1_{\ac\bc\ch\dh}=+N^1_{\bc\ac\ch\dh}=-N^1_{\ac\bc\dh\ch}$. That is,
$N^1_{\@\@\@\@}\to N^1_{(\ac\bc)[\ch\dh]}$.
Also, the Reader may prefer to combine some of the terms, such as
\eqn\eXXX{{\eqalign{
 &\inv2\L^\pp(D_+\Db_+\L^\pp)N^1_{\@\@}
  +\inv2\L^\pp(\Db_+D_+\L^\pp)N^2_{\@\@}\cr
 &\]9=\frc i4\L^\pp(\vd_\pp\L^\pp)(N^1_{\@\@}{+}N^2_{\@\@})
  +\inv4\L^\pp([D_+,\Db_+]\L^\pp)(N^1_{\@\@}{-}N^2_{\@\@})~.\cr}}}

Similarly for rightons,
\eqn\eLR{\eqalign{L_R&=\inv2\int\rd^2\sR~\Nt^\mm\big(\Y;\3\big)+\hc\cr
                     &=\inv4\big[[D_-,\Db_-]\Nt^\mm(\Y;\3)\big]\Zp+\hc\cr}}
where $N^\mm$ may depend on all of $\Y,\Y^\mm$, $(D_-\Y)$,
$(\Db_-\Y)$ and $(D_-\Db_-\Y)$. Expanding this function in the
superderivatives and the $\Y^\mm$'s, and truncating the expansion so as to
prohibit higher derivative terms and non-negative canonical dimension
coefficients, we find:
 $${\eqalign{ \Nt^\mm=~
 &\Y^\mm \Nt_\@
  +\inv2\Y^\mm(D_-\Db_-\Y^\mm)\Nt^1_{\@\@}
  +\inv2\Y^\mm(\Db_-D_-\Y^\mm)\Nt^2_{\@\@}\cr
 &+\inv2(D_-\Y^\mm)(D_-\Y^\mm)\Nt^3_{\@\@}
  +\inv2(D_-\Y^\mm)(\Db_-\Y^\mm)\Nt^4_{\@\@}\cr
 &+\inv2(\Db_-\Y^\mm)(\Db_-\Y^\mm)\Nt^5_{\@\@}
  +\inv2\Y^\mm\Y^\mm(D_-\Db_-\Y)\Nt^1_{\@\@\@}\cr
 &+\inv2\Y^\mm\Y^\mm(\Db_-D_-\Y)\Nt^2_{\@\@\@}
  +\inv2\Y^\mm(D_-\Y^\mm)(D_-\Y)\Nt^3_{\@\@\@}\cr
 &+\inv2\Y^\mm(D_-\Y^\mm)(\Db_-\Y)\Nt^4_{\@\@\@}
  +\inv2\Y^\mm(\Db_-\Y^\mm)(D_-\Y)\Nt^5_{\@\@\@}\cr
 &+\inv2\Y^\mm(\Db_-\Y^\mm)(\Db_-\Y)\Nt^6_{\@\@\@}
  +\inv2\Y^\mm\Y^\mm(D_-\Y)(D_-\Y)\Nt^1_{\@\@\@\@}\cr
 &+\inv2\Y^\mm\Y^\mm(D_-\Y)(\Db_-\Y)\Nt^2_{\@\@\@\@}
  +\inv2\Y^\mm\Y^\mm(\Db_-\Y)(\Db_-\Y)\Nt^3_{\@\@\@\@}~, \cr}} \eqno\eThN{b}
 $$
where the $\Nt_\@,\3,\Nt_{\@\@\@\@}$'s are all arbitrary functions of the
$\Y$.

\subsec{$\j$-terms}\noindent
\subseclab\ssJ
Finally, we come to the terms with the simplest $\vs,\vsb$-dependence. As
listed in~\eDSF{g}, $\F_\LL,(\Db_+\L)$ and $(\Db_+D_+\L)$ satisfy the
chiral lefton pair of superconstraints~\eQSF{a}, \ie, they depend only on
$\vs^-$. Thus
\eqn\eXXX{ L_{\J\LL} = \int\rd\vs^-\J^+(\F_\LL;\Db_+\L;\Db_+D_+\L) +\hc }
is a hermitian supersymmetric Lagrangian density term, given any analytic
spin $+\inv2$ function $\J^+(\F_\LL;\Db_+\L;\Db_+D_+\L)$ and its hermitian
conjugate\ft{Recall that the unidexterous superfields, $\L$ and $\Y$, while
in general complex, {\it may be chosen\/} to be real or imaginary.}.

The superfields $\F_\LL,\Db_+\L$ and $\Db_+D_+\L$, as defined above~\eQSF{}
and~\eHSF{}, contain no positive spin component fields, which is why the
$\L^\pp$ have been introduced. For a Lagrangian density without higher
derivative and negative dimensional coefficient functions, we truncate the
expansion of $\J^+$ in $\Db_+\L$ and $\Db_+D_+\L$, and obtain the list:
\eqn\eThJ{ (\Db_+\L^{\ac\pp})\J_\ac~,\quad
 \inv4(\Db_+\L^{\ac\pp})(\Db_+D_+\L^{\bc\pp})\J_{\ac\bc}~,\quad
  +\inv4(\Db_+\L^{\ac\pp})(\Db_+\L^{\bc\pp})(\Db_+\L^\ch)
   \J_{\ac\bc\ch}~, }
where the $J_\ac,\J_{\ac\bc}$ and $\J_{\ac\bc\ch}$'s are arbitrary analytic
functions of only the $\F_\LL$. However, all of these turn out to be
special cases of various terms from~\eThN{a}. For example,
\eqn\eXXX{ (\Db_+\L^{\ac\pp})\J_\ac }
may be rewritten unambiguously (up to total derivatives) and without any
loss of generality, as a unidexterous F-term:
\eqn\eXXX{ \inv2\big[D_+(\Db_+\L^{\ac\pp})\J_\ac(\F_\LL)\big]\Zp+\hc
 \cong\inv4\big[[D_+,\Db_+]\L^{\ac\pp}\J_\ac(\F_\LL)\big]\Zp+\hc }
The same of course holds for its conjugate, and their parity mirrors,
$\J^-(\F_\RR;\Db_-\Y;\Db_-D_-\Y)$ and its conjugate.

Note that it was necessary to introduce the superfields $\L^\pp$ and
$\Y^\mm$ for there to be any unidexterous F-terms,~\eLL\ and~\eLR.
 The Readers interested only in the ambidextrous haploid
superfields~\eHSF{a{-}d} need not concern themselves with~\eLL\ and~\eLR,
and the appropriate terms from the rest of~\eMGL{}.

\subsec{Physical degrees of freedom}\noindent
\subseclab\ssPhDeg
Note that {\it each term\/} in Eqs.~\eLD, \eLQ{}, \eLW, \eLS, \eLL\
and~\eLR\ leads to a {\it separately supersymmetric term\/} in the
Lagrangian density~\eMGL{}. The plentiful of these expressions, if not
already~\eMGL{} and~\eSGL{}, make it clear that 2-dimensional
(2,2)-supersymmetric models are far more diverse than $N{=}1$
supersymmetric models in 4-dimensional spacetime.

\topic{Interdependent kinetic terms}
In models in higher dimensional spacetimes, the kinetic terms come
entirely from the `D-terms'~\eLD. In 2-dimensional spacetime, however,
we find kinetic terms in all other contributions to the Lagrangian density,
$L_W$ in~\eLW, $L_\S$ in~\eLS, $L_L$ in~\eLL, $L_R$ in~\eLR\
and $L_Q$ in~\eLQ{}. All of these then modify the `standard' (D-term)
kinetic terms.

In particular and as discussed already in \SS\,\ssL, the
$\l$-terms~\eSGL{f,g}, and in fact {\it all\/} the nine terms in the
Lagrangian density~\eSGL{}, contain terms that directly modify the kinetic
terms for the components of the various superfields. This type of mechanism
was used in Ref.~\rMarjD\ to prove that kinetic terms in fact can be
invariant under renormalization flow; indeed, the kinetic terms emerging
from all except $L_D$ in~\eSGL{h} are protected by the usual
nonrenormalization theorems (see, \eg, p.358 of Ref.~\rGGRS\ for an
important caveat).

A minor comment may not be out of place here. Educated on simple models with
`flat' kinetic terms such as
$\int\rd^4\vs\,\F\FB\ni(\vd\f){\cdot}(\vd\fb)$, one is easy to skim over
terms like $(\vd_\pp\ell^\ah)(\vd_\mm r^\ih)$ as being trivial. Indeed, the
simple D-term $\int\rd^4\vs\,\L\Y$
\eqn\eXXX{{\eqalign{
 &\[4[(\vd_\pp\ell)-2iL_\pp][(\vd_\mm r)-2iR_\mm]\cr
 &=(\vd_\pp\ell)(\vd_\mm r)-4L_\pp R_\mm
  -2i(\vd_\pp\ell)R_\mm-2iL_\pp(\vd_\mm r)~,\cr
 &=\vd_\pp(\ell\vd_\mm r)-4L_\pp R_\mm
  -2i\vd_\pp(\ell R_\mm)-2i\vd_\mm(L_\pp r)~,\cr
 }}}
turns into a total derivative, except for the non-derivative $L_\pp R_\mm$
term. This clearly carries no dynamics for the spin-0 fields $\ell$ and
$r$. However, consider
$\int\rd^2\vs(\Db_+\L^\ah)(\Db_-\Y^\ih)M_{\ah\ih}(\F)$, which is a
special case of the third term in~\eThQ{b}. Upon fermionic
integration, we obtain the `kinetic' terms:
\eqn\eXXX{ -\inv4[(\vd_\pp\ell^\ah)-2iL^\ah_\pp]
                 [(\vd_\mm r^\ih)-2iR^\ih_\mm]M_{\ah\ih}(\f)~. }
These definitely do not turn into total derivatives, as long as the
`metric' $M_{\ah\ih}(\f)$ is not constant; in fact, these provide
non-trivial $\f$-dependent kinetic (mixing) terms for the oppositely handed
bosons $\ell^\ah$ and $r^\ih$.

\topic{Auxiliary fields and (non)linearity}
After having expanded the Lagrangian density into component fields by means
of fermionic integration, we notice that a number of fields have algebraic
equations of motion. By contrast, fields with differential equations of
motion are called physical. For example, consider using the
$(\Db_-\F^\nb)M_\nb$ by itself for a Lagrangian:
\eqn\eXXX{ L_1 = \inv8[D_+,\Db_+]D_-(\Db_-\f^\nb)M_\nb\Zp+\hc }
The full component expansion is given in Appendix~B.2. From there, we find
\eqn\eXXX{{\eqalign{0={\d L_1\over\d F^\nb}
 &=F^\m(M_{\nb\1\m}+\Mb_{\m\1\nb})+
 \j^\r_-(\j^\m_+M_{\nb\1\m\r}+\l^\ah_+M_{\nb\1\ah\r})\cr
 &\]2-\j^\r_-(\j^\mt_{\LL+}\Mb_{\r\1\mt\nb}
               +\c^\bb_+\Mb_{\r\1\bb\nb}
               +\bar\bl^\ah_+\Mb_{\r\1\ah\nb})~,\cr }}}
and the conjugate relation by varying with respect to $F^\m$. These are the
equations of motion for $F^\m$ and $F^\nb$. Wherever the hermitian matrix
$(M_{\nb\1\m}+\Mb_{\m\1\nb})$ is invertible, the equations of motion can be
solved for all the $F$'s, and these solutions may be substituted back into
the Lagrangian density. Writing $M^{\m\nb}$ for the matrix inverse of
$(M_{\nb\1\m}+\Mb_{\m\1\nb})$, we'd have
\eqn\eEoM{ F^\m 
 = M^{\m\nb}\big[\j^\r_-(\j^\mt_{\LL+}\Mb_{\r\1\mt\nb}
                        +\c^\bb_+\Mb_{\r\1\bb\nb}
                        +\bar\bl^\ah_+\Mb_{\r\1\ah\nb})
                -\j^\r_-(\j^\s_+M_{\nb\1\s\r}
                        +\l^\ah_+M_{\nb\1\ah\r})\big]~, }
and a similar expression for $\F^\nb$. Clearly, this is nonlinear in the
fields: roughly, $F$ is quadratic in the fermions, but the details depend
on the choice of the Lagrangian density. Its substitution back in the
Lagrangian density produces additional cubic and quartic terms, and so
affects the interactions between the various fields.

The relation~\eEoM\ also enters in another important place. Prior to
elimination of the $F$'s, the supersymmetry transformation of $\j_-$
is
\eqn\eXXX{ \d_\e\j_-=\inv{\sqrt2}D_-(\e{\cdot}Q+\bar\e{\cdot}\Qb)\F|
                    =-\sqrt2\e^+F+\3 }
which is linear in the fields (including the omitted terms). After
substituting~\eEoM, this transformation rule becomes nonlinear in fields,
the details of this nonlinearity depending on the choice of the Lagrangian
density. We will refer to these two incarnations of the supersymmetry
transformation rules as `linear' and `nonlinear'.

It is therefore logically backward to postulate a (nonlinear) set of
supersymmetry transformations involving only physical fields, and then seek
an invariant Lagrangian density. Although the {\it linear\/} supersymmetry
transformation rules for a superfield only depend on its definition, \ie,
which superconstraints it might satisfy, the {\it nonlinear\/}
transformation rules depend on the choice of the Lagrangian density through
the equations of motion for the auxiliary fields. In particular,
modifications of the Lagrangian density (such as through varying moduli)
also modify the non-linear supersymmetry transformation rules.
 In contradistinction, the linear transformation rules, with the auxiliary
fields uneliminated, remain unchanged\ft{May the erudite Reader forgive
this decidedly un-English but accurate agglutination.}, determined
entirely by the nature of the superfield itself.

\newsec{Explicit Constraining}\noindent
\seclab\sExCn
The Lagrangian density~\eMGL{} has been obtained as a sum of Berezin
integrals over suitable subsets of the fermionic coordinates. The
superderivative superconstraints~\eHSF{} and~\eQSF{} (and easily
also~\eNMF{}, should one desire) were {\it implicitly\/} and {\it
automatically\/} used in the process.

Of course, the Reader may instead wish to include such (super)constraints
explicitly. To that end, and as disclaimed above, here we examine some
simple Lagrangians which enable an explicit inclusion of (super)constraints,
and so a more standard treatment of the unidexterous superfields. The
present treatment, while inspired by that of Ref.~\rCBJiWa, is hopefully a
self-sufficient introduction to the topic.

\subsec{Chiral bosons}\noindent
One standard Lagrangian density for a `chiral boson' is~\rCBSieg:
\eqn\eLCB{ L_{CB} = (\vd_\pp\f)(\vd_\mm\f)+Z^\mm_\pp(\vd_\mm\f)^2~. }
The inclusion of the constraint through a linear term, $Z^\mm(\vd_\mm\f)$
would have left the dynamical degrees of freedom unhalved. Although
$\f(\s^\mm)$ is gauged away, besides $\f(\s^\pp)$ we also end up with
$\Tw{Z}^\mm(\s^\pp)$, where $\Tw{Z}^\mm\define(Z^\mm{+}\vd_\pp\f)$,
at least classically~\refs{\rAuxWS,\rSriCB}. Note that upon imposing the
constraint, $\vd_\mm\f=0$, the physical field $\f(\s^\pp)$ and the Lagrange
multiplier, $Z^\mm_\pp$, drop out of the Lagrangian density~\eLCB,
indicating that this Lagrangian density is invariant under a gauge symmetry.
A straightforward calculation verifies that~\eLCB\ transforms into a total
derivative under the Siegel symmetry transformation:
\eqn\eWSS{{\eqalign{
 \d_S\f &= \k^\mm(\vd_\mm\f)~,\cr
 \d_S Z^\mm_\pp &= -(\vd_\pp\k^\mm) - (\vd_\mm\k^\mm)Z^\mm_\pp
  + \k^\mm(\vd_\mm Z^\mm_\pp)~.\cr }}}
The Lagrangian density~\eLCB\ is also invariant with respect to a global
inhomogeneous (axial) symmetry:
\eqn\eAxS{ \d_\a\f = \a~,\qquad \d_\a Z^\mm_\pp = 0~,\qquad
           \a=\hbox{\it const}~, }
which allows the interpretation of $\f$ as a `self-dual gauge scalar'.
(Self-duality here means that $\vd_\mm\f=(\s_3^{ij}{-}\e^{ij})\vd_j\f=0$,
where $i,j=0,1$and $\e^{01}=1$.)

The following two quite obvious observations will guide the construction of
the (2,2)-supersymmetric version:
\item{1.} the second term enforces the constraint(s) via the Lagrange
          multiplier;
\item{2.} the remaining part of the Lagrangian (here, the first term)
          pertains to the unconstrained field.
 \par\noindent
These two features will then be required of the (2,2)-supersymmetric
version.

\subsec{Chiral lefton}\noindent
Consider a simple model, with one chiral superfield, $\F$, governed by
the following simple Lagrangian density:
\eqn\eLCL{L_{CL}
 =\inv4\int\rd^4\vs~\Big[\F\FB+\inv2\BM{Z}^\mm|D_-\F|^2\Big]~.}
Variation with respect to the Lagrange multiplier (general) superfield,
$\BM{Z}^\mm_\pp$, then produces the quadratic superconstraint,
$|D_-\F|^2{=}0$, which restricts the chiral superfield into a chiral lefton
superfield~\eQSF{a}. This one time we did include a term which ultimately
leads to higher derivatives in the Lagrangian density.
 Defining the components of $\BM{Z}^\mm$ as:
\eqna\eZc
 $$\twoeqsalignno{
 Z^\mm_{D}
 &=\rlap{$-\inv8(D_-\Db_-\Db_+D_++D_+\Db_+\Db_-D_-)\BM{Z}^\mm\Zp~,
 \quad Z^\mm=\BM{Z}^\mm\Zp~,$\hss}                         &&&\eZc{a}\cr
 Z^\mm_{\bl+}
 &= -\inv{4\sqrt2}[D_+,\Db_+]\Db_-\BM{Z}^\mm\Zp~, \quad&\quad
 Z^\mm_{\l+}
 &= -\inv{4\sqrt2}[D_+,\Db_+]D_-\BM{Z}^\mm\Zp~,              &\eZc{b}\cr 
 Z^\mm_{\bl-}
 &= -\inv{4\sqrt2}[D_-,\Db_-]\Db_+\BM{Z}^\mm\Zp~, &
 Z^\mm_{\l-}
 &= -\inv{4\sqrt2}[D_-,\Db_-]D_+\BM{Z}^\mm\Zp~,              &\eZc{c}\cr 
 Z^\mm_{\bar F}
 &= -\inv2\Db_+\Db_-\BM{Z}^\mm\Zp~, &
 Z^\mm_{F}
 &= -\inv2D_-D_+\BM{Z}^\mm\Zp~,                              &\eZc{d}\cr 
 Z^\mm_{\pp}
 &= -\inv4[D_+,\Db_+]\BM{Z}^\mm\Zp~, &
 Z^\mm_{\mm}
 &= -\inv4[D_-,\Db_-]\BM{Z}^\mm\Zp~,                         &\eZc{e}\cr 
 Z^\mm_{\bar X}
 &= -\inv2D_+\Db_-\BM{Z}^\mm\Zp~, &
 Z^\mm_{X}
 &= -\inv2D_-\Db_+\BM{Z}^\mm\Zp~,                            &\eZc{f}\cr 
 Z^\mm_{\j\a}
 &= -\inv{\sqrt2}D_\a\BM{Z}^\mm\Zp~, &
 Z^\mm_{\bar \j\a}
 &= -\inv{\sqrt2}\Db_\a\BM{Z}^\mm\Zp~,                       &\eZc{g}\cr 
 }$$
where $\a=-,+$. Note that all sixteen component fields of $\BM{Z}^\mm$ act
as Lagrange multipliers.

The simple Lagrangian density~\eLCL\ contains, among many, the following
terms:
\eqna\eXXX
 $$\eqalignno{ L_{CL}=
 &\inv2[(\vd_\mm\f)(\vd_\pp\fb)+(\vd_\pp\f)(\vd_\mm\fb)]-F\Fb
  +\frc i2[(\j_-\dvd_\pp\jb_-)+(\j_+\dvd_\mm\jb_+)]            &\eXXX{a}\cr
 &+Z^\mm_D[\j_-\jb_-]                                          &\eXXX{b}\cr
 &+Z^\mm_\pp[i(\vd_\mm\j_-)\jb_- - (\vd_\mm\f)(\vd_\mm\fb)]
  -Z^\mm_\mm[i(\vd_\mm\j_-)\jb_- - F\Fb]                       &\eXXX{c}\cr
 &+\inv2Z^\mm[(\vd_\pp\vd_\mm\j_-)\jb_-+\j_-(\vd_\pp\vd_\mm\jb_-)
            -2(\vd\j_+)(\vd\jb_+)\cr
 &\]5       +i(\vd_\pp\vd_\mm\f)(\vd_\mm\fb)-i(\vd_\mm\f)(\vd_\pp\vd_\mm\fb)
             +i(\vd_\mm F)\Fb-iF(\vd_\mm\Fb)]~~~               &\eXXX{d}\cr
 &+\ldots\cr }$$
where $(A\dvd B)\define[A(\vd B)-(\vd A)B]$.

Variation with respect to $Z^\mm_D$ imposes $|\j_-|^2=0$ (so that
$\j_-=0=\jb_-$), whereupon the subsequent constraints simplify. Variation
with respect to $Z^\mm_\pp$ and $Z^\mm_\mm$ then imposes the
constraints $|\vd_\mm\f|^2=0$ and $|F|^2=0$  (so that
$\vd_\mm\f=0=\vd_\mm\fb$ and $F=0=\Fb$), respectively. This further
simplifies the last shown constraint, so that variation with respect to
$Z^\mm$ imposes the constraint
$|\vd_\mm\j_+|^2=0$ (so that $\vd_\mm\j_+=0$). All the sixteen component
fields of $\BM{Z}^\mm$ drop out of the Lagrangian upon enforcing these
constraints, indicating the presence of a generalization of the Siegel
symmetry~\eWSS. The net effect of the superconstraint term in~\eLCL\ then is
to impose the constraints (and their conjugates):
\eqna\eCst
 $$\twoeqsalignno{
 \j_- &= 0~, \quad&\quad F &= 0~; &\eCst{a}\cr
 \vd_\mm\f &= 0~, \quad&\quad \vd_\mm\j_+ &= 0~; &\eCst{b}\cr
 }$$
The first two of these, Eqs.~\eCst{a}, halve the number of component fields
in the chiral superfield~\eHSF{a} to those of the chiral
lefton~\eQSF{a}; the second two, Eqs.~\eCst{b}, ensure that the remaining
component fields are left-movers. This completes the (initial, primary)
constraining of the chiral superfield to a lefton chiral; the other
quartoid fields~\eQSF{} can be treated in the same fashion, {\it mutatis
mutandis\/}. Furthermore, a similar treatment of the unidexterous haploid
superfields~\eHSF{e,f} seems equally straightforward, albeit rather more
beset with technical detail.

The choice~\eLCL\ is not necessarily the most convenient. The simple
alternative
\eqn\eLcL{L_{CL}
 =\inv4\int\rd^4\vs~\F\FB +\inv4\int\rd\sR\BM{Z}^\mm_\pp|D_-\F|^2~, }
produces a subtly different result.
 Defining a partial list of component fields of $\BM{Z}^\mm_\pp$ as:
\eqna\eZc
 $$\twoeqsalignno{
 Z^\mm_\pp &=\BM{Z}^\mm_\pp\Zp~,                     \quad&\quad
 Z^\mm     &=\inv4[D_-,\Db_-]\BM{Z}^\mm_\pp\Zp~,          &\eZc{a,b}\cr
 \z^\mm_+  &= \inv{\sqrt2}D_-\BM{Z}^\mm_\pp\Zp~,     \quad&\quad
 \bz^\mm_+ &= -\inv{\sqrt2}\Db_-\BM{Z}^\mm_\pp\Zp~,       &\eZc{c,d}\cr 
 }$$
the constraint term becomes
\eqn\eXXX{{\eqalign{\inv8&[D_-,\Db_-]{\bf Z}^\mm_\pp|D_-\F|^2\Zp\cr
 &=Z^\mm|\j_-|^2-Z^\mm_\pp\big[|\vd_\mm\f|^2+\frc i2(\j_-\dvd_\mm\jb_-)\big]
  +i\z^\mm_+(\vd_\mm\f)\jb_-+i\bz^\mm_+\j_-(\vd_\mm\fb)~.
 }}}
Variation with respect to $Z^\mm$ annihilates the fermions, $\j_-,\jb_-$.
Thereupon, variation with respect to $Z^\mm_\pp$ sets
$\vd_\mm\f=0=\vd_\mm\fb$. The auxiliary component field $F$ and the
fermions $\j_+,\jb_+$ appear only in the expansion of the first (`kinetic')
term. It is then their equation of motion that sets $F=0=\Fb$ and
$\vd_\mm\j_+=0=\vd_\mm\jb_+$. Thus, we are again left with the component
fields of a (quartoid) chiral lefton superfield, and the sixteen components
fields of ${\bf Z}^\mm_\pp$ which disappear from the action upon enforcing
the constraints.

So, whereas the superconstraint term in~\eLCL\ by itself reduces the chiral
superfield to a chiral lefton and annihilates the `kinetic' term in the
process, the superconstraint term in~\eLcL\ does only `half' of that, the
rest of the constraining is effected by the equations of motion, derived
from the `kinetic' term in~\eLcL. Having demonstrated these two
possibilities, we leave the application- and taste-dependent choice between
these (and possibly other) alternatives to the Reader.

\subsec{The general model}\noindent
The skeptical Reader might inquire as to the relevancy of the above simple
model~\eLCL\ to the vastly more complicated Lagrangian density~\eMGL{},
which are quite freely interspersed with the sundry unidexterous haploid
and quartoid superfields. To that end, note the following simple facts
about this Lagrangian density:
\item{1.} Terms that would involve any of the constraints~\eCst{} are
          {\it implicitly\/} projected out of the Lagrangian
          density~\eMGL{}, during the fermionic integration---by
          its very definition.
\item{2.} The Lagrangian density~\eMGL{} does not include any of the
          Lagrange multiplier superfields enforcing the
          superconstraints~\eHSF{} and~\eQSF{}, because of item~1.
\item{3.} The Siegel transformations~\eWSS\ turn unidexterous field(s) into
          a multiple of the unidexterity superconstraint. That is, the
          (unidexterity) superconstraints are the generators of the Siegel
          symmetry.
\par\noindent
The Lagrangian density~\eMGL{}, being {\it implicitly\/} restricted to the
constrained (super)field space, is then trivially an invariant of the
Siegel symmetry. So, at least at the classical level, there should be no
obstacle to using the unidexterous superfields in all the ways they appear
herein. 

One could instead relax the unidexterity superconstraints~\eHSF{e,f} on the
$\L$'s and $\Y$'s, and impose these superconstraints explicitly through
adding suitable Lagrangian multiplier terms to (a suitable subset of) the
Lagrangian density~\eMGL{}. Finding thereupon a suitable generalization of
the Siegel symmetry for this modified Lagrangian density, should guarantee
a well-defined interactive quantum theory. This, or any other of the many
approaches to constrained quantization, is however beyond our present
scope. The Reader weary of the unidexterous quantization issues may of
course safely use the Lagrangian density~\eMGL{} with the unidexterous
superfields carefully extracted.

On the other hand, the axial symmetry analogue of~\eAxS\ for each of the
unidexterous superfields is explicitly broken by each term where an
inhomogeneously transforming component field of a unidexterous superfield
appears without a spacetime derivative acting on it. Many, however, do not
break the axial symmetries: consider the term
$\int\rd^2\vs(\Db_+\L^\ah)(\Db_-\Y^\ih)M_{\ah\ih}(\F)$, obtained as a
special case from~\eThQ{b}, which upon fermionic integration yields
\eqn\eXmp{{\eqalign{
 &-\inv4[(\vd_\pp\ell^\ah)-2iL^\ah_\pp]
        [(\vd_\mm r^\ih)-2iR^\ih_\mm]M_{\ah\ih}(\f)
 +\j^\m_-\j^\n_+\bl^\ah_+\vr^\ih_-M_{\ah\ih\1\m\n}(\f)\cr
 &+\big\{F^\m\bl^\ah_+\vr^\ih_-
       +\frc i2\j^\m_-[(\vd_\pp\ell^\ah)-2iL^\ah_\pp]\vr^\ih_-
       +\frc i2\j^\m_-\bl^\ah_+[(\vd_\mm r^\ih)-2iR^\ih_\mm]
         \big\}M_{\ah\ih\1\m}(\f)~.\cr }}}
This involves $\ell^\ah$ and $r^\ih$, the lowest component fields of the
lefton and righton superfields, only through their spacetime derivatives.
Consequently, it is invariant under the two inhomogeneous `axial'
transformations: $\d_\a\ell^\ah=\a^\ah$ (where $\vd_\pp\a^\ah{=}0$), and
$\d_\a r^\ih=\a^\ih$ (where $\vd_\mm\a^\ih{=}0$), with respect to which all
other fields are held inert.

The inquisitive Reader is invited to sift through the multitude of
individually supersymmetric parts of the Lagrangian density~\eMGL{} and
identify all the terms with such property, should this be a requirement of
a desired application.

Finally, the constraining Lagrangian density~\eLCL\ could serve as a
prototype of an {\it excplicit\/} constraining of the various constrained
superifields~\rCBSieg. For example, given an unconstrained complex
(2,2)-superfield, $\W$, and two unconstrained Lagrange multiplier
superfields, $\BM{Z}^\mm$ and $\BM{Z}^\pp$, the Lagrangian term
\eqn\eXXX{ \inv8\int\rd^4\vs~\big[\BM{Z}^\mm(\Db_-\W)(D_-\ba\W)
                                  +\BM{Z}^\pp(\Db_+\W)(D_+\ba\W)\big]~, }
will constrain $\W$ into a chiral superfield. Again, the maintenance of the
generalization of the Siegel symmetry off the constrained superfield
subspace and quantum issues are beyond our present scope.

\newsec{Wave-Function(al)s and Geometry}\noindent
\seclab\sWFnG
The Hilbert space of supersymmetric field theories is in many regards much
simpler than the one in models without
supersymmetry~\refs{\rSuSyM,\rWitInd,\rEliSch,\rClaHal,\rCec,\rCeGiPa}.
Although many of the known results have been derived for 2-dimensional
supersymmetric field theories, the lack of present generality warrants
another look at these issues.

\subsec{Vacua and Wave-Function(al)s}\noindent
The wave-function(al)s in supersymmetric theories are functions of the
superfields and so admit a formal expansion in the fermionic fields. Local
such wave function(al)s (in the 2-dimensional spacetime sense) will then
have a terminating Taylor series owing to the finite number of fermionic
fields and their anticommutivity, and so must be multinomials in fermionic
fields.

A hallmark of fermionic fields is the fact that their kinetic
terms are first order in time derivatives, so half the fermionic
fields turn out to be conjugate momenta to the other half. The exact
pairing of course depends on the exact choice of the Lagrangian
density~\eMGL{}. On the other hand, the identification of the `canonical
coordinate' and the `canonical momentum' within any one canonically
conjugate pair of fermion fields is completely arbitrary.

In a second-quantized theory, we define a (Dirac) `vacuum' state,
$\ket{0}$, which is annihilated by precisely one half of the fermionic
fields. Except, of course, the choice of precisely which half will
annihilate $\ket{0}$ remains free. As well known, Eqs.~\eSusyQ\ imply
\eqn\eSuSyQ{
 \big\{\, Q_- \,,\, \Qb_- \,\big\}+\big\{\, Q_+ \,,\, \Qb_+ \,\big\}~
             = ~4H~, }
and so
\eqn\eXXX{{\eqalign{\V{H}
 &=\inv4\V{\{Q_-,\Qb_-\}} + \inv4\V{\{D_+,\Db_+\}}\cr
 &=\inv4\big[|Q_-|^2+|\Qb_-|^2+|Q_+|^2+|\Qb_+|^2\big]\geq0~, }}}
the equality reached precisely for supersymmetric states, \ie, those
annihilated by all four supercharges. That is, in supersymmetric models,
the Hamiltonian is positive definite, and its global minima (of zero
energy, hence {\it the true vacua\/}) are the supersymmetric states.

It is then the supersymmetric states, {\it the true vacua\/}, that we seek
in a supersymmetric model, so as to erect the full Hilbert space by acting
on these vacua with creation operators. In  turn, the vacua, \ie,
supersymmetric states are easily constructed from the `vacua' $\ket0$.

\topic{Single species models}
Even if considering just a single species of fermionic fields, such as
$\j^\m_\pm,\j^\nb_\pm$ which appear in chiral superfields and their
conjugates (as used in Wess-Zumino model), respectively, one easily
finds six {\it logically possible\/} choices of such a $\ket{0}$:
\eqna\eSVac
$$ \twoeqsalignno{
 \hbox{\bf\BM{(c,c)} vacuum}:&\qquad
 &\j^\m_+\ket{0}_{(c,c)}=&0=\j^\m_-\ket{0}_{(c,c)}~; &\eSVac{a} \cr
 \hbox{\bf\BM{(a,a)} vacuum}:&\qquad
 &\j^\mb_+\ket{0}_{(a,a)}=&0=\j^\mb_-\ket{0}_{(a,a)}~; &\eSVac{b} \cr
 \hbox{\bf\BM{(a,c)} vacuum}:&\qquad
 &\j^\mb_+\ket{0}_{(a,c)} =&0 = \j^\m_-\ket{0}_{(a,c)}~; &\eSVac{c} \cr
 \hbox{\bf\BM{(c,a)} vacuum}:&\qquad
 &\j^\m_+\ket{0}_{(c,a)} =&0 = \j^\mb_-\ket{0}_{(c,a)}~; &\eSVac{d} \cr
 \hbox{\bf Left vacuum}:&\qquad
 &\j^\m_-\ket{0}_\LL =&0 = \j^\mb_-\ket{0}_\LL~. &\eSVac{e} \cr
 \hbox{\bf Right vacuum}:&\qquad
 &\j^\m_+\ket{0}_\RR =&0 = \j^\mb_+\ket{0}_\RR~. &\eSVac{f} \cr
 }$$
We have assumed here that a degree of uniformity is to be required of the
chiral superfields and their conjugates, so that all (un)conjugate fermions
appear in the defining equations~\eSVac{} on the same footing. In
particular, the definitions~\eSVac{} are invariant with respect to any
invertible linear transformation among the chiral superfields $\F^\m$ (which
commutes with the Lorentz symmetry, and so preserves spin), accompanied by
the conjugate transformation among their antichiral conjugates, $\F^\mb$.
Clearly, if (global) invariance is to be required only with respect to a
subset of these transformations, several additional choices of $\ket{0}$
will be possible.

For any one choice of $\ket{0}$, the half of the fermionic fields which do
not annihilate $\ket{0}$ then act as creation operators, and states are
constructed as multinomials in these `creation' fermions, acting on
$\ket{0}$. Naturally, these wave function(al)s (states) built upon these
`vacua' will be called by the same names: that is, a $(c,c)$ state will be
built from a $(c,c)$ vacuum, $\ket{0}_{(c,c)}$, an $(a,c)$ state from
$\ket{0}_{(a,c)}$, and so on.

Now, the kinetic term for the fermions is typically something like
\eqn\eXXX{ {\eqalign{L_{\rm kin.}^{(F)} 
 &=~ iG_{\m\nb}\big[\j_+^\m(\vd_\mm\j_+^\nb)
                   +\j_-^\m(\vd_\pp\j_-^\nb)\big]~, \cr
 &=~ iG_{\m\nb}\big(\j_+^\m\dot\j_+^\nb
                   +\j_-^\m\dot\j_-^\nb
                   -\j_+^\m\acute\j_+^\nb
                   +\j_-^\m\acute\j_-^\nb\big)~, \cr
}}}
where $\dot\j_\pm\define\vd_0\j_\pm$ is the time-derivative of $\j_\pm$,
$\acute\j_\pm\define\vd_1\j_\pm$ its space-derivative, and $G_{\m\nb}$ is
the metric determined by the choice of the Lagrangian density~\eMGL{}.
Therefore, $i\j_{\pm\m}\define iG_{\m\nb}\j^\nb_\pm$ is the canonical
conjugate of $\j^\m_\pm$, so that
\eqn\eCCR{
 \big\{\,i\j_{\pm\m}\,,\,\j^\n_\pm\,\big\}~=~i\d^\n_\m~,
 \quad\hbox{\it i.e.}\quad
 \big\{\,\j^\nb_\pm\,,\,\j^\m_\pm\,\big\}~=~G^{\m\nb}~. }

Notice now that the fifth and sixth logical possibilities, tentatively
called the `Left' and `Right' vacua in fact must vanish, since the two
conditions~\eSVac{e} may be combined to yield
\eqn\eXXX{ 0~=~\big\{\,\j^\m_-\,,\,\j^\nb_-\,\big\}\ket{0}_\LL
 ~=~-iG^{\m\nb}\ket{0}_\LL~, }
whereupon $\ket{0}_\LL\id0$ is inevitable. So, unlike (2,2)-superfields in
2-dimensional spacetime, there are only four types  (two conjugate
pairs) of (2,2)-vacua --- for a single species of fermions.

In a $\s$-model, the bosonic coordinates $\f^\m,\f^\nb$ are identified
as target space coordinate-valued world-sheet scalars. That is, they map the
world-sheet ({\it domain\/}) into $X$ (the {\it target space\/}). In a
Fourier expansion (locally in $X$), the 0-wavevector modes may be
identified with local coordinates $z,\bar z$ in $X$. Then, half of the
fermions $\j_\pm,\jb_\pm$ can be identified with the differentials
$\rd z,\rd\bar z$, and the other half with the partials $\vd,\bar\vd$. We
adopt:
\eqn\eAsgn{\twoeqsalign{
 G_{\m\nb}\j^\m_-&\mapsto\vd_\nb~,&\qquad \j^\nb_-&\mapsto\rd z^\nb~,\cr
 \j^\m_+&\mapsto\rd z^\m~,&\qquad G_{\m\nb}\j^\nb_+&\mapsto\vd_\m~,\cr }}
That the anticommutivity of $\j^\m_+$ and $\j^nb_-$ matches the
anticommutivity of $\rd z,\rd\bar z$, standard in differential geometry.
However, whereas the `momenta' $G_{\m\nb}\j^\m_-$ and $G_{\m\nb}\j^\nb_+$
also anticommute, the basis elements of the tangent bundle, $\vd_\nb$ and
$\vd_\m$ {\it commute\/}. Nevertheless, the above identification has proven
of great use in the analysis of the geometry of supersymmetric field theory
models.

With the identifications~\eAsgn, the $(c,c)$ wave function 
\eqn\eCCS{ \ket{p,q;h}_{(c,c)}
 =h^{\m_1{\cdots}\m_p}_{\nb_1{\cdots}\nb_q}(\f,\fb)
  \j_{+\m_1}{\cdots}\j_{+\m_p}\j^{\nb_1}_-{\cdots}\j^{\nb_q}_-
   \ket{0}_{(c,c)}}
corresponds to the $\wedge^pT_X$-valued $q$-form
 $$
   h^{\m_1{\cdots}\m_p}_{\nb_1{\cdots}\nb_q}(\f,\fb)
   \vd_{\m_1}{\wedge}{\cdots}{\wedge}\vd_{\m_p}{\wedge}
    \rd z^{\nb_1}{\wedge}{\cdots}{\wedge}\rd z^{\nb_q}~.
 $$
Similarly, the $(a,c)$ wave function
\eqn\eCCS{ \ket{p,q;\w}_{(a,c)}
 =\w_{\m_1{\cdots}\m_p\nb_1{\cdots}\nb_q}(\f,\fb)
  \j^{\m_1}_+{\cdots}\j^{\m_p}_+\j^{\nb_1}_-{\cdots}\j^{\nb_q}_-
   \ket{0}_{(a,c)}}
corresponds to the $\wedge^pT_X^*$-valued $q$-form
 $$
   \w_{\m_1{\cdots}\m_p\nb_1{\cdots}\nb_q}(\f,\fb)
   \rd z^{\m_1}{\wedge}{\cdots}{\wedge}\rd z^{\m_p}{\wedge}
    \rd z^{\nb_1}{\wedge}{\cdots}{\wedge}\rd z^{\nb_q}~.
 $$

\topic{Supersymmetric states}
States which are invariant with respect to supersymmetry transformations
are of very special interest~\rSuSyM; we recall a few important facts here.
Representing a supersymmetric state, let $\ket{*}$ stand for a wave-function
annihilated by all supercharges. Then
\eqn\eXXX{ \big\{\,Q_\pm\,,\,\Qb_\pm\,\big\}\ket{*}~=~0
           \quad\Iff\quad
           (H\pm p)\ket{*}~=~0~, }
which is equivalent to
\eqn\eXXX{ p\ket{*}~=~0~=~H\ket{*}~. }
That is, translationally invariant states of zero energy are
supersymmetric, and {\it vice versa\/}.

Now, the precise form of $H$, $p$, $Q_\pm$ and $\Qb_\pm$ (as operators on
the field space) depends on the precise choice of the Lagrangian
density~\eMGL{}. Restricting to translationally invariant (world-sheet
space-independent) states, is in many ways equivalent to completely
disregarding the spatial extension of the world-sheet, and reducing from
the 2-dimensional (supersymmetric) field theory to (supersymmetric) quantum
mechanics. In practice, this allows us to {\it ignore\/} the momentum
density $p$ and all the terms which involve world-sheet spatial
derivatives; $H$ correspondingly simplifies to the Fokker-Planck type
Hamiltonian.
 For the case of Wess-Zumino models (those containing only $n$ chiral
superfields and their conjugates), these operators have been identified
many years ago~\refs{\rEliSch,\rClaHal}.

For the more general case involving more than just one species of
superfields~\eHSF{}, \eQSF{}, \eNMF{}, \etc, this remains unexplored,
albeit straightforward.

\topic{Three remarks}
Before we turn to a cursory look at multiple-species models, three general
remarks are in order. These are based on the fact that the adjoint pair of
(Dolbeault) differential operators $\db\define\rd z^\m\vd_\m\wedge$ and
$\db\con$, with the only non-zero anticommutation relation
$\{\db,\db\con\}=\triangle_{\bar\vd}$, where $\triangle_{\bar\vd}$ is the
$\db$-Laplacian.
 Indeed, the correspondence between supersymmetric and cohomology
theories, first noted in Ref.~\rSuSyM, stemms from the (formal) isomorphism
between these and the pair of anticommuting differential operators
$D_\pm,\Db_\pm$ with the only non-zero anticommutation relation~\eSusyD.
 
\item{1.} Corresponding to the two pairs $\vd,\vd\con$, and $\db,\db\con$
in differential geometry, we have $Q_-,\Qb_-$ and $Q_+,\Qb_+$.
 Then, the supersymmetric analogue of complex conjugation in differential
geometry is not hermitian conjugation but {\it parity}, \BM{P}. Of course,
{\it upon Euclideanization\/}, parity does act as complex conjugation on the
(bosonic `body' of the) world-sheet Riemann surface.
 On the other hand, the differential-geometric analogue of hermitian
conjugation in supersymmetric models is the formal exchange
$\db\iff\db\con$, not complex conjugation.

\item{2.} The analogue of $\triangle_{\bar\vd}$ is $2(H+p)$, \ie,
$\vd_\pp$; the analogue of $\triangle_{\vd}$ $2(H-p)$, \ie, $\vd_\mm$.
 When complex conjugation is a symmetry of the target space (as it happens,
but not exclusively, on K\"ahler manifolds),
$\triangle_{\bar\vd}=\triangle_{\vd}$. The analogous situation with the
supersymmetry algebra and its representations happens when $2(H+p)=2(H-p)$:
in the sector of zero-momenum states, where $p\ket*=0$. Equivalently, this
implies $\vd_\pp=\vd_\mm$, which holds on $\s^1$-independent functions,
\ie, upon dimensional reduction to (supersymmetric) quantum mechanics.

\item{3.} Being annihilated by both $D_+$ and $\Db_+$, the righton
superfields, $\Y$~\eHSF{f}, are isomorphic to $\db$-closed and co-closed
forms, that is, to $\db$-{\it harmonic} forms. Similarly, the lefton
superfields, $\L$~\eHSF{f\e}, are isomorphic to $\vd$-closed and co-closed
forms, that is, to $\vd$-{\it harmonic} forms. This correspondence has, to
the best of my knowledge, not been observed before and its utility remains
unexplored.

\topic{Several species models}
With more types of superfields, and so more species of fermions, the number
of different types (sectors) of `vacua' and wave-function(al)s increases
very quickly.

 Even if restricting to just the `minimal' haploid superfields,
there are many more species of fermions: $\j^\m_\pm,\j^\nb_\pm$ from
(anti)chiral superfields, $\x^\a_-,\c^\a_+$ and $\x^\bb_-,\c^\bb_+$ from
twisted-(anti)chiral superfields, $\l^a_+,\bl^a_+$ (and, if complex, their
conjugates) from lefton superfields and $\r^i_-,\vr^i_-$  (and, if complex,
their conjugates) from righton superfields. Half of these are canonically
conjugate variables to the other half (the precise pairing depending on the
precise choice of the Lagrangian density), so we in general expect six
types of fermionic `creation' operators, not just two as above. Clearly,
there will also be many more choices of $\ket{0}$.

Consider a simple toy example with a single `minimal' haploid superfield of
each kind~\eHSF{} and the Lagrangian density
\eqna\eToy
 $$\eqalignno{ L_{\rm toy}
 &=\inv4\int\rd^4\vs~(\F\FB-\X\XB)                              &\eToy{a}\cr
 &\]2+\inv4\Big[\int\rd\sL\L^\pp(\Db_+D_+\L^\pp)+\hc\Big]~,
                                                                &\eToy{b}\cr
 &\]2+\inv4\Big[\int\rd\sR\Y^\mm(\Db_-D_-\Y^\mm)+\hc\Big]~,
                                                                &\eToy{c}\cr
 }$$
where, for simplicity, we require $\L^\pp,\Y^\mm$ to be real. Then, the
first two terms produce the standard kinetic terms for the scalar fields
$\f$ and $x$, their fermionic superpartners, $\j_\pm$ and $\x_-,\c_+$,
respectively, the auxiliary fields $F,X$, and their hermitian conjugates.
The scalars and have their usual canonically conjugate momenta, and
$\j_-,\j_+$ and $\x_-,\c_+$ are canonically conjugate pairs.

We then focus on the second part~\eToy{b}, which produces (up to total
derivatives):
\eqn\eXXX{L^{(\L)}_{\rm toy}=\inv2\big[(\vd_\pp\ell^\pp)+2iL^\pp_\pp\big]^2
                            -i(\bl^\pp_+\dvd_\pp\l^\pp_+)~. }
The equations of motion for $L^\pp_\pp,\l_+^\pp$ and $\bl_+^\pp$,
respectively, set:
\eqn\eXXX{ L^\pp_\pp=\frc i2(\vd_\pp\ell^\pp)~,\qquad
           \vd_\pp\bl^\pp_+=0~,\quad \vd_\pp\l^\pp_+=0~. }
Upon substituting the first of these back into the Lagrangian density, both
$L^\pp_\pp$ and $\ell^\pp$ disappear from it, indicating a gauge symmetry.
 As for the fermions, {\it on shell\/}, $\bl^\pp_+,\l^\pp_+=\hbox{\it
const.}$ (Recall that $\vd_\mm\L^\pp=0$.)

More importantly, however, the boson $\ell^\pp$ has a well-defined
canonically conjugate ($L^\pp_\pp$-dependent) momentum, and $\l^\pp_+$ and
$\bl^\pp_+$ are canonically conjugate to each other. Similarly, the last
part~\eToy{c} dictates that $\r^\mm_-$ and $\vr^\mm_-$ are canonically
conjugate to each other.

Thus, we can define {\it sixty-four\/} different (Dirac) vacua $\ket0$,
annihilated by a choice of six fermions:
\eqn\eXXX{
 \bigg\{\matrix{\j_-\cr\jb_-\cr}\bigg\},
 \bigg\{\matrix{\j_+\cr\jb_+\cr}\bigg\},
 \bigg\{\matrix{\x_-\cr\xb_-\cr}\bigg\},
 \bigg\{\matrix{\c_+\cr\cb_+\cr}\bigg\},
 \bigg\{\matrix{\l^\pp_+\cr\bl^\pp_+\cr}\bigg\},
 \bigg\{\matrix{\r^\mm_-\cr\vr^\mm_-\cr}\bigg\}~, }
where each braced pair indicates a free choice.
 Building upon these vacua, there are sixty-four (super)sectors of the
Hilbert space---vastly generalizing the $(c,c)$, $(a,c)$, $(c,a)$ and
$(a,a)$ sectors familiar from Landau-Ginzburg orbifolds, but present in any
(2,2)-supersymmetric theory.

\subsec{Haploid correlation functions}\noindent
In a model with the Lagrangian density~\eMGL{}, it is easy to define
correlation functions for the various terms in~\eMGL{}. For example, let us
denote by $\d_iW$ the $i^{th}$ term that can be added to the superpotential
$W$ in~\eMGL{b}. This means that each $\d_iW$ is a chiral superfield, and
so satisfies Eqs.~\eHSF{a}. Then, the correlation function
\eqn\eXXX{ \V{0|\,\d_1W\cdots\d_nW\,|0} }
is also called chiral, since the product $\d_1W{\cdots}\d_nW$ is chiral.
Similarly, a twisted-chiral correlation function is the correlation
function of a (product of) twisted-chiral object(s), a lefton correlation
function of a (product of) lefton object(s), \etc

The study of correlation functions simplifies on noting that in
theories with unbroken supersymmetry, the vacua must be invariant under
supersymmetry, and so annihilated by the supercharges $Q_\pm,\Qb_\pm$.
This induces a number of `topological' results~\refs{\rCec,\rCeGiPa,\rNSEW},
which will herein be refered to as `rigidity', not unrelated to the
mathematically standard notion of `rigidity', and to avoid confusion with
the mathematically standard notion of `topological'.

The `rigidity' results of the work on chiral correlation functions
in Wess-Zumino models~\refs{\rNSVZ,\rAKMRV,\rCec,\rCeGiPa} has a direct
analogue for correlation functions of ambixterous haploid (and all
quartoid) superfield operators. Given any two chiral superfields,
$\F(\s^\pp_1,\s^\mm_1)$ and $\F(\s^\pp_2,\s^\mm_2)$, which depend on two
different points on the world-sheet, $\s^\pp_i,\s^\mm_i$ for $i=1,2$, the
basic result is as follows:
\eqna\eRgd
 $$\eqalign{\pd{}{\s_1^\mm}\V{0|\F_1\F_2|0}
 &=\inv{2i}\V{0|\{\Qb_-,(Q_-\F_1)\}\F_2|0}\cr
 &=\inv{2i}\underbrace{\bra0\Qb_-}_{=0}(Q_-\F_1)\F_2\ket0
  +\inv{2i}\bra0(Q_-\F_1)\F_2\underbrace{\Qb_-\ket0}_{=0}~=~0~,\cr }
 \eqno\eRgd{a}
 $$
since $[\Qb_\pm,\F_i]=0$, and
 $$ \pd{}{\s_1^\pp}\V{0|\F_1\F_2|0}
 =\inv{2i}\V{0|\{\Qb_+,(Q_+\F_1)\}\F_2|0}~=~0~, \eqno\eRgd{b}
 $$
in a similar fashion. The same, upon $Q_\pm{\iff}\Qb_\pm$, is of
course true for $\V{0|\FB_1\FB_2|0}$.
 
By the same token,
 $$\eqalignno{
 \pd{}{\s_1^\mm}\V{0|\X_1\X_2|0}
 &=\inv{2i}\V{0|\{\Qb_-,(Q_-\X_1)\}\X_2|0}~=~0~, &\eRgd{c}\cr
 \pd{}{\s_1^\pp}\V{0|\X_1\X_2|0}
 &=\inv{2i}\V{0|\{Q_+,(\Qb_+\X_1)\}\X_2|0}~=~0~, &\eRgd{d}\cr }$$
and the same for $\V{0|\XB_1\XB_2|0}$.
 The same rigidity will hold for the quartoid superfields~\eQSF{}, as they
all satisfy the superconstraints of some ambidexterous haploid superfield.
This can be remembered easily by regarding $\F_\LL$ and $\F_\RR$ as two
`halves' of $\F$, and $\FB_\LL$ and $\FB_\RR$ as two `halves' of $\FB$.

The situation is quite different for the unidexterous fields:
 $$ \pd{}{\s_1^\mm}\V{0|\L_1\L_2|0}
 =\inv{2i}\V{0|\{\Qb_-,(\underbrace{Q_-\L_1}_{=0})\}\L_2|0}~=~0~,
 \eqno\eRgd{b}
 $$
holds again, but now
 $$\pd{}{\s_1^\pp}\V{0|\L_1\L_2|0}
 =\inv{2i}\V{0|\{\Qb_+,(Q_+\L_1)\}\L_2|0}
 =\inv{2i}\bra0(Q_+\L_1)(\Qb_-\L_2)\ket0~\neq~0~. \eqno\eRgd{f}
 $$
Similarly,
 $$\eqalignno{
 \pd{}{\s_1^\mm}\V{0|\Y_1\Y_2|0}
 &=\inv{2i}\V{0|\{\Qb_-,(Q_-\Y_1)\}\Y_2|0}~\neq~0~, &\eRgd{h}\cr
 \pd{}{\s_1^\pp}\V{0|\Y_1\Y_2|0}
 &=\inv{2i}\V{0|\{Q_+,(\Qb_+\Y_1)\}\Y_2|0}~=~0~. &\eRgd{h}\cr }$$

That is:
\item{1.} Corellation functions of (products of) any one type of
          ambidexterous haploid superfields~\eHSF{a{-}d} (and any quartoid
          superfields~\eQSF{} that obey the same superconstraint) are
          independent of the superfields' positions on the world-sheet. 
\item{2.} Corellation functions of (products of) unidexterous haploid
          superfields~\eHSF{e,f} are correspondingly unidexterous functions
          of the superfields' positions on the world-sheet.

\noindent It should be clear that only the constrained superfields defined
by the simple superconstraints~\eHSF{} and~\eQSF{}, which are linear in both
superderivatives and superfields, posess this remarkable property of
`rigidity'.

\subsec{Target space geometry}\noindent
The field (target) space for our model is spanned by the physical component
fields in the collection of superfields $\F,\FB,\X,\XB,\L,\Y$.
 The entire Lagrangian density~\eMGL{} depends on a grand total of
3 rank-0, 24 rank-1, 32 rank-2, 12 rank-3 and 6 rank-4 (tensor-like)
`coefficient functions' (and their complex conjugates) defined in~\eThW{},
\eThS{}, \eThN{}, \eThQ{} and~\eLD.
 Even if the unidexterous superfields, $\L,\Y,\F_\LL,\F_\RR,\FB_\LL,\F_\RR$
and the non-minnimal superfields, $\Q,\QB,\P,\PB$, were to be dropped, the
Lagrangian density depends on 3 rank-0 and 4 rank-1 coefficient functions,
defined in~\eThW{}, \eThS{}, \eThQ{} and~\eLD:
\eqn\eNUD{{\cmath{ K(\f,\fb,x,\bx)~,\quad W(\f)~,\quad \S(x)~,\cr
 M_\nb(\f,x)~,\quad M_\bb(\f,x)~,\quad
 \Mt_\nb(\f,\bx)~,\quad \Mt_\a(\f,\bx)~,\cr  }}}
and the conjugates of all but $K$, which is real. Each term from
Eqs.~\eThW{}, \eThS{}, \eThQ{} and~\eLD\ featuring one of these functions
produces, upon the appropriate fermionc integration, a {\it separately\/}
and {\it manifestly supersymmetric\/} term in the Lagrangian~\eMGL{}.
Therefore, supersymmetry imposes no further condition on the
coefficient functions~\eNUD\ and their complement, as defined in~\eThW{},
\eThS{}, \eThQ{} and~\eLD.

Recall that the last four displayed coefficient functions in~\eNUD\
contribute both to a variation in the D-term (\ie, a change of $K$) and a
`topological' term such as~\eTpT. Four other terms,
\eqn\eXXX{\cmath{
\inv8[D_+,\Db_+]D_-\big[(D_-\F^\m)M^\LL_\m+(D_-\X^\a)M^\LL_\a\big]+\hc\cr
\hbox{and}\quad\inv8
[D_-,\Db_-]D_+\big[(D_+\F^\m)\Mt^\RR_\m+(D_+\X^\bb)\Mt^\RR_\bb\big]+\hc\cr
 }}
vanish on retaining only the chiral and twisted-chiral superfields: the
coefficient functions $M^\LL_\@,M^\RR_\@$ become constants, whereupon the
repeated superderivative annihilates the term.

Clearly, these rank-0\34, `coefficient functions' are rank-0\34 tensor-like
objects over the field space. However, it must be emphasized that these
{\it need not be global tensors\/} of such rank! Consider for example the
term $(\Db_-\F^\nb)M_\nb(\FB_\LL,\F,\X,\L)$ from Eq.~\eThQ{b}. The
(co)vector $M_\nb$ need not be a global (co)vector field over the space
coordinatized by the (lowest components of) $\F^\m,\F^\nb$. In particular,
as used in Eq.~\eExpl\ and below, the (co)vector $M_\nb$ may well be a
contraction
\eqn\eXXX{ \F^\m M_{\m\nb} + \X^\a M_{\a\nb} + \L^\ah M_{\ah\nb} + \3 }
Thus, the fact that in some application, the space coordinatized by the
(lowest components of) $\F^\m,\F^\nb$ does not admit a global 1-form
$M_\nb\rd\f^\nb$ does not preclude the inclusion in the Lagrangian
density of the first $\l$-term in Eq.~\eThQ{b}.

Furthermore, the coefficient functions in~\eThW{}, \eThS{}, \eThQ{}
and~\eLD\ all admit some degree of arbitrariness, generalizing the
well-known `K\"ahler symmetry', $K\simeq K+f(\F)+\bar{f}(\FB)$, where
$f(\F)$is an arbitrary function of the chiral superfields, and $\bar{f}$
its conjugate. In point of fact, the function $K$ in~\eLD\ is defined only
up to the very general reparametrization:
\eqn\eXXX{ K(\F,\FB,\X,\3)\simeq
           K + f_{(-)}+ f_{(+)} + \bar{f}_{(-)}+ \bar{f}_{(+)}~, }
where
\eqn\eXXX{ \Db_-f_{(-)}=0=\Db_+f_{(+)}~,\quad
            D_-\bar{f}_{(-)}=0=D_+\bar{f}_{(+)}~. }
That is, $f_{(\pm)}$ and $\bar{f}_{(\pm)}$ are {\it arbitrary\/} functions
of any of the (superderivative) superfields annihilated by any one of the
superderivatives; see Eqs.~\eKerD{}. For the other coefficient functions,
the reparametrizing functions (such as the $f$'s above) must be annihilated
by at least one more superderivative than the coefficient function that
they reparametrize. For example, the superpotential obeys
\eqn\eXXX{ W\simeq W + g_\LL + g_\RR~, }
where $g_\LL,g_\RR$ are a chiral lefton and a chiral righton function.
Clearly, $g_\LL,g_\RR$ are nonzero only in models that involve chiral
leftons and chiral rightons. 

 It is then patently clear that the geometry of the field (target) space of
general (2,2)-supersymmetric models is considerably more general than just
K\"ahler manifolds with an assortiment of (holomorphic) vector bundles (or
even sheaves). As it was known since Ref.~\rGHR, even a model depending on
just the first one of the seven functions~\eNUD\ provides more general
geometries, further augmented by the two `superpotentials', $W(\f)$ and
$\S(\f)$. The continued extension of the geometrical structure relying on
the inclusion of rank-2 coefficient functions $W_{\a\bb}$ and $\S_{\m\nb}$
was noted in Ref.~\rGGW. These appear through the F-terms
\eqn\eXXX{ \inv2\int\rd^2\vs~(\Db_+\X^\a)(\Db_-\X^\bb)W_{\a\bb}(\F)~,
 \quad\hbox{and}\quad
 \inv2\int\rd^2\sT~(D_+\F^\m)(\Db_-\F^\nb)\S_{\m\nb}(\X)~,}
which are in fact special cases of the first two $\l$-terms~\eThQ{b}.
 To the best of my knowledge, this is the first time the four 1-forms
in~\eNUD\ have made their public appearance. Of course, if one insists on
restricting to only chiral, twisted-chiral superfields and their
conjugates, the seven objects~\eNUD\ admittedly exhausts the list (subject
to our general restrictions), with the latter four `merely' modifying the
D-terms and contributing `topological' terms such as~\eTpT.

Note that, except for $K$ which is real, all the coefficient functions are
analytic of their arguments, and so holomorphic functions of their complex
arguments. The Lagrangian terms in which they appear are not
$\int\rd^4\vs$-integrated terms, whence they fall under the scope of the
non-renormalization theorems~\refs{\rWB,\rGGRS}. More precisely, their
contribution to the D-terms is subject to renormalization, but the
complement of these contributions is not. That is, the Lagrangian terms
involving all but the first of the functions~\eNUD\ (and all those omitted
in~\eNUD!) are expected to provide marginal operators (and so deformations)
in a quantum theory.

\newsec{Summary, Outlook and Conclusions}\noindent
In an attempt to provide a comprehensive but comprehensible intrinsic study
of (2,2)-supersymmetric models in 2-dimensional spacetime, we purposefully
avoid constructions based on dimensional reduction from 4-dimensions.

A self-contained discussion of types of constrained superfields was given
in \SS\,\sHSF. These include {\it haploid\/} ({\it quartoid\/}) superfields,
which depend on a half (quarter) of the fermionic coordinates. The
`minimal' haploid superfields are defined by a simple pair of first order
superconstraints~\eHSF{}, while their non-minimal brethern are defined by
one simple second order superconstraint~\eNMF{}. Quartoid superfields are
defined by a simple triplet of first order superconstraints~\eQSF{}. All
quartoid superfields~\eQSF{}, and two of six `minimal' haploid
superfields, those defined in~\eHSF{e,f}, are unidexterous (depend only on
one world-sheet light-cone coordinate). Numerous other types of constrained
superfields are discussed, albeit not as systematically, in \SS\,\ssOthr.

The possible choices for a Lagrangian density for the haploid and quartoid
superfields are discussed in \SS\,\sLag. The otherwise infinite number of
choices is curbed by demanding that the resulting equations of motion
contain no higher than second derivatives on bosons, and no higher than
first derivatives on fermions, and that no coefficient function turns out
to have negative canonical dimension. This leads to the general form of the
Lagrangian density~\eSGL{}, with terms supplied in~\eLD, \eThQ{}, \eThW{},
\eThS{} and~\eThN{}. The exhaustive construction and verification of
supersymmetry of these Lagrangian density terms was facilitated by creating
lists of (superderivative) superfields which are annihilated by some
subset of the four superderivatives~\eDSF{} and~\eKerD{}. \SS\,\sExCn\
discusses some issues of quantization of constrained superfields.

Upon choosing a preferred Lagrangian density, one can turn to a study of
the Hilbert space, the structure of the supersymmetric wave-functions,
correlation functions, and the relations to the geometry of the target
space. This is illustrated in \SS\,\sWFnG, where some of the basic results
are recalled and generalized.

\subsec{Further topics}\noindent
{\it En route\/}, a nuber of issues have been deferred to a later study,
and we recall some of these, for the sake of the Reader. 

\topic{Transformations and `dualities'}
In \SS\,\ssSuSpSym, the group of discrete transformations of the `soul' of
the (2,2)-superspace (generated by the four fermionic coordinates) has been
studied. Two of these transformations have been identified as the root of
two of the dualities in string theory:
\item{1.} left-handed conjugation, $\CBp$, as being the root of mirror
          symmetry, and
\item{2.} unconjugate parity, \BM{p}, as being the root of the
          Type~IIA$\iff$Type~IIB duality~\rJoeD.
 \par\noindent
 Rather conspicuously, mirror symmetry is an actual symmetry of the total
Hilbert space, while the Type~IIA$\iff$Type~IIB duality is not. At the same
time, $\CBp$ commutes with Lorentz symmetry while \BM{p} does not; more
importantly, $\CBp$ is a symmetry of the supersymmetry algebra~\eSusyD,
while \BM{p} is not. It is tempting to conjecture that this latter property
of \BM{p} makes it the root of a duality between two {\it different\/}
string theories, rather than a symmetry of a single (super)string theory, as
is the case with $\CBp$ and mirror symmetry.
 In any case, these operations are but two elements of a discrete group,
denoted in \SS\,\ssSuSpSym\ as $2S_4$. The remaining 45 non-trivial
operations may similarly generate additional `duality' (`triality', and
even `quadriality') relations among (super)string (and related) theories,
and are well worth exploring. In particular, $S_4$ is generated by one more
independent simple swap. This third simple transformation (involution, \ie,
self-inverse) should at least be of interest in (super)string theory.
 As noted, there is however a continuous group of transformations,
$GL(4,\IC)$, at least a subgroup of which could provide further insight
into (super)string (and related) dynamics. This too seems to warrant a
closer study.

\topic{Dimensional reduction}
Supersymmetric quantum mechanics provides the underlying framework for
understanding the general features of any supersymmetric field
theory~\rSuSyM. The present analysis of (2,2)-supersymmetric models in
2-dimensional spacetime can of course be dimensionally reduced to
(2,2)-supersymmetric quantum mechanics. Since the Lorentz symmetry in
2-dimensional spacetimes is already abelian, it is not clear if any
generality would be lost in this way. That is, it remains unclear whether a
similarly intrinsic ({\it ab initio\/}) and comprehensive study of
supersymmetric quantum mechanic would allow for possibilities not obtained
by dimensional reduction from the results presented and indicated herein.
The study in Refs.~\rAleph\ deos seem to suggest this, in fact,
$\aleph_0$-fold so.

 On the other hand, with no continuous Lorentz symmetry in supersymmetric
quantum mechanics, the notion of chirality is now lost. Then, the spin
labels ($\pm$ sub- and superscripts) as used in this article simply count
the extended supersymmetries. This change has to be taken into account when
translating the present results to various applications of N$\geq$4
supersymmetric quantum mechanics.

\topic{Quantization}
Some of the issues of quantization of constrained superfields have been
mentioned in \SS\,\sExCn. A complete and detailed Hamiltonian (Dirac, BRST,
\etc) treatment would seem to be well worth, especially in view of the
apparent difficulties with unidexterous (super)fields; see above, and
Ref.~\rCBAbWo\ for a recent account and further references. Suffice it here
to note that both general approaches outlined in \SS\,\sExCn\ have both
advantages and drawbacks. While the use of unconstrained superfields
allowes for a more straightforward quantization, it also requires many more
component fields and a complementing set of gauge symmetries (to gauge away
the extra components introduced by passing to unconstrained superfields).

On the other hand, quantization of constrained superfields requires a
detailed study of the full content of the constraints, and a proper
enforcement by means of suitable Lagrange multipliers. As the component
fields of the Lagrange multiplier superfields ultimately must disappear
from the action, the proper Siegel ($\k$-) symmetry must be identified and
ensured. 

\topic{Gauge symmetries}
As noted from the outset, gauge symmetry issues (in a comparably
comprehensive setting) are completely ignored herein, and are deferred to a
later effort. Partial results may be found in Refs.~\refs{\rTwJim,\rGGW}.

However, quite unrelated to usual gauge symmetries, Eqs.~\eSgg{} present
a rather peculiar type of inhomogeneous `gauge' transformations, where the
supercoordinates themselves serve as generators.

These are perhas not unrelated to the relation:
\eqn\eXXX{{\eqalign{
 &\{\F:\Db_\pm\F{=}0\}\simeq
 \{\0\F\simeq\0\F+\vsb^+\0\F^{(1)}_++\vsb^-\0\F^{(1)}_-\}~,\cr
 &\hbox{where}~~
   \0\F^{(1)}_+\simeq\0\F^{(1)}_++\vsb^-\0\F^{(2)}~,\quad
   \0\F^{(1)}_-\simeq\0\F^{(1)}_-+\vsb^+\0\F^{(2)}~.\cr }}}
Here, as throughout the article, the circle atop indicates that the
superfield is unconstrained. The supefields $\F^{(i)}$ may be regarded as
$i^{th}$ order gauge parameters. That is, component fields of the originally
unconstrained superfield $\0\F$ are being elliminated by means of the (1st
order) gauge transformation
\eqna\eSGg
 $$
   \d\0\F = \vsb^+\0\F^{(1)}_++\vsb^-\0\F^{(1)}_-~. \eqno\eSGg{a}
 $$
This, however, would attempt to gauge away some of the component fields
twice (recall the doubly shaded region in Figs.~4 and~6), and so the gauge
parameters, $\0\F^{(1)}_\pm$, are themselves subject to a (2nd order) gauge
transformation,
 $$
   \d\0\F^{(1)}_-=\vsb^+\0\F^{(2)}~,\quad\hbox{and}\quad
           \d\0\F^{(1)}_+=\vsb^-\0\F^{(2)}~, \eqno\eSGg{b}
 $$
with respect to which the first order gauge transformation~\eSGg{a} is
invariant. The component fields remaining in this sequential double quotient
are the same as those in a chiral superfield. This sequential double
quotient representation of the chiral superfield is obviously rather more
involved technically. However, it also gives the chiral superfield the
r\^ole of a gauge field (having an inhomogeneous transformation), which is
certainly not the usual case. The non-minimal chiral superfield also admits
a similar description:
\eqn\eXXX{ \{\Q:\Db_+\Db_-\Q{=}0\}\simeq
            \{\0\Q\simeq\0\Q+\vsb^-\vsb^+\0\Q^{(1)}\}~, }
which is a simple quotient by the inhomogeneous transformation
\eqn\eSGG{ \d\0\Q = \vsb^-\vsb^+\0\Q^{(1)}~, }
as compared to the sequential double quotient, with respect to the
cascading pair of inhomogeneous transformations~\eSGg{}, of the `minimal'
chiral superfield. Of course, the other haploid superfields, `minimal' and
non-minimal, admit similar quotient representations. Quite generally, the
constrained and the quotient representations are `dual' in a well-defined
sense: the former represents the kernel, the latter the cokernel of a
mapping.

The utility of this quotient representation as well as the duality between
the quotient and the constrained representations remains unclear, and will
require a study in its own right. It is however clear that it will be
linked to the possible r\^oles of the `peculiar' inhomogeneous
transformations, such as~\eSgg{}, \eSGg{} and~\eSGG. Note also that the
quotient description has a direct implication in a BRST descrition:
Eq.~\eSGG\ induces a single set of ghost superfields, whereas
Eqs.~\eSGg{a,b} induce a set of ghost superfields, and a set of
ghost-for-ghost superfields, respectively.

\topic{Other superconstraints}
Finally, as shown in \SS\,\ssOthr, the simple superconstraints~\eHSF{},
\eQSF{} and~\eNMF{} barely skratch the surface of the incredibly rich
category of superconstraint theory. While the abelian nature of Lorentz
symmetry in $\leq$2-dimensional spacetime is the major reason of the much
larger number of possibilities for defining superconstraints, several of
the truly novel superconstraints of \SS\,\ssOthr\ are equally well possible
in $>$2-dimensional spacetimes! Unfortunately, no natural or systematic way
of enlisting and studying these non-simple, non-homogeneous or non-linear
superconstraints seems to present itself, and it is not even clear that any
{\it finite\/} form of enlisting is even possible or whether such efforts
are worth the trouble.

Clearly, however, some of these do deserve a closer study, such as the
family of superconstraints~\eCTC, which interpolates continuously between
chiral and twisted-chiral superfields, and hence between models the IR
regime of which is believed to correspond to mirror compactifications in
string theory. A similar family interpolating between \BM{p}-reflected
models does not seem to be possible in 2-dimensions, as \BM{p} does not
commute with Lorentz symmetry. Even in supersymmetric quantum mechanics, a
construction of such an interpolation is obstructed because \BM{p} does not
leave the superalgebra~\eSusyD\ invariant. This seems to further deepen the
chasm between mirror symmetry, as generated by $\CBp$, and the
Type~IIA$\iff$Type~IIB duality, as generated by \BM{p}.
 These and related issues are left for a subsequent study.

 %
\appendix{A}{Notational Nitpicking}\noindent 
Following Wess and Bagger's definitions~\rWB:
\eqn\eSpRL{
 \j^\a\define\e^{\a\b}\j_\b~,\quad
 \jb^\ad\define\e^{\ad\bd}\jb_\bd~,\quad
 \j_\a=\e_{\a\b}\j^\b~,\quad
 \jb_\ad=\e_{\ad\bd}\jb^\bd~, }
and
\eqn\eXXX{ \j{\cdot}\c\define\j^\a\c_\a~,\quad\hbox{but}\quad
           \jb{\cdot}\cb\define\jb_\ad\cb^\ad~. }
It follows that
\eqna\eXXX
 $$ \twoeqsalignno{
 \vs^2  &\define +\e_{\a\b}\vs^\a\vs^\b~=~2\vs^+\vs^-~, &\qquad
 D^2    &\define \e^{\a\b}D_\b D_\a~=~2D_+D_-~,  &\eXXX{a}\cr
 \noalign{\vskip-3mm\noindent but\vskip-2mm}
 \vsb^2 &\define +\e_{\ad\bd}\vsb^\bd\vsb^\ad~=~2\vsb^-\vsb^+~, &\quad
 \Db^2  &\define \e^{\ad\bd}\Db_\ad \Db_\bd~=~2\Db_-\Db_+~.  &\eXXX{b}\cr
 }$$
Therefore (recall Eqs.~\eDs):
\eqn\eXXX{ D_\a\vs^\b=\d^\b_\a~,\quad D^2\,\vs^2 ~=~-4~,\qquad
 \Db_\ad\vsb^\bd=-\d^\bd_\ad~,\quad \Db^2\,\vsb^2~=~-4~. }
Notice that Eqs.~\eSpRL\ imply that
\eqn\eXXX{ \j^-=\j_+~,\quad \j^+=-\j_-~,\qquad
           \bar\j^-=\bar\j_+~,\quad \bar\j^+=-\bar\j_-~, }
We say that $\j^-=\j_+$ (and their conjugates) have spin $-\inv2$, and
$\j^+=-\j_-$ (and their conjugates) have spin $+\inv2$.

Fermionic Dirac delta-functions are utterly simple: \eg,
$\d(\vs^+)\id\vs^+$. For multiple delta-functions, we must merely choose an
order. Fixing $\d^4(\vs)\define\vsb^-\vsb^+\vs^+\vs^-$ so that
$\int\rd^4\vs\>\d^4(\vs)=1$, we find:
\eqn\eXXX{ \d^2(\vs)=\vs^+\vs^-~,\quad
            \d^2(\sT)=\vsb^+\vs^-~,\quad
             \d^2(\sR)=\vsb^-\vs^-~, }
\eqn\eXXX{ \d^2(\vsb)=\vsb^-\vsb^+~,\quad
            \d^2(\sW)=\vsb^-\vs^+,\quad
             \d^2(\sL)=\vsb^+\vs^+~.}

The following {\it operatorial\/} identities may be of help:
\eqn\eXXX{ [D_+,\Db_+]\id2(D_+\Db_+ -i\vd_\pp)~,\qquad
           [D_-,\Db_-]\id2(D_-\Db_- -i\vd_\mm)~, }
whence
\eqn\eXXX{{\eqalign{ [D_-,\Db_-][D_+,\Db_+]
 &\id2[D_-,\Db_-](D_+\Db_+ -i\vd_\pp)~,\cr
 &\id2[D_-,\Db_-]D_+\Db_+ -4i\vd_\pp D_-\Db_- -4\vd_\pp\vd_\mm~,\cr
 &\id2[D_+,\Db_+](D_-\Db_- -i\vd_\mm)~,\cr
 &\id2[D_+,\Db_+]D_-\Db_- -4i\vd_\mm D_+\Db_+ -4\vd_\pp\vd_\mm~.\cr }}}
Also,
\eqn\eXXX{{\cmath{[D_-,\Db_-]D_-\id2i\vd_\mm D_-~,\qquad
                  [D_-,\Db_-]\Db_-\id-2i\vd_\mm\Db_-~,\cr
                  [D_+,\Db_+]D_+\id2i\vd_\pp D_+~,\qquad
                  [D_+,\Db_+]\Db_+\id-2i\vd_\pp\Db_+~,\cr }}}
and
\eqn\eXXX{ \big([D_1,D_2]AB\big) =\big([D_1,D_2]A\big)B
           + A\big([D_1,D_2]B\big) + (-)^A 2(D_{[1}A)(D_{2]}B)~, }
for any two superderivatives $D_1,D_2$ and any two superfields $A,B$, and
where
\eqn\eXXX{(-)^A\define\cases{+1\cr-1\cr}\quad\hbox{if $A$ is }
                       \cases{\hbox{commuting,}\cr\hbox{anticommuting.}\cr}}

\appendix{B}{Expounding Expansions}\noindent 
For the benefit of the Reader unskilled in projecting onto component
fields by means of the superderivatives in \SS\,\ssCFlds, basic
definitions and some simple consequences are listed here.

\subsec{Component fields}\noindent
\topic{Chiral superfield: $\Db_\pm\F=0$}
$\F|=\f$, $D_\pm\F|=\sqrt2\j_\pm$, $D_-D_+\F|=2F$,\\
$[D_-,\Db_-]\F|=-\Db_-D_-\F|=-2i(\vd_\mm\f)$,\\
$[D_+,\Db_+]\F|=-\Db_+D_+\F|=-2i(\vd_\pp\f)$,\\
$[D_-,\Db_-]D_+\F|=-\Db_-D_-D_+\F|=-2i(\vd_\mm\j_+)$,\\
$[D_+,\Db_+]D_-\F|=-\Db_+D_+D_-\F|=-2i(\vd_\pp\j_-)$,\\
$[D_-,\Db_-][D_+,\Db_+]\F|=-[D_-,\Db_-]\Db_+D_+\F|
 =-[D_+,\Db_+]\Db_-D_-\F=-4(\vd_\pp\vd_\mm\f)$.

\topic{Antichiral superfield: $D_\pm\FB=0$}
$\FB|=\f$, $\Db_\pm\FB|=\sqrt2\jb_\pm$, $\Db_+\Db_-\FB|=2\Fb$,\\
$[D_-,\Db_-]\FB|=D_-\Db_-\FB|=2i(\vd_\mm\fb)$,\\
$[D_+,\Db_+]\FB|=D_+\Db_+\FB|=2i(\vd_\pp\fb)$,\\
$[D_-,\Db_-]\Db_+\FB|=D_-\Db_-\Db_+\FB|=2i(\vd_\mm\jb_+)$,\\
$[D_+,\Db_+]\Db_-\FB|=D_+\Db_+\Db_-\FB|=2i(\vd_\pp\jb_-)$,\\
$[D_-,\Db_-][D_+,\Db_+]\FB|=[D_-,\Db_-]D_+\Db_+\FB|
 =[D_+,\Db_+]D_-\Db_-\FB=-4(\vd_\pp\vd_\mm\fb)$.

\topic{Twisted-chiral superfield: $\Db_-\X=0=D_+\X$}
$\X|=x$, $D_-\X|=\sqrt2\x_-$, $\Db_+\X|=\sqrt2\c_+$, $D_-\Db_+\X|=2X$,\\
$[D_-,\Db_-]\X|=-\Db_-D_-\X|=-2i(\vd_\mm x)$,\\
$[D_+,\Db_+]\X|=D_+\Db_+\X|=2i(\vd_\pp x)$,\\
$[D_-,\Db_-]\Db_+\X|=-\Db_-D_-\Db_+\X|=-2i(\vd_\mm\c_+)$,\\
$[D_+,\Db_+]D_-\X|=D_+\Db_+D_-\X|=2i(\vd_\pp\x_-)$,\\
$[D_-,\Db_-][D_+,\Db_+]\X|=[D_-,\Db_-]D_+\Db_+\X|
 =-[D_+,\Db_+]\Db_-D_-\X=4(\vd_\pp\vd_\mm\f)$.

\topic{Twisted-antichiral superfield: $D_-\XB=0=\Db_+\XB$}
$\XB|=\f$, $\Db_-\XB|=\sqrt2\xb_-$, $D_+\XB|=\sqrt2\cb_+$,
$D_+\Db_-\XB|=2\Xb$,\\
$[D_-,\Db_-]\XB|=D_-\Db_-\XB|=2i(\vd_\mm\bx)$,\\
$[D_+,\Db_+]\XB|=-\Db_+D_+\XB|=-2i(\vd_\pp\bx)$,\\
$[D_-,\Db_-]D_+\XB|=D_-\Db_-D_+\XB|=2i(\vd_\mm\cb_+)$,\\
$[D_+,\Db_+]\Db_-\XB|=-\Db_+D_+\Db_-\XB|=-2i(\vd_\pp\xb_-)$,\\
$[D_-,\Db_-][D_+,\Db_+]\XB|=-[D_-,\Db_-]\Db_+D_+\XB|
 =[D_+,\Db_+]D_-\Db_-\XB=4(\vd_\pp\vd_\mm\bx)$.

\topic{Lefton superfield: $D_-\L=0=\Db_-\L$}
$\L|=\ell$, $D_+\L|=\sqrt2\l_+$, $\Db_+\L|=\sqrt2\bl_+$,\\
$[D_+,\Db_+]\L|=4L_\pp$, $D_+\Db_+\L|=i(\vd_\pp\ell)+2L_\pp$,
$\Db_+D_+\L|=i(\vd_\pp\ell)-2L_\pp$.

\topic{Righton superfield: $D_+\Y=0=\Db_+\Y$}
$\Y|=x$, $D_-\Y|=\sqrt2\x_-$, $\Db_+\Y|=\sqrt2\c_+$,\\
$[D_-,\Db_-]\Y|=4R_\mm$, $D_-\Db_-\Y|=i(\vd_\mm r)+2R_\mm$,
$\Db_-D_-\Y|=i(\vd_\mm r)-2R_\mm$.

\topic{Non-minimal chiral superfield: $\Db_+\Db_-\Q=0$}
$\Q|=t$, $D_\pm\Q|=\sqrt2\q_\pm$, $D_-D_+\Q|=2T$,\\
$[D_-,\Db_-]\Q|=4T_\mm$, $D_-\Db_-\Q|=i(\vd_\mm t)+2T_\mm$,
$D_-\Db_+\Q|=2T_\mp$,\\
$[D_+,\Db_+]\Q|=4T_\pp$, $D_+\Db_+\Q|=i(\vd_\pp t)+2T_\pp$,
$D_+\Db_-\Q|=2T_\pm$,\\
$[D_-,\Db_-]D_+\Q|=4\sqrt2\t_-$,
$[D_-,\Db_-]\Db_+\Q|=-2\sqrt2i(\vd_\mm\vq_+)$,\\
$[D_+,\Db_+]D_-\Q|=4\sqrt2\t_+$,
$[D_+,\Db_+]\Db_-\Q|=-2\sqrt2i(\vd_\pp\vq_-)$,\\
$[D_-,\Db_-][D_+,\Db_+]\Q|=4[(\vd_\mm\vd_\pp t)-2i(\vd_\mm T_\pp)
                                               -2i(\vd_\pp T_\mm)]$.

\topic{Non-minimal antichiral superfield: $D_-D_+\QB=0$}
$\QB|=\bt$, $D_\pm\QB|=\sqrt2\qb_\pm$, $\Db_+\Db_-\QB|=2\Tb$,\\
$[D_-,\Db_-]\QB|=4\Tb_\mm$, $D_-\Db_-\QB|=i(\vd_\mm\bt)+2\Tb_\mm$,
$D_-\Db_+\QB|=2\Tb_\mp$,\\
$[D_+,\Db_+]\QB|=4\Tb_\pp$, $D_+\Db_+\QB|=i(\vd_\pp\tb)+2\Tb_\pp$,
$D_+\Db_-\QB|=2\Tb_\pm$,\\
$[D_-,\Db_-]\Db_+\QB|=4\sqrt2\tb_-$,
$[D_-,\Db_-]D_+\QB|=2\sqrt2i(\vd_\mm\bar\vq_+)$,\\
$[D_+,\Db_+]\Db_-\QB|=4\sqrt2\tb_+$,
$[D_+,\Db_+]D_-\QB|=2\sqrt2i(\vd_\pp\bar\vq_-)$,\\
$[D_-,\Db_-][D_+,\Db_+]\QB|=4[(\vd_\mm\vd_\pp\tb)+2i(\vd_\mm\Tb_\pp)
                                                +2i(\vd_\pp\Tb_\mm)]$.

\topic{Non-minimal twisted-chiral superfield: $D_+\Db_-\P=0$}
$\P|=p$, $D_\pm\P|=\sqrt2\p_\pm$, $\Db_\pm\P|=\sqrt2\vp_\pm$,
$D_-D_+\P|=2P$,\\
$[D_-,\Db_-]\P|=4P_\mm$, $D_-\Db_-\P|=i(\vd_\mm p)+2P_\mm$,
$D_-\Db_+\P|=2P_\mp$,\\
$[D_+,\Db_+]\P|=4P_\pp$, $D_+\Db_+\P|=i(\vd_\pp p)+2P_\pp$,
$\Db_+\Db_-\P|=2\FF{P}$,\\
$[D_-,\Db_-]D_+\P|=-2\sqrt2i(\vd_\mm\p_+)$,
$[D_-,\Db_-]\Db_+\P|=4\sqrt2\Tw\vf_-$,\\
$[D_+,\Db_+]D_-\P|=4\sqrt2\vf_+$,
$[D_+,\Db_+]\Db_-\Q|=2\sqrt2i(\vd_\pp\vp_-)$,\\
$[D_-,\Db_-][D_+,\Db_+]\P|=-4[(\vd_\mm\vd_\pp p)-2i(\vd_\mm P_\pp)
                                                +2i(\vd_\pp P_\mm)]$.

\topic{Non-minimal twisted-antichiral superfield: $D_-\Db_+\PB=0$}
$\PB|=\bp$, $D_\pm\PB|=\sqrt2\pb_\pm$, $\Db_\pm\PB|=\sqrt2\ba\vp_\pm$,
$D_-D_+\PB|=2\Pb$,\\
$[D_-,\Db_-]\PB|=4\Pb_\mm$, $D_-\Db_-\PB|=i(\vd_\mm\bp)+2\Pb_\mm$,
$D_-\Db_+\PB|=2\Pb_\mp$,\\
$[D_+,\Db_+]\PB|=4\Pb_\pp$, $D_+\Db_+\PB|=i(\vd_\pp\bp)+2\Pb_\pp$,
$\Db_+\Db_-\PB|=2\skew{-4}\bar{\FF{P}}$,\\
$[D_-,\Db_-]D_+\PB|=4\sqrt2\skew3\Tw{\bar\vf}_-$,
$[D_-,\Db_-]\Db_+\PB|=2\sqrt2i(\vd_\mm\pb_+)$,\\
$[D_+,\Db_+]D_-\PB|=-2\sqrt2i(\vd_\pp\ba\vp_-)$,
$[D_+,\Db_+]\Db_-\PB|=4\sqrt2\bar\vf_+$,\\
$[D_-,\Db_-][D_+,\Db_+]\PB|=-4[(\vd_\mm\vd_\pp\bp)-2i(\vd_\mm\Pb_\pp)
                                                  +2i(\vd_\pp\Pb_\mm)]$.

\subsec{Projection technique}\noindent
As an example of the projection technique, below is the detailed expansion
of the $\l$-term $\inv8[D_+,\Db_+]D_-(\Db_-\F^\nb)M_\nb|$, which has been
discussed in \SS\,\ssL. The uninitiated Reader should be able to follow
each step and then apply this technique for any of the other parts of the
Lagrangian density~\eSGL{}.
\eqna\eXXX
 $$\eqalignno{ \inv8[D_+,\Db_+]D_-(\Db_-\F^\nb)M_\nb\Zp\[{20}\cr
 &=\inv8[D_+,\Db_+]
    \big[(2i\vd_\mm\F^\nb)M_\nb
        +(D_-\F^\m)(\Db_-\F^\nb)M_{\nb\1\m}\big]\Zp~,           &\eXXX{a}\cr
 &=\inv8\Big\{
      (2i\vd_\mm[D_+,\Db_+]\F^\nb)M_\nb
     +(2i\vd_\mm\F^\nb)([D_+,\Db_+]M_\nb)\cr
 &\]4+2(2i\vd_\mm\2{D_+\F^\nb})(\Db_+M_\nb)
     -2(2i\vd_\mm\Db_+\F^\nb)(D_+M_\nb)\cr
 &\]4+\big[[D_+,\Db_+](D_-\F^\m)(\Db_-\F^\nb)\big]M_{\nb\1\m}
     +(D_-\F^\m)(\Db_-\F^\nb)([D_+,\Db_+]M_{\nb\1\m})\cr
 &\]4+2\big[D_+(D_-\F^\m)(\Db_-\F^\nb)\big](\Db_+M_{\nb\1\m})
     -2\big[\Db_+(D_-\F^\m)(\Db_-\F^\nb)\big](D_+M_{\nb\1\m})
                                              \Big\}\Big|~,~~~~ &\eXXX{b}\cr
 &=\inv8\Big\{
      \big(2i\vd_\mm(2i\vd_\pp\F^\nb)\big)M_\nb\cr
 &\]4+(2i\vd_\mm\F^\nb)
      \big[[D_+\Db_+(\FB_\LL+\X)-(\Db_+D_+\F)+[D_+,\Db_+]\L]M'_\nb\cr
 &\]{13}+[D_+(\F+\L),\Db_+(\FB_\LL+\X+\L)]M''_\nb\big]\cr
 &\]4-4i(\vd_\mm\Db_+\F^\nb)[D_+(\F+\L)]M'_\nb\cr
 &\]4+\big(([D_+,\Db_+]D_-\F^\m)(\Db_-\F^\nb)
     +(D_-\F^\m)([D_+,\Db_+]\Db_-\F^\nb)\cr
 &\]6-2(D_+D_-\F^\m)(\Db_+\Db_-\F^\nb)
     +2(\2{\Db_+D_-\F^\m})(\2{D_+\Db_-\F^\nb})\big)M_{\nb\1\m}\cr
 &\]4+(D_-\F^\m)(\Db_-\F^\nb)
      \big[[D_+\Db_+(\FB_\LL+\X)-(\Db_+D_+\F)+[D_+,\Db_+]\L]M'_{\nb\1\m}\cr
 &\]{18}+[D_+(\F+\L),\Db_+(\FB_\LL+\X+\L)]M''_{\nb\1\m}\big]\cr
 &\]4+2\big[(D_+D_-\F^\m)(\Db_-\F^\nb)-(D_-\F^\m)(\2{D_+\Db_-\F^\nb})\big]
                                       [\Db_+(\FB_\LL+\X+\L)]M'_{\nb\1\m}\cr
 &\]4-2\big[(\2{\Db_+D_-\F^\m})(\Db_-\F^\nb)
                                        -(D_-\F^\m)(\Db_+\Db_-\F^\nb)\big]
 [D_+(\F+\L)]M'_{\nb\1\m}                     \Big\}\Big|~,~~~~ &\eXXX{c}\cr
 &=\inv8\Big\{-4(\vd_\mm\vd_\pp\f^\nb)M_\nb\cr
 &\]4+(2i\vd_\mm\f^\nb)
      \big[[2i\vd_\pp(\fb_\LL{+}x{-}\f)+4L_\pp]M'_\nb
      +4(\j_+{+}\l_+)(\jb_{\LL+}{+}\c_+{+}\bl_+)M''_\nb\big]\cr
 &\]4-8i(\vd_\mm\j^\nb_+)(\j_++\l_+)M'_\nb\cr
 &\]4+\big[2(-2i\vd_\pp\j^\m_-)(\j^\nb_-)+2(\j^\m_-)(2i\vd_\pp\j^\nb_-)
     -2(-2F^\m)(2F^\nb)\big]M_{\nb\1\m}\cr
 &\]4+2\j^\m_-\j^\nb_-
      \big[[2i\vd_\pp(\fb_\LL{+}x{-}\f)+4L_\pp]M'_{\nb\1\m}
      +4(\j_+{+}\l_+)(\jb_{\LL+}{+}\c_+{+}\bl_+)M''_{\nb\1\m}\big]\cr
 &\]4+4(-2F^\m)\j^\nb_-(\jb_{\LL+}{+}\c_+{+}\bl_+)M'_{\nb\1\m}
      +8\j^\m_-(F^\nb)(\j_+{+}\l_+)M'_{\nb\1\m}\Big\}~,         &\eXXX{d}\cr
}$$
We have immediately used that $(D_-\Db_-\FB)=(2i\vd_\mm\FB)$, and
underlined the terms that vanish because of the definitions $\Db_\pm\F=0$
and $D_\pm\FB=0$. Terms such as $[D_+(\F+\L)]M'_\nb$ are simply an
abbreviation for
\eqn\eXXX{ [D_+(\F+\L)]M'_\nb \to
           [(D_+\F^\r)M_{\nb\1\r}+(D_+\L^\ah)M_{\nb\1\ah}]~. }
Next, use that
\eqn\eXXX{ -4(\vd_\mm\vd_\pp\f^\nb)M_\nb=-4\vd_\pp[(\vd_\mm\f^\nb)M_\nb]
 +4(\vd_\mm\f^\nb)[\vd_\pp(\fb_\LL{+}x{+}\f)]M'_\nb~, }
and drop a few total derivative terms to obtain (after some
straightforward algebra):
\eqna\eXpl
 $$ \eqalignno{ \inv8[D_+,\Db_+]D_-(\Db_-\F^\nb)M_\nb\Zp\[{19}\cr
 &{\cong}\big[(\vd_\mm\f^\nb)(\vd_\pp\f^\m)+F^\m F^\nb
  +\frc i2(\j^\nb_+\dvd_\mm\j^\m_+)
  +\frc i2(\j^\nb_-\dvd_\pp\j^\m_-)\big]M_{\nb\1\m}             &\eXpl{a}\cr
 &+\big[(\vd_\mm\f^\nb)[\inv2(\vd_\pp\ell^\ah)+iL_\pp]
      +\frc i2(\j^\nb_+\dvd_\mm\l^\ah_+)\big]M_{\nb\1\ah}       &\eXpl{b}\cr
 &+i(\vd_\mm\f^\nb)\big[\j^\r_+(\jb^\mt_{\LL+}M_{\nb\1\mt\r}
                               +\c^\a_+M_{\nb\1\a\r}
                               +\bl^\ah_+M_{\nb\1\ah\r})        &\eXpl{c}\cr
 &\]8                  +\l^\bh_+(\jb^\mt_{\LL+}M_{\nb\1\mt\bh}
                                +\c^\a_+M_{\nb\1\a\bh}
                                +\bl^\ah_+M_{\nb\1\ah\bh})\big] &\eXpl{d}\cr
 &+\frc i2\j^\r_-\j^\nb_-\big[(\vd_\pp\fb^\mt_\LL)M_{\nb\1\mt\r}
                             +(\vd_\pp x^\a)M_{\nb\1\a\r}
                             -(\vd_\pp\f^\m)M_{\nb\1\m\r} \big]
  +\j^\r_-\j^\nb_-L^\ah_\pp M_{\nb\1\ah\r}~~~~~~~~~             &\eXpl{e}\cr
 &-F^\r\j^\nb_-(\jb^\mt_{\LL+}M_{\nb\1\mt\r}
               +\c^\a_+M_{\nb\1\a\r}
               +\bl^\ah_+M_{\nb\1\ah\r})
  +\j^\r_-F^\nb(\j^\m_+M_{\nb\1\m\r}
               +\l^\ah_+M_{\nb\1\ah\r})                         &\eXpl{f}\cr
 &+\j^\s_-\j^\nb_-\big[\j^\r_+(\jb^\mt_{\LL+}M_{\nb\1\mt\r\s}
                              +\c^\a_+M_{\nb\1\a\r\s}
                              +\bl^\ah_+M_{\nb\1\ah\r\s})       &\eXpl{g}\cr
 &\]7                 +\l^\bh_+(\jb^\mt_{\LL+}M_{\nb\1\mt\bh\s}
                              +\c^\a_+M_{\nb\1\a\bh\s}
                             +\bl^\ah_+M_{\nb\1\ah\bh\s})\big]~.&\eXpl{h}\cr
 }$$
Of course, the Lagrangian density must include also the hermitian conjugate
terms.

Besides the contribution to the `standard' D-terms, leading with the terms
in~\eXpl{a}, notice the mixed kinetic terms in~\eXpl{b}, the mixed 4-fermion
(Fermi or Thirring) terms in~\eXpl{g{-}h}, and the terms in~\eXpl{f} which
change the (still algebraic, \ie, non-dynamical) equations of motion for
$F,\Fb$. Upon ellimination of these, the supersymmetry transformation rules
become non-linear in the fermionic fields, as discussed in the last part of
\SS\,\ssPhDeg, and the terms in~\eXpl{f} modify this non-linearity.

 %
\vfill
\bigskip\noindent{\it Acknowledgments\/}:
I am indebted to S.James Gates, Jr., George Mini\'c and Joe Polchinski for
helpful and encouraging discussions, and to the generous support of the
Department of Energy through the grant DE-FG02-94ER-40854. Many thanks to
the Institute for Theoretical Physics at Santa Barbara, where this work was
completed, and the National Science Foundation for their support under Grant
No.~PHY94-07194.

\vfill\eject
\listrefs

 %
\bye